\documentclass[twocolumn]{aastex63}
\def\figureloc{./}
\begin{document}
\title{Star Formation Timescales of the Halo Populations from Asteroseismology and Chemical Abundances
\footnote{Based on data collected with the Subaru Telescope, which is operated by the National Astronomical Observatory of Japan.}
}
\author{Tadafumi Matsuno}
\affiliation{Kapteyn Astronomical Institute, University of Groningen, Landleven 12, 9747 AD Groningen, The Netherlands}
\email{matsuno@astro.rug.nl}
\affiliation{Department of Astronomical Science, School of Physical Sciences, SOKENDAI, The Graduate University for Advanced Studies, 2-21-1 Osawa, Mitaka, Tokyo 181-8588, Japan}
\affiliation{National Astronomical Observatory of Japan (NAOJ), 2-21-1 Osawa, Mitaka, Tokyo 181-8588, Japan}
\author{Wako Aoki}
\affiliation{Department of Astronomical Science, School of Physical Sciences, SOKENDAI, The Graduate University for Advanced Studies, 2-21-1 Osawa, Mitaka, Tokyo 181-8588, Japan}
\affiliation{National Astronomical Observatory of Japan (NAOJ), 2-21-1 Osawa, Mitaka, Tokyo 181-8588, Japan}
\author{Luca Casagrande}
\affiliation{Research School of Astronomy and Astrophysics, Australian National University, Canberra, ACT 2611, Australia}
\author{Miho N. Ishigaki}
\affiliation{Astronomical Institute, Tohoku University, Sendai 980-8578, Japan}
\author{Jianrong Shi}
\affiliation{Key Lab of Optical Astronomy, National Astronomical Observatories, Chinese Academy of Sciences, Beijing 100101, People's Republic of China}
\author{Masao Takata}
\affiliation{Department of Astronomy, School of Science, The University of Tokyo, 7-3-1 Hongo, Bunkyo-ku, Tokyo 113-0033, Japan}
\author{Maosheng Xiang}
\affiliation{Max-Planck Institute for Astronomy, K\"{o}nigstuhl 17, D-69117 Heidelberg, Germany}
\author{David Yong}
\affiliation{Research School of Astronomy and Astrophysics, Australian National University, Canberra, ACT 2611, Australia}
\author{Haining Li}
\affiliation{Key Lab of Optical Astronomy, National Astronomical Observatories, Chinese Academy of Sciences, Beijing 100101, People's Republic of China}
\author{Takuma Suda}
\affiliation{The Open University of Japan, Wakaba 2-11, Mihama-ku, Chiba 261-8586, Japan}
\affiliation{Research Center for the Early Universe, University of Tokyo, Hongo 7-3-1, Bunkyo-ku, Tokyo 113-0033, Japan}
\affiliation{Department of Liberal Arts, Tokyo University of Technology, Nishi Kamata 5-23-22, Ota-ku, Tokyo 144-8535, Japan}
\author{Qianfan Xing}
\affiliation{Key Lab of Optical Astronomy, National Astronomical Observatories, Chinese Academy of Sciences, Beijing 100101, People's Republic of China}
\author{Jingkun Zhao}
\affiliation{Key Lab of Optical Astronomy, National Astronomical Observatories, Chinese Academy of Sciences, Beijing 100101, People's Republic of China}

\begin{abstract}

We combine asteroseismology, optical high-resolution spectroscopy, and kinematic analysis for 26 halo red giant branch stars in the \textit{Kepler} field in the range of $-2.5<[\mathrm{{Fe}/{H}}]<-0.6$.
After applying theoretically motivated corrections to the seismic scaling relations, we obtain an average mass of $0.97\pm 0.03\,\mathrm{M_{\odot}}$ for our sample of halo stars.
Although this maps into an age of $\sim 7\,\mathrm{Gyr}$, significantly younger than independent age estimates of the Milky Way stellar halo, we considerer this apparently young age is due to the overestimation of stellar mass in the scaling relations.
There is no significant mass dispersion among lower red giant branch stars ($\log g>2$), which constrains a relative age dispersion to $<18\%$, corresponding to $<2\,\mathrm{Gyr}$.
The precise chemical abundances allow us to separate the stars with [{Fe}/{H}]$>-1.7$ into two [{Mg}/{Fe}] groups.
While [$\alpha$/{Fe}] and [{Eu}/{Mg}] ratios are different between the two subsamples, [$s$/Eu], where $s$ stands for Ba, La, Ce, and Nd, does not show a significant difference.  
These abundance ratios suggest that the chemical evolution of the low-Mg population is contributed by type~Ia supernovae, but not by low-to-intermediate mass asymptotic giant branch stars, providing a constraint on its star formation timescale as $100\,\mathrm{Myr}<\tau<300\,\mathrm{Myr}$. 
We also do not detect any significant mass difference between the two [{Mg}/{Fe}] groups, thus suggesting that their formation epochs are not separated by more than 1.5 Gyr. 

\end{abstract}
\section{Introduction}

The environment of the birth place of stars is encoded in their surface chemical composition and their orbital characteristics. 
The history of the Milky Way can potentially be recovered from such information obtained through extensive measurements of stellar chemical abundances \citep[e.g.,][]{Freeman2002a,Frebel2015} and stellar kinematics \citep[e.g.,][]{Helmi1999a} by astrometric and spectroscopic observations \citep[e.g., ][]{Nordstroem2004a,Steinmetz2006a,Yanny2009,Zhao2012,Perryman1997a,GaiaCollaboration2016a,Majewski2017,Martell2017a}.
However, measurements of stellar age, another fundamental parameter in the study of the Milky Way history, is challenging \citep{Soderblom2010a}.  
Although ages have been estimated from their locations on the Hertzsprung-Russell diagram (HRD) for nearby turn-off stars, the same method is not applicable to intrinsically more luminous red giant stars since their locations do not depend on the age. 
The lack of age estimates for distant stars prevents us from directly sketching the history of the Galaxy as a function of time.

The emergence of satellite missions that carried out long-term monitoring of stellar oscillations through their variation in luminosity with high-precision, such as \textit{Kepler} \citep{Koch2010a} and CoRoT \citep{Auvergne2009a}, has opened a new window to estimate the mass of a large number of red giant stars \citep{Kallinger2010a,Kallinger2010b,Hekker2011b,Stello2013a,Pinsonneault2014a,Mathur2016a,Yu2018a,Pinsonneault2018a}, which is otherwise almost inaccessible \citep[however see also ][]{Feuillet2016,Martig2016a,Ness2016a,Das2020a}.
Stars with a thick convective envelope show a particular type of oscillations (solar-like oscillations), which can be used to infer their structure including the mass \citep[Asteroseismology; ][]{Aerts2010a,Chaplin2013a}.
Having the mass of a red giant is of paramount importance in estimating its age since the time spent on the red giant branch is short compared to that on the main-sequence phase and since the mass is the fundamental parameter that controls the evolution of a star including its main-sequence lifetime, although mapping into an age scale is problematic since it depends on the adopted stellar models and assumed mass-loss \citep[e.g.,][]{Casagrande2016a}.
Since red giants are intrinsically luminous, the development of asteroseismology allows us to explore stellar ages over a wide volume in the Galaxy. 
These stellar ages obtained through asteroseismology have revealed the evolution of the Galactic disk \citep[e.g.,][]{Casagrande2016a,Anders2017a,SilvaAguirre2018a,Wu2018a,Wu2019a}.

However, application of asteroseismology to halo stars has been limited, despite the importance of determining their ages.
The Galactic halo contains rich information about the merging history of the Milky Way.
Most halo stars are affected by galaxy mergers; they are accreted onto the Milky Way after having formed in satellite dwarf galaxies, or heated from the Galactic disk at the time of major mergers \citep{Helmi1999b,Bullock2005a,McCarthy2012a,Jean-Baptiste2017a,Belokurov2019a}. 
For stars known to have originated from accretion or mergers, their ages constrain the epoch of those events \citep{Gallart2019a}.
The Gaia mission has provided major recent progress in our understanding of the Galactic halo \citep[e.g.,][]{Belokurov2018a,Helmi2018a} by enabling us comprehensive studies of stellar kinematics, which leads to the increasing demand for ages of halo stars.
Although the bulk of the stellar halo has known to be composed of old stars \citep[$\gtrsim 10\,\mathrm{Gyr}$ old; ][]{Iben1967a,Schuster2012}, investigation of age variation among them has been limited.

\citet{Nissen2010} showed that in a photometrically selected sample of nearby halo stars there are two major stellar populations among halo stars; one has high [{$\alpha$}/{Fe}] and the other has low [{$\alpha$}/{Fe}] abundance ratios; hereafter referred to as the high-$\alpha$ and low-$\alpha$ populations, respectively.
The two populations also differ in stellar kinematics and ages \citep{Schuster2012}, indicating different origins.
The low-$\alpha$ population exhibits younger age on average by 2--3 Gyr with a tendency towards being on a retrograde orbit, which is consistent with an accretion origin. 
This result is also confirmed by \citet{Hawkins2014a}, who use the color of $\alpha$-rich and $\alpha$-poor halo turn-off stars in the low-resolution spectroscopic survey, SEGUE/SDSS. 
The chemo-dynamical analysis of halo stars by \citet{Helmi2018a} and \citet{Mackereth2019a} revealed that this low-$\alpha$ population would correspond to stars from a relatively massive dwarf galaxy that was accreted onto the Milky Way \citep[Gaia Enceladus a.k.a. Gaia Sausage; ][]{Belokurov2018a}\footnote{Note that there is a discussion that the Gaia-Enceladus is not exactly the same as the Gaia Sausage \citep[e.g.][]{Evans2020a}.}. 
The other population with high-$\alpha$ abundance is suggested to be formed in the Milky Way (in-situ formation).
\citet{Belokurov2019a} proposed that this population would have formed in the ancient Milky Way disk and later been heated to the halo from chemo-dynamical analysis and age distribution of halo and disk stars. 

The age of halo stars has been examined in several studies for stars in globular clusters \citep[e.g., ][]{MarinFranch2009a,Forbes2010a,Dotter2011a} and for field stars \citep[e.g., ][]{Jofre2011a,Schuster2012,Hawkins2014a,Carollo2016a,Kilic2019a,Das2020a}. 
Although the ages of globular clusters have provided a wealth of information about the accretion history of the Milky Way \citep[e.g., ][]{Kruijssen2018a,Massari2019a}, there is no guarantee that they are the representative of the halo stellar population.  
Previous studies that utilized the combination of chemical abundances and stellar ages for field stars estimate age of turn-off stars from their locations on the HRD \citep{Schuster2012}.
However, they are intrinsically faint, so explorations have been limited to the solar neighbourhood. 
Considering the fact that the Galactic halo extends to a large distance, the combination of ages and abundances for luminous stars is highly desired.
\citet{Das2020a} approached this problem by estimating ages from infrared photometry and Gaia parallax measurements using a Bayesian framework, showing the prolonged star formation history of the Gaia Enceladus. 
However, since the color and luminosity of a red giant are not very sensitive to its age, their method requires a large enough sample for statistical treatment and a good prior knowledge about the age distribution of the population for the Bayesian approach. 
Asteroseismology, on the other hand, directly provides stellar mass with high precision, which enables discussion based on a smaller number of stars.

There have been two main challenges in the application of asteroseismology to halo stars.
The first problem is that asteroseismic mass estimates at low metallicity have not been tested or calibrated well compared to solar-metallicity stars.
In asteroseismology, mass estimates rely on the following two scaling relations:
\begin{eqnarray}
\Delta\nu &\propto \sqrt{\bar\rho} \propto M^{0.5}R^{-1.5} \label{eq:delnu}\\
\nu_{\rm max} &\propto g/\sqrt{T_{\rm eff}} \propto MR^{-2}T_{\rm eff}^{-0.5}\label{eq:numax},
\end{eqnarray}
where $\Delta\nu$ is the mean large frequency separation between adjacent radial modes, $\nu_{\rm max}$ is the frequency at which the oscillation amplitude becomes maximum, and $g$, $\bar\rho$, $M$, $R$, and $T_{\rm eff}$ are surface gravity, mean density, mass, radius, and effective temperature, respectively.
We immediately obtain $M\propto \nu_{\rm max}^3\Delta\nu^{-4}T_{\rm eff}^{1.5}$ from these two scaling relations, which is used for mass estimates. 
The scaling relations, Eq. \ref{eq:delnu} and Eq. \ref{eq:numax}, have been tested extensively for a large number of red giants \citep{Huber2011a,Huber2012a,Baines2014a,Brogaard2016a,Gaulme2016a,Themesl2018a,Hekker2020a}.
We note that \citet{Brogaard2016a}, \citet{Gaulme2016a}, and \citet{Themesl2018a} compared radii and masses for eclipsing binaries and reported small but significant offsets. 
The majority of the red giants used for the testing are, however, solar metallicity stars.
There is no guarantee that the scaling relations work similarly for low metallicity stars, and, in fact, a number of observational and theoretical studies have suggested the necessity of applying corrections.

Observationally, \citet{Epstein2014a} pointed out that simple use of the scaling relations tends to over-estimate masses for low metallicity stars based on nine low metallicity stars ([{Fe}/{H}]$<-1.0$) in the Apache Point Observatory Galactic Evolution Experiment (APOGEE) of the Sloan Digital Sky Survey (SDSS).
\citet{Casey2018a} also reached the same conclusion using their photometric selection of metal-poor stars (three stars).
These studies motivated theoretical efforts to make corrections to the scaling relations to obtain accurate masses for low-metallicity stars \citep[e.g.,][]{Sharma2016a,Guggenberger2016a}, which lead to lower mass estimates for stars in \citet{Epstein2014a}.
\citet{Miglio2016a} and \citet{Valentini2019a} analysed metal-poor stars observed by the K2 mission \citep{Howell2014a} in the globular cluster M4 (eight stars) and those in the field \citep[four $+$ nine stars from ][]{Epstein2014a}, respectively. 
Their mass estimates are based on a theoretical correction to the Eq. \ref{eq:delnu} and in good agreement with independent estimates, suggesting that theoretical $\Delta \nu$ estimates work well.

The other difficulty is in finding halo stars with asteroseismic information.
Since asteroseismology requires high quality continuous light curves over a long period, dedicated observations by space telescopes are desired\footnote{There are studies that observe stellar oscillations by radial velocity monitoring. However, space photometric observations have a clear advantage in terms of the precision of measurements of $\Delta\nu$ and $\nu_{\rm max}$, the multiplicity of targets and the depth of observations.}.
The \textit{Kepler} mission has observed the same field near the Galactic plane (so-called \textit{Kepler} field) over about four years and obtained the highest quality light curves that are suitable for asteroseismology for about 16000 stars \citep{Yu2018a}.
Because of the low galactic latitude, most of the stars in the \textit{Kepler} field are disk stars, while the fraction of halo stars increases at faint magnitudes \citep{Mathur2016a}.
Therefore, we need a way to efficiently select halo stars for follow-up high-resolution spectroscopy.

Halo stars are characterized by large relative velocity to the Sun and low metallicity. 
Therefore, radial velocity and metallicity measurements by spectroscopic surveys provide an efficient way to select halo stars. 
Fortunately, spectroscopic surveys such as the Large Sky Area Multi-Object Fibre Spectroscopic Telescope (LAMOST) and APOGEE have extensively observed stars in the \textit{Kepler} field and have provided radial velocity and metallicity measurements \citep{Pinsonneault2014a,Pinsonneault2018a,DeCat2015a,Ren2016a,Zong2018a}.

In this study, we obtain high-$S/N$, high-resolution spectra for 26 halo star candidates with asteroseismic information ($S/N\sim100-200$ and $R>60,000$).
The spectra provide us with precise measurements of stellar parameters including effective temperature, which is needed for asteroseismic mass estimates. 
Chemical abundances are also measured with high-precision to investigate the two separate halo populations (high-$\alpha$ and low-$\alpha$), and to investigate possible correlation between abundances and stellar masses, such as those seen among solar metallicity stars \citep[e.g.,][]{Nissen2015a,Spina2018a,Bedell2018a,DelgadoMena2019a,Feuillet2018a,Buder2019a}.
Compared to previous studies of halo stars with asteroseismology, our study can be characterized by the sample size and the high-precision stellar parameters and chemical abundances.
All of the stars in our sample are in the \textit{Kepler} field, and hence precise oscillation frequency measurements are available from about the four year observation \citep{Yu2018a}.

Apart from the formation history of the Galaxy, asteroseismic information of halo stars would constrain formation mechanism of stars with peculiar abundances.
High-resolution spectroscopic surveys of low-metallicity stars have revealed metal-poor stars whose abundance ratio is very different from solar composition (e.g., carbon-enhanced metal-poor stars; \cite{Beers2005}, r-process enhanced stars; \cite{Sneden1994a}, Li-rich stars; \cite{Li2018a}). 
Chemical peculiarity can be caused by mass accretion from a binary companion, internal nucleosynthesis, or inhomogeneity in the early Universe. 
Asteroseismic constraints on mass and evolutionary status would constrain the detail of these processes and provides a clue to the origin of stars whose formation mechanism is not yet well understood.

We start this paper with target selection and observation in Section \ref{sec:observation}, which is followed by the use of asteroseismic data in Section \ref{sec:seismology}, kinematic analysis in Section \ref{sec:kinematics}, and abundance analysis in Section \ref{sec:abundance}.
We discuss the results in Section \ref{sec:discussion} combining these three methods.
Summary and future prospects are provided in Section \ref{sec:summary}.

\section{Observation}\label{sec:observation}

\subsection{Target selection}

\begin{figure}
  \plotone{\figureloc 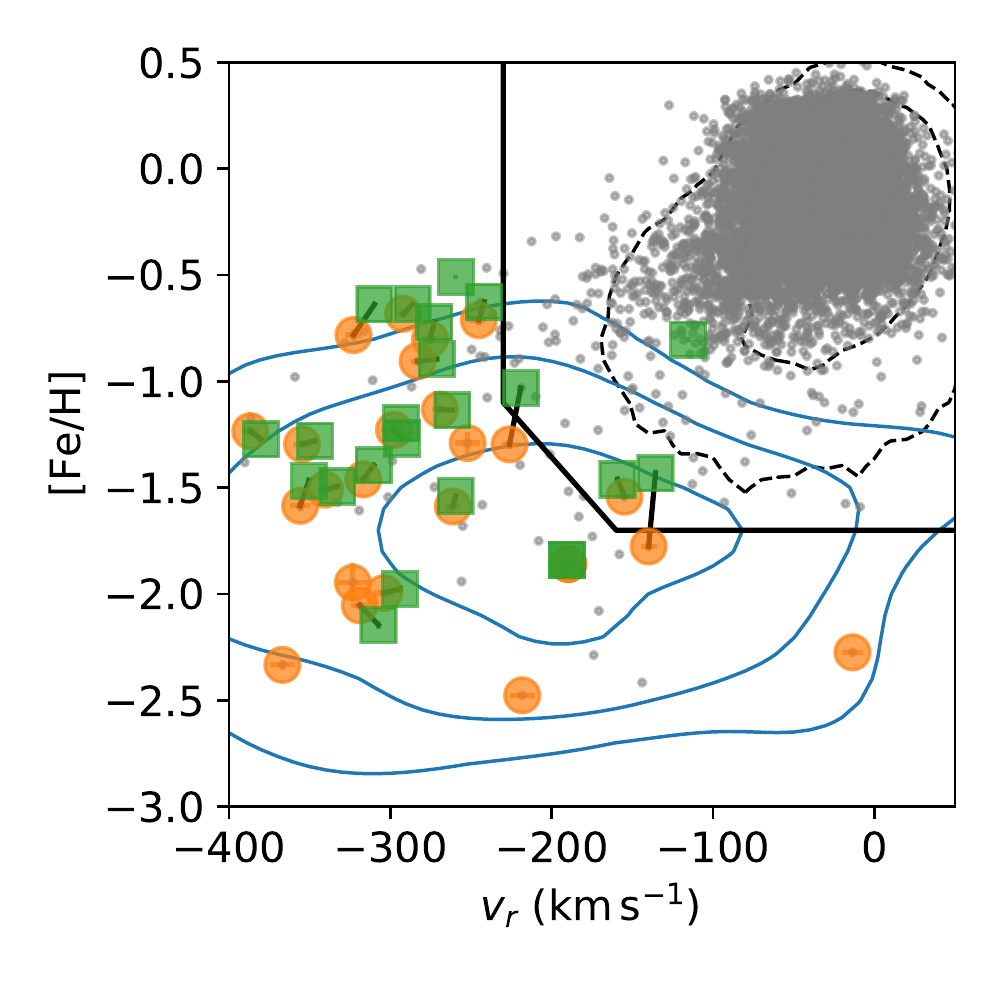}
  \caption{Target selection in the radial velocity -- [{Fe}/{H}] plane. The contours are disk (black dashed) and halo (blue solid) stars from the Besan\c{c}on model (each contour contains 39.3, 86.5, and 98.9 \% of the population.). Stars outside of the black solid line are primarily selected as halo star candidates. Targets are shown with orange circles (LAMOST DR5) and green squares (APOKASC2). If a target is both in LAMOST DR5 and APOKASC2, two points are connected by a black line. Grey points are all the stars that are both in \citet{Yu2018a} and in LAMOST DR5. Note that while LAMOST DR4 was used at the time of selection, here we use updated measurements in LAMOST DR5.  \label{fig:selection}}
\end{figure}

Targets are selected from LAMOST DR4 and APOKASC2 catalogs based on radial velocity ($V_r$) and metallicity measurements (Figure \ref{fig:selection}). 
To make sure all the targets have frequency measurements, we only select stars in the catalog of asteroseismic analysis of $\sim$16,000 red giants by \citet{Yu2018a}.
The following selection criteria are invoked to minimize the contamination of disk stars based on the distributions of disk and halo stars in a mock catalog of \textit{Kepler} field generated from the Besan\c{c}on galaxy model \citep{Robin2003a}. 
The boundary used of the selection (the black solid line in Figure~\ref{fig:selection}) connects $V_r=-230\,\mathrm{km\,s^{-1}}$ for $[\mathrm{Fe/H}]>-1.1$, $(V_r,[\mathrm{Fe/H}])=(-230,\,-1.1),\,(-160,\,-1.7)$, and $[\mathrm{Fe/H}]=-1.7$ for $V_r>-170\,\mathrm{km\,s^{-1}}$.

There are two stars that do not exactly satisfy the criteria presented above.
The more extreme one is KIC5858947 and the other is KIC9696716.
These stars are observed since the observing condition does not allow us to study fainter stars that satisfy the criteria.
While APOGEE measurements of other two stars (KIC5184073 and KIC8350894) do not satisfy the criteria, these stars are selected based on the LAMOST measurements.
The radial velocity of KIC7693833 is very different from other halo stars and more similar to disk stars; however, this star is kept in the sample considering its low metallicity.

\subsection{Observations}

\begin{deluxetable*}{lrrrrrrrrrrr}
  \tablecaption{Observation log \label{tab:obs}}
  \tablehead{ 
    \colhead{Object} & \colhead{Date\tablenotemark{a}} & \colhead{Exposure} & \colhead{$S/N$\tablenotemark{b}} & \colhead{$V_r$ (HDS)}   & \colhead{$V_{r, L}$\tablenotemark{c}}    & \colhead{[Fe/H]$_{\rm L}$\tablenotemark{c}} & \colhead{$V_{r, A}$\tablenotemark{d}}    & \colhead{[Fe/H]$_{\rm A}$\tablenotemark{d}} & \colhead{$V_{r, G}$\tablenotemark{e}} & \colhead{$\sigma(V_{r, G})$\tablenotemark{e}} & \colhead{$M_{G}$\tablenotemark{e}}\\
                     &                             & \colhead{(s)}        &                                  & \colhead{(km s$^{-1}$)} & \colhead{(km s$^{-1}$)} &                            & \colhead{(km s$^{-1}$)} &  & \colhead{(km s$^{-1}$)} & \colhead{(km s$^{-1}$)} & (mag)                   \\
            }
  \startdata
KIC5184073 &   July 11, 2018 &        5400 &  107.0 & -136.21 &    -139.92 &       -1.78 &     -135.65 &    -1.43 &          \nodata &                \nodata &            13.21  \\
KIC5439372 &   July 10, 2018 &        1200 &  135.0 & -213.57 &    -218.27 &       -2.48 &     -211.94 & \nodata  &          -212.89 &                   0.58 &            11.82  \\
KIC5446927 &  August 4, 2017 &        1200 &  115.0 & -272.89 &    -275.41 &       -0.80 &     -272.76 &    -0.72 &          -272.90 &                   0.39 &            11.71  \\
KIC5698156 &July 9\&11, 2018 &        1200 &  246.0 & -380.25 &    -386.80 &       -1.23 &     -380.12 &    -1.27 &          -379.28 &                   0.72 &            10.45  \\
KIC5858947 &   July 10, 2018 &        1800 &  135.0 & -104.35 &    \nodata &     \nodata &     -115.43 &    -0.80 &          -106.91 &                   2.09 &            11.72  \\
KIC5953450 &   July 11, 2018 &        2400 &  265.0 & -286.53 &    -292.09 &       -0.68 &     -286.19 &    -0.64 &          -284.63 &                   0.82 &            12.67  \\
KIC6279038 &    July 9, 2018 &        2200 &  118.0 & -308.10 &    -318.93 &       -2.05 &     -307.36 &    -2.15 &          -308.15 &                   0.75 &            12.43  \\
KIC6520576 &   July 11, 2018 &        5400 &  171.0 & -365.09 &    -366.93 &       -2.33 &     \nodata &  \nodata &          \nodata &                \nodata &            13.18  \\
KIC6611219 &   July 11, 2018 &        1800 &  113.0 & -293.95 &    \nodata &     \nodata &     -293.35 &    -1.20 &          -293.47 &                   0.72 &            12.06  \\
KIC7191496 &  August 2, 2017 &        1200 &  167.0 & -290.55 &    -303.87 &       -2.00 &     -294.03 &    -1.98 &          -297.34 &                   1.58 &            11.82  \\
KIC7693833 &  August 2, 2017 &        1200 &  179.0 &   -7.10 &     -13.53 &       -2.27 &       -7.02 & \nodata  &            -6.69 &                   0.44 &            11.74  \\
KIC7948268 &    July 9, 2018 &        2200 &  155.0 & -292.58 &    -297.72 &       -1.23 &     -292.76 &    -1.27 &          -290.87 &                   1.79 &            12.44  \\
KIC8350894 &   July 11, 2018 &        3600 &  198.0 & -219.94 &    -226.13 &       -1.30 &     -219.21 &    -1.03 &          -219.83 &                   1.61 &            12.76  \\
KIC9335536 &   July 10, 2018 &        5400 &  114.0 & -350.52 &    -355.91 &       -1.59 &     -350.49 &    -1.47 &          -351.49 &                   1.21 &            12.97  \\
KIC9339711 &    July 9, 2018 &        1800 &  178.0 & -332.81 &    -340.41 &       -1.51 &     -332.85 &    -1.49 &          -332.68 &                   0.56 &            12.08  \\
KIC9583607 &   July 10, 2018 &        2400 &  169.0 & -309.74 &    -322.84 &       -0.78 &     -309.81 &    -0.64 &          -309.86 &                   0.37 &            11.46  \\
KIC9696716 &    July 9, 2018 &        1200 &  171.0 & -146.71 &    -154.83 &       -1.54 &     -159.20 &    -1.46 &          -148.23 &                   4.28 &            11.73  \\
KIC10083815 &  August 4, 2017 &        1800 &  136.0 & -270.98 &    -283.10 &       -0.90 &     -271.17 &    -0.90 &          -270.22 &                   1.08 &            12.26  \\
KIC10096113 &   July 11, 2018 &        4800 &  160.0 & -241.44 &    -245.16 &       -0.71 &     -241.83 &    -0.63 &          -240.37 &                   2.64 &            13.21  \\
KIC10328894 &    July 9, 2018 &        3600 &  188.0 & -315.65 &    -323.23 &       -1.95 &     \nodata &  \nodata &          -316.37 &                   1.89 &            13.02  \\
KIC10460723 &  August 2, 2017 &        1800 &  152.0 & -346.84 &    -354.70 &       -1.29 &     -346.67 &    -1.28 &          -347.34 &                   0.62 &            12.32  \\
KIC10737052 &   July 11, 2018 &        5400 &  200.0 & -242.77 &    -252.10 &       -1.29 &     \nodata &  \nodata &          -237.99 &                   1.04 &            13.20  \\
KIC10992126 &   July 10, 2018 &        1200 &  142.0 & -259.55 &    \nodata &     \nodata &     -259.38 &    -0.51 &          -259.84 &                   0.41 &            10.85  \\
KIC11563791 &   July 11, 2018 &         600 &  164.0 & -262.45 &    -269.27 &       -1.13 &     -261.56 &    -1.14 &          -262.20 &                   0.46 &            10.95  \\
KIC11566038 &   July 10, 2018 &        1800 &  122.0 & -310.67 &    -316.79 &       -1.46 &     -310.34 &    -1.40 &          -309.94 &                   2.11 &            12.04  \\
KIC12017985 &  August 2, 2017 &         600 &  225.0 & -190.24 &    -189.72 &       -1.86 &     -190.52 &    -1.84 &          -184.34 &                   2.03 &            10.14  \\
            &    July 9, 2018 &         600 &  202.0 &         &            &             &             &          &                  &                        &                   \\
KIC12253381 &    July 9, 2018 &        2200 &  121.0 & -272.44 &    -261.25 &       -1.59 &     -259.55 &    -1.54 &          -261.64 &                   1.20 &            12.51  
  \enddata 
  \tablenotetext{a}{Observations in 2017 are conducted with $R\sim 60,000$, while those in 2018 are with $R\sim 80,000$.}
  \tablenotetext{b}{$S/N$ are measured from continuum around 5765 $\mathrm{\AA}$ per 0.024 $\mathrm{\AA}$ pixel.}
  \tablenotetext{c}{From LAMOST DR5 catalog.}
  \tablenotetext{d}{From APOGEE DR16 catalog.}
  \tablenotetext{e}{From Gaia DR2 catalog.}
\end{deluxetable*}

Observations were conducted with the High Dispersion Spectrograph \citep[HDS; ][]{Noguchi2002} on the Subaru Telescope.
Most of the targets were observed in July 2018, during which we occasionally had thin clouds.
Spectra were taken with a standard setup of HDS with 2$\times$2 CCD binning which covers from $\sim 4000\,\mathrm{\AA}$ to $\sim 6800\,\mathrm{\AA}$ (StdYd).
Exposures taking longer than 40 minutes are split into shorter exposures, each of which typically takes 20--30 minutes. 
Image slicer \#2 was used to achieve high signal-to-noise ratio ($S/N\gtrsim100$) with $R\sim 80,000$ \citep{Tajitsu2012a}.
A subset of spectra were taken in 2017 as a back-up program for other proposals when the sky was covered with thin clouds.
These spectra were taken with the same wavelength coverage but with $R\sim 60,000$ without the image slicer.
KIC12017985 was observed both in 2017 and in 2018 to confirm the consistency of our analysis.
We mainly adopt parameters from the 2018 observation in figures for this star, and denote the 2017 observation as KIC 12017985--17.
We note that KIC10992126 was observed but not included in the following analysis, since it turned out to have large uncertainties in the measured frequencies.

Spectra were reduced with an IRAF\footnote{IRAF is distributed by the National Optical Astronomy Observatory, which is operated by the Association of Universities for Research in Astronomy (AURA) under a cooperative agreement with the National Science Foundation.} script, \texttt{hdsql}\footnote{\url{https://www.subarutelescope.org/Observing/Instruments/HDS/hdsql-e.html}} including linearity correction \citep{Tajitsu2010a}, cosmic ray rejection, scattered light subtraction, flat fielding, aperture extraction, wavelength calibration using ThAr lamp, heliocentric velocity correction, and continuum placements.
Spectra are combined just after the heliocentric radial velocity correction by simply adding fluxes in each frame. 
The $S/N$ was estimated from a continuum region around 5765 $\mathrm{\AA}$ and the radial velocity was estimated from wavelengths of iron lines.
Details of the observation, as well as radial velocity and metallicity measurements in surveys, are shown in Table \ref{tab:obs}.
We note that Table~\ref{tab:obs} shows single values for radial velocities and metallicities of individual objects for each of LAMOST and APOGEE. We adopt measurements from the highest $S/N$ spectrum of each star for LAMOST values and those from the combined spectra for APOGEE values. 

\begin{figure*}
\plotone{\figureloc 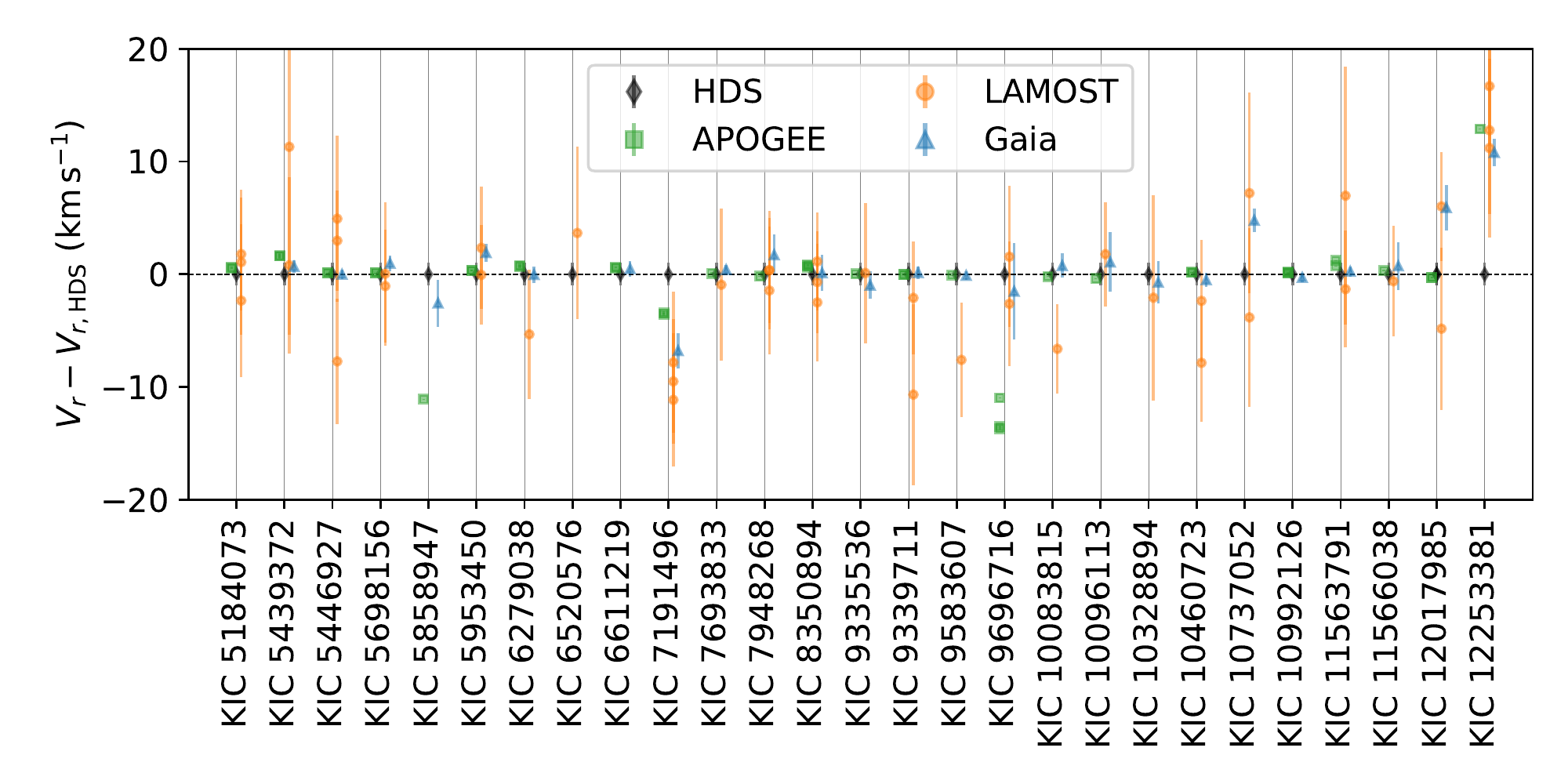}
\caption{
Comparison of measured radial velocities relative to the HDS measurements. \label{fig:rv}}
\end{figure*}  

The uncertainty in radial velocity measurements is estimated to be $1\,\mathrm{km\,s^{-1}}$ considering the stability of the instrument.
The measured radial velocities are compared with previous measurements in Figure~\ref{fig:rv}, where all the radial velocity measurements in surveys are plotted.
KIC5439372, KIC5858947, KIC7191496, KIC9696716, and KIC12253381 show signatures of radial velocity variation between APOGEE data and our observation, suggesting the possibility of the existence of binary companions.  
Most of these stars show large \texttt{radial\_velocity\_error} in Gaia DR2, which is shown as $\sigma(V_{r,G})$ in Table~\ref{tab:obs}, for their magnitude, which supports the likelihood of radial velocity variation. 
In addition, KIC10737052 shows large offsets between our observation and Gaia measurements, which suggests binarity of these objects\footnote{We note that this might also be the case for KIC12017985. However, since all the observations other than the Gaia, including our two observations and two APOGEE measurements, report consistent radial velocities, we do not conclude that this star is likely in a binary system.}.

\section{Asteroseismology}\label{sec:seismology}

\begin{deluxetable*}{lrrrrrrrrrr}
  \tablecaption{Asteroseismic parameters, mass, and radius \label{tab:seis}}
  \tablehead{ \colhead{Object} & \colhead{$\nu{\rm max}$\tablenotemark{a}} & \colhead{$\sigma_{(\nu_{\rm max})}$\tablenotemark{a}} & \colhead{$\Delta \nu$\tablenotemark{a}} & \colhead{$\sigma_{(\Delta \nu)}$\tablenotemark{a}} & \colhead{Evo. stage\tablenotemark{a}} & \colhead{$M$\tablenotemark{b}}                    & \colhead{$\sigma_{(M)}$\tablenotemark{b}}              & \colhead{$R$\tablenotemark{b}}                    & \colhead{$\sigma_{(R)}$\tablenotemark{b}}            & \colhead{$M_{\rm sc}$\tablenotemark{c}}           \\
                               & \colhead{($\mu$Hz)}      & \colhead{($\mu$Hz)}               & \colhead{($\mu$Hz)}    & \colhead{($\mu$Hz)}            &                               & \colhead{($\mathrm{M_{\odot}}$)} & \colhead{($\mathrm{M_{\odot}}$)}   & \colhead{($\mathrm{R_{\odot}}$)} & \colhead{($\mathrm{R_{\odot}}$)} & \colhead{($\mathrm{M_{\odot}}$)} 
  }
  \startdata
KIC5184073 &    9.25 &     0.22 &   1.659 &    0.031 &      RGB &  0.78 &   0.08 &  16.77 &     0.69 &    0.92 \\
KIC5439372 &    6.39 &     0.22 &   1.186 &    0.054 &      RGB &  0.98 &   0.23 &  22.53 &     2.47 &    1.17 \\
KIC5446927 &   21.88 &     0.38 &   2.874 &    0.048 &       RC &  1.49 &   0.13 &  14.97 &     0.58 &    1.44 \\
KIC5698156 &    9.73 &     0.31 &   1.677 &    0.031 &      RGB &  0.76 &   0.11 &  16.38 &     0.94 &    0.96 \\
KIC5858947 &  168.93 &     0.89 &  14.533 &    0.019 &      RGB &  0.98 &   0.05 &   4.36 &     0.09 &    1.01 \\
KIC5953450 &  140.87 &     0.83 &  12.715 &    0.021 &      RGB &  0.99 &   0.05 &   4.81 &     0.10 &    1.01 \\
KIC6279038 &    5.59 &     0.28 &   1.018 &    0.047 &      RGB &  1.18 &   0.34 &  26.50 &     3.29 &    1.42 \\
KIC6520576 &   17.91 &     0.50 &   2.693 &    0.016 &  unknown &  0.94 &   0.11 &  13.19 &     0.64 &    0.99 \\
KIC6611219 &    6.96 &     0.21 &   1.330 &    0.028 &      RGB &  0.70 &   0.10 &  18.61 &     1.12 &    0.89 \\
KIC7191496 &   16.23 &     0.24 &   2.455 &    0.021 &      RGB &  0.88 &   0.06 &  13.50 &     0.35 &    1.04 \\
KIC7693833 &   31.73 &     0.32 &   4.046 &    0.014 &      RGB &  1.03 &   0.05 &  10.35 &     0.18 &    1.12 \\
KIC7948268 &  120.45 &     0.82 &  11.450 &    0.015 &      RGB &  0.93 &   0.04 &   5.03 &     0.08 &    0.97 \\
KIC8350894 &   12.69 &     0.29 &   2.005 &    0.024 &      RGB &  0.90 &   0.08 &  15.47 &     0.54 &    1.08 \\
KIC9335536 &   11.08 &     0.82 &   1.861 &    0.095 &      RGB &  0.85 &   0.28 &  15.84 &     2.10 &    1.02 \\
KIC9339711 &   20.51 &     0.31 &   2.825 &    0.019 &      RGB &  1.05 &   0.07 &  13.05 &     0.36 &    1.21 \\
KIC9583607 &   25.25 &     0.51 &   3.816 &    0.055 &       RC &  0.76 &   0.05 &   9.93 &     0.25 &    0.70 \\
KIC9696716 &   24.09 &     0.53 &   3.305 &    0.022 &      RGB &  0.92 &   0.07 &  11.30 &     0.33 &    1.05 \\
KIC10083815 &   17.99 &     0.38 &   2.605 &    0.015 &      RGB &  0.91 &   0.07 &  13.06 &     0.39 &    1.08 \\
KIC10096113 &   36.31 &     0.59 &   4.168 &    0.042 &       RC &  1.44 &   0.10 &  11.50 &     0.32 &    1.42 \\
KIC10328894 &   30.72 &     0.40 &   3.965 &    0.016 &      RGB &  0.96 &   0.05 &  10.19 &     0.20 &    1.07 \\
KIC10460723 &   22.97 &     0.45 &   3.146 &    0.013 &       RC &  1.08 &   0.07 &  12.56 &     0.28 &    1.10 \\
KIC10737052 &   26.57 &     0.40 &   3.518 &    0.017 &       RC &  1.10 &   0.05 &  11.72 &     0.18 &    1.11 \\
KIC11563791 &   43.03 &     0.51 &   5.005 &    0.019 &      RGB &  1.01 &   0.05 &   8.85 &     0.19 &    1.15 \\
KIC11566038 &   31.36 &     0.32 &   3.954 &    0.023 &      RGB &  1.03 &   0.05 &  10.44 &     0.23 &    1.15 \\
KIC12017985 &   18.24 &     0.29 &   2.620 &    0.018 &      RGB &  1.00 &   0.07 &  13.54 &     0.36 &    1.15 \\
KIC12017985-17&  \nodata&   \nodata& \nodata & \nodata  &  \nodata &  0.99 &   0.07 &  13.48 &     0.36 &    1.15 \\
KIC12253381 &   22.00 &     0.40 &   3.032 &    0.014 &      RGB &  0.96 &   0.06 &  12.07 &     0.31 &    1.12 \\
  \enddata
  \tablenotetext{a}{From \citet{Yu2018a}.}
  \tablenotetext{b}{Obtained from the scaling relations with correction.}
  \tablenotetext{c}{Obtained from the scaling relations without correction.}
\end{deluxetable*}

All the targets are in \citet{Yu2018a} catalog, which provides results of asteroseismic analysis of red giant stars using about four years of photometric data in the \textit{Kepler} field using the SYD asteroseismic pipeline \citep{Huber2009a}.
The long baseline enabled them to precisely measure $\nu_{\rm max}$ and $\Delta\nu$ and to utilize the evolutionary status in literatures.
Although they also derived mass and radius, their stellar parameters are based on a collection of values from various literature sources.
Therefore, those mass and radius estimates need to be revised using updated stellar parameters from our analysis of high-resolution spectra.

We derive the mass and radius using \texttt{Asfgrid} \citep{Sharma2016a}, which includes a correction to the $\Delta\nu$ scaling relation taking evolutionary status into account.
While $\nu_{\rm max}$ and $\Delta\nu$ are taken from \citet{Yu2018a}, the other necessary input parameters, $T_{\rm eff}$ and [Fe/H], are taken from our spectroscopic measurements described in Section \ref{sec:abundance}. 
For the solar values, we adopt $\nu_{\rm max}=3090\,\mathrm{\mu Hz}$, $\Delta\nu=135.1\,\mathrm{\mu Hz}$, and $T_{\rm eff}=5777\,\mathrm{K}$ \citep{Huber2011a}.

In Table \ref{tab:seis}, we also provide masses obtained from the simple scaling relations without any correction.
The magnitude of the correction can be important (as large as $0.24\,\mathrm{M_{\odot}}$, corresponding to $20\%$; see the values of KIC6279038 in Table~\ref{tab:seis}).

The upper mass limit of metal-poor main-sequence stars has been estimated to be $\sim0.8\,\mathrm{M_{\odot}}$ \citep[e.g.,][]{Iben1967a,Melendez2010,VandenBerg2014a}.
We expect similar masses for our sample, since the timescale of the evolution after the turn-off is very short.
Two stars are obviously more massive than others (Table~\ref{tab:seis}), which would be treated as outliers (KIC5446927 and KIC10096113; separately discussed in Section~\ref{sec:massive}). 
The weighted average of the masses excluding the two outliers is $0.96\pm 0.02\,\mathrm{M_{\odot}}$ ($\sigma=0.10\,\mathrm{M_{\odot}}$).
Although this value is closer to the expected mass than the mass obtained from the scaling relations without a correction ($1.04\pm 0.02\,\mathrm{M_{\odot}}$, $\sigma=0.10$), it is still significantly higher than the canonical $0.8\,\mathrm{M_{\odot}}$ expected for old halo stars.
More discussion is presented in Section~\ref{sec:dis_seis}.

\begin{figure}
  \plotone{\figureloc 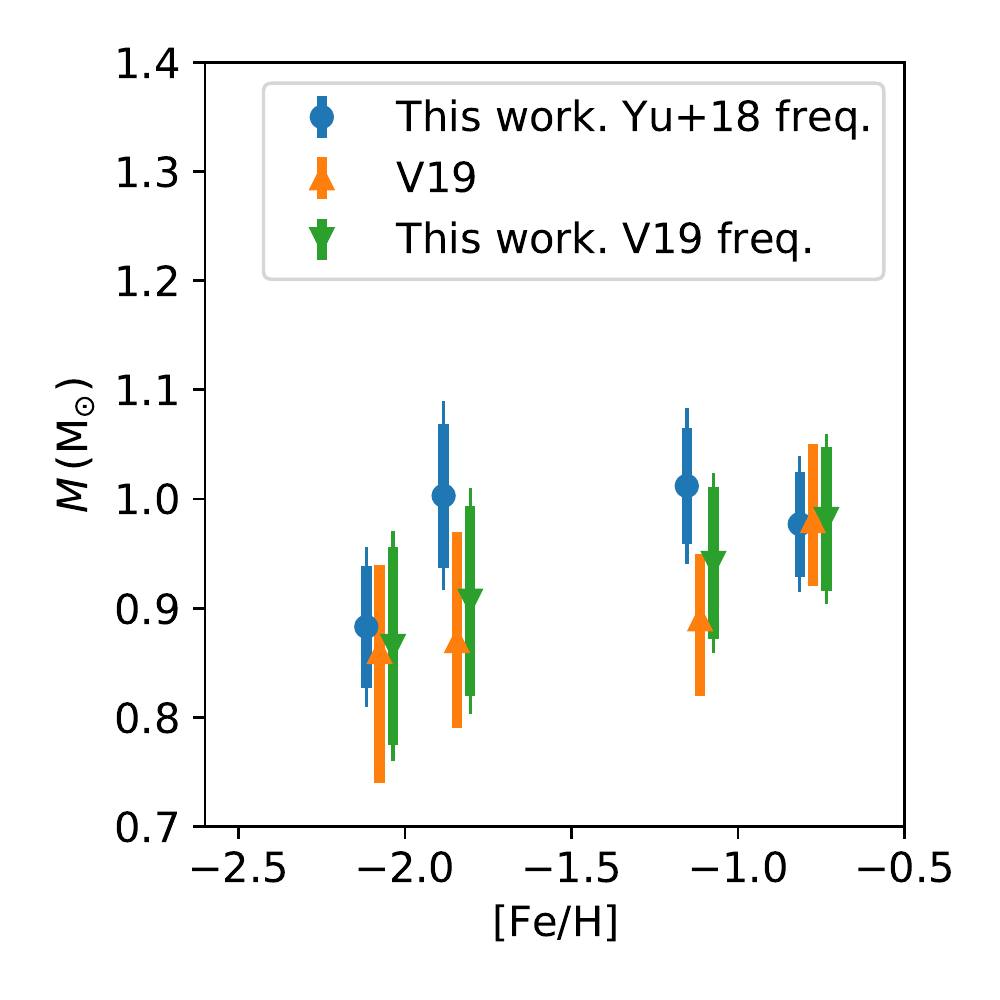}
  \caption{Comparison of derived masses with \citet{Valentini2019a} for four stars (KIC7191496, KIC12017985, KIC11563791, and KIC5858947, from left to right). Each star has three data points: (1) the mass derived in this work, (2) the mass derived in \citet{Valentini2019a}, and (3) the mass derived with our method but with $\nu_{\rm max}$ and $\Delta \nu$ values provided in \citet{Valentini2019a}.The effect of 1\% systematic uncertainties in $\Delta\nu$ and $\nu_{\rm max}$ measurements are also indicated with thin error bars for the (1) and (3) cases. Data points are horizontally shifted for visualization purposes. \label{fig:mass_comp}}
\end{figure}

\begin{figure}
  \plotone{\figureloc 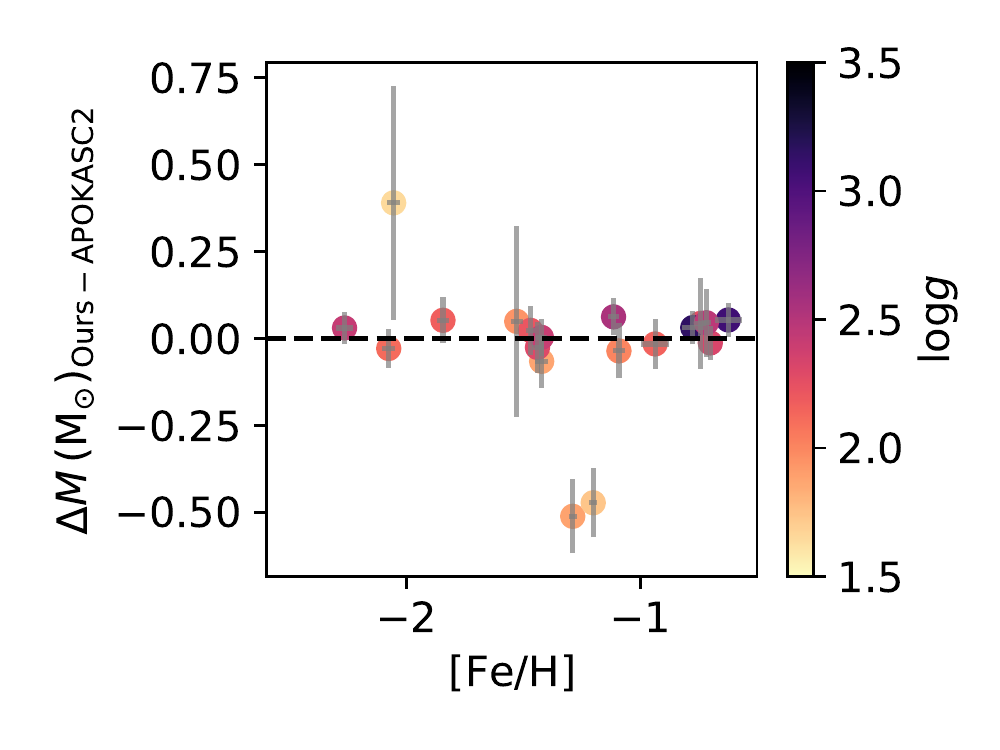}
  \caption{Comparison of derived masses with the APOKASC2 catalog \citep{Pinsonneault2018a} for common stars. Note that the errorbars reflect only our measurement uncertainties. The median uncertainties provided in the APOKASC2 catalog are 7.2\% and 4.3\% for random and systematic uncertainties, respectively.\label{fig:mass_comp_apokasc}}
\end{figure}

Four stars in our sample are common with \citet{Epstein2014a}, which are re-analysed by \citet{Valentini2019a}, who use frequencies derived with the COR asteroseismic pipeline \citep{Mosser2009a} and a Bayesian approach for mass (and age) estimation.
We compare our derived masses with \citet{Valentini2019a} for the four stars in Figure~\ref{fig:mass_comp}.
There is a good agreement between our study and \citet{Valentini2019a}.
Figure~\ref{fig:mass_comp} also includes masses that are derived using the same procedure as in this study but with frequencies adopted in \citet{Valentini2019a}.
The agreement for KIC12017985 and KIC11563791 becomes better, indicating the small difference would be due to the use of different pipelines for frequency analysis.
Since systematic offsets in measured frequencies should not depend on stellar metallicity, further discussions on the effect of asteroseismic pipelines are beyond the scope of this study. 
Such discussion is provided in \citet{Pinsonneault2018a}.
We note that they quantified the systematic $\nu_{\rm max}$ or $\Delta \nu$ differences between COR and SYD pipelines as $\sim 1$\% at most.
The effect of these systematic uncertainties are also indicated in Figure~\ref{fig:mass_comp}.

Figure~\ref{fig:mass_comp_apokasc} compares our mass estimates with those in the APOKASC2 catalog \citep{Pinsonneault2018a}.
There is a good agreement between the two mass estimates except for the most luminous (lowest $\log g$) stars, for which frequency measurements become more difficult.
After excluding stars with $\log g<2$, the average mass difference is only $0.016\,\mathrm{M_{\odot}}$ (scatter is $0.033\,\mathrm{M_{\odot}}$).
The two stars for which we obtain significantly lower mass compared to the APOKASC-2 catalog are KIC5698156 and KIC6611219, which also have $\log g<2.0$ (1.89 and 1.74 respectively).
There is a one star for which we obtain a higher mass than the APOKASC-2 catalog (KIC6279038), although the difference is barely above 1-sigma measurement uncertainty.
This star also has a low surface gravity ($\log g=1.65$), and has large uncertainty in the obtained mass.
This comparison also confirms that our mass scale is not significantly different from previous studies. 

\section{Kinematics}\label{sec:kinematics}

\begin{deluxetable*}{lrrrrr}
  \tablecaption{Parallax and extinction \label{tab:gaia}}
  \tablehead{ \colhead{Object} & \colhead{$\pi_{\rm Gaia}$} & \colhead{$\sigma_{(\pi_{\rm Gaia})}$} & \colhead{$\pi_{\rm seis}$} & \colhead{$\sigma_{(\pi_{\rm seis})}$} & \colhead{$E(B-V)$\tablenotemark{a}}  \\ 
                               & \colhead{(mas)}            & \colhead{(mas)}                    & \colhead{(mas)}            & \colhead{(mas)}                                     & \colhead{(mag)}   
  }
  \startdata
KIC5184073 &     0.117 &           0.015 &             0.202 &                   0.009 &   0.06 \\
KIC5439372 &     0.289 &           0.020 &             0.274 &                   0.032 &   0.06 \\
KIC5446927 &     0.331 &           0.029 &             0.379 &                   0.016 &   0.06 \\
KIC5698156 &     0.663 &           0.024 &             0.776 &                   0.048 &   0.06 \\
KIC5858947 &     1.156 &           0.028 &             1.277 &                   0.046 &   0.07 \\
KIC5953450 &     0.791 &           0.025 &             0.755 &                   0.027 &   0.06 \\
KIC6279038 &     0.145 &           0.027 &             0.177 &                   0.024 &   0.04 \\
KIC6520576 &     0.202 &           0.013 &             0.230 &                   0.011 &   0.05 \\
KIC6611219 &     0.272 &           0.025 &             0.334 &                   0.021 &   0.06 \\
KIC7191496 &     0.424 &           0.019 &             0.421 &                   0.013 &   0.05 \\
KIC7693833 &     0.636 &           0.024 &             0.592 &                   0.015 &   0.12 \\
KIC7948268 &     0.762 &           0.024 &             0.748 &                   0.019 &   0.04 \\
KIC8350894 &     0.193 &           0.020 &             0.259 &                   0.010 &   0.05 \\
KIC9335536 &     0.203 &           0.015 &             0.233 &                   0.032 &   0.05 \\
KIC9339711 &     0.415 &           0.020 &             0.404 &                   0.013 &   0.07 \\
KIC9583607 &     0.570 &           0.024 &             0.627 &                   0.016 &   0.04 \\
KIC9696716 &     0.429 &           0.025 &             0.506 &                   0.018 &   0.04 \\
KIC10083815 &     0.367 &           0.026 &             0.409 &                   0.017 &   0.07 \\
KIC10096113 &     0.280 &           0.015 &             0.321 &                   0.011 &   0.18 \\
KIC10328894 &     0.271 &           0.014 &             0.302 &                   0.007 &   0.05 \\
KIC10460723 &     0.370 &           0.020 &             0.363 &                   0.010 &   0.04 \\
KIC10737052 &     0.255 &           0.013 &             0.262 &                   0.006 &   0.06 \\
KIC11563791 &     0.975 &           0.025 &             0.953 &                   0.029 &   0.05 \\
KIC11566038 &     0.495 &           0.020 &             0.471 &                   0.012 &   0.04 \\
KIC12017985 &     0.799 &           0.027 &             0.887 &                   0.026 &   0.04 \\
KIC12017985-17&     0.799 &           0.027 &             0.892 &                   0.028 &   0.04 \\
KIC12253381 &     0.354 &           0.023 &             0.344 &                   0.011 &   0.04 \\
\enddata
\tablenotetext{a}{Obtained from the relation $E(B-V)=0.884\alpha$, where $\alpha$ and its relation to $E(B-V)$ are defined and provided in \citet{Green2018a}. Note that $\alpha$ is proportional to the amount of reddening and normalized to give $E(g-r)=0.901$ at $\alpha=1$.}
\end{deluxetable*}

\begin{deluxetable*}{lrrrrrr}
  \tablecaption{Kinematic properties\label{tab:kinematics}}
  \tablehead{Object & \colhead{$v_R$} & \colhead{$\sigma_{(v_R)}$}& \colhead{$v_\phi$}&\colhead{$\sigma_{(v_\phi)}$} & \colhead{$v_z$}& \colhead{$\sigma_{(v_z)}$} \\
                    &\colhead{($\mathrm{km\,s^{-1}}$)} & \colhead{($\mathrm{km\,s^{-1}}$)}& \colhead{($\mathrm{km\,s^{-1}}$)}& \colhead{($\mathrm{km\,s^{-1}}$)}& \colhead{($\mathrm{km\,s^{-1}}$)}&\colhead{($\mathrm{km\,s^{-1}}$)}   
            }
  \startdata
KIC5184073 &   158.3 &     5.4 &      83.6 &       4.3 &  -119.6 &     4.6 \\
KIC5439372 &   -71.5 &    16.6 &      -5.5 &       1.9 &   165.5 &    24.1 \\
KIC5446927 &  -152.0 &     9.4 &     -18.2 &       2.8 &   -73.8 &     1.4 \\
KIC5698156 &    46.0 &     3.7 &    -136.2 &       1.8 &   -29.6 &     3.0 \\
KIC5858947\tablenotemark{a} &  -155.7 &     6.3 &      96.7 &       1.7 &     7.9 &     1.0 \\
KIC5953450 &   334.3 &     9.6 &      -3.3 &       1.4 &    -6.1 &     1.8 \\
KIC6279038 &   -34.5 &    15.4 &     -73.5 &       4.5 &    37.6 &    14.3 \\
KIC6520576 &  -101.3 &     8.5 &     -79.5 &       5.1 &  -130.7 &     3.3 \\
KIC6611219 &  -267.1 &    20.6 &     -23.2 &       8.4 &   -22.8 &     1.6 \\
KIC7191496 &    38.0 &     1.0 &     -43.6 &       1.1 &   -50.9 &     0.5 \\
KIC7693833 &    66.9 &     2.1 &     240.1 &       1.1 &    -1.3 &     0.4 \\
KIC7948268 &   120.8 &     1.4 &     -41.1 &       1.2 &   -18.5 &     1.4 \\
KIC8350894 &  -169.2 &     7.8 &      61.1 &       4.7 &   -45.2 &     0.8 \\
KIC9335536 &   251.6 &    17.1 &    -241.9 &      41.9 &   151.4 &    36.7 \\
KIC9339711 &    52.4 &     0.5 &     -61.6 &       1.0 &  -158.7 &     2.7 \\
KIC9583607 &  -139.4 &     5.1 &     -83.9 &       1.0 &    26.6 &     2.6 \\
KIC9696716 &  -116.3 &     4.8 &     107.2 &       1.4 &   -49.2 &     0.6 \\
KIC10083815 &   358.3 &    12.7 &     -59.9 &       5.8 &   -60.9 &     0.8 \\
KIC10096113 &   -54.8 &     2.4 &      25.9 &       1.7 &   -59.0 &     1.0 \\
KIC10328894 &   -48.6 &     2.7 &     -81.8 &       1.0 &    20.4 &     2.5 \\
KIC10460723 &   106.8 &     1.3 &    -118.5 &       2.0 &   -34.1 &     1.7 \\
KIC10737052 &  -334.1 &     7.8 &      79.9 &       5.3 &    75.2 &     2.8 \\
KIC11563791 &   -28.6 &     1.7 &     -42.1 &       1.3 &    86.1 &     4.2 \\
KIC11566038 &  -158.6 &     4.8 &     -43.6 &       1.7 &   -32.9 &     0.8 \\
KIC12017985 &  -140.6 &     4.4 &      76.3 &       1.4 &   -88.1 &     1.6 \\
KIC12253381 &  -373.6 &    11.9 &      49.5 &       6.1 &   -37.8 &     1.4 \\
  \enddata
\tablenotetext{a}{The \texttt{ruwe} is $\sim 1.7$, and hence the values presented here must be considered with a caution for this star.}
\end{deluxetable*}

\begin{figure}
  \plotone{\figureloc 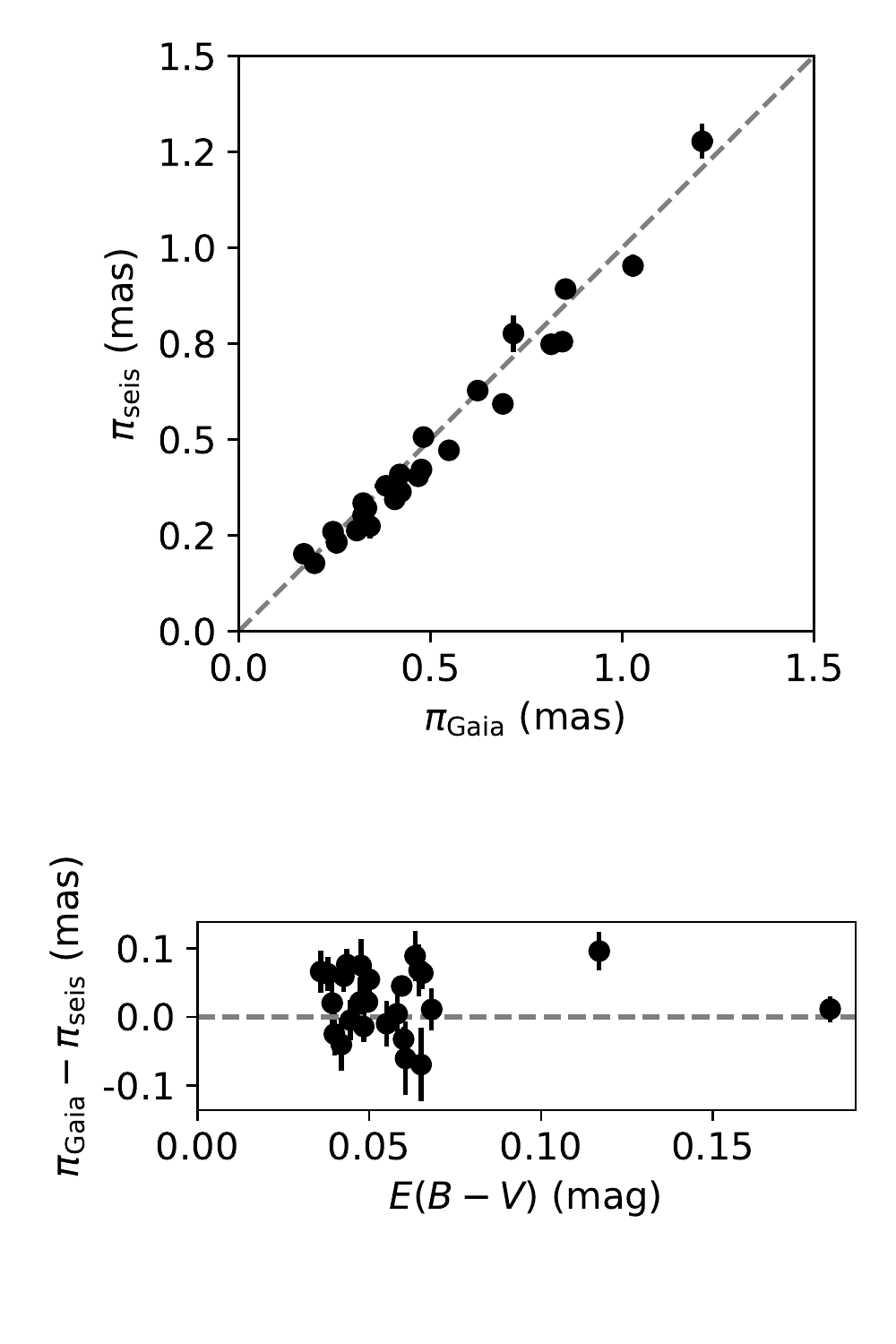}
  \caption{Comparison of Gaia and asteroseismic parallaxes. The zero point offset of Gaia parallax is corrected. The lower panel shows the difference as a function of reddening. \label{fig:distance}}
\end{figure}

Although our selection of halo stars is based on radial velocity and metallicity measurements without taking astrometric measurements into account (Figure~\ref{fig:selection}), we can confirm that most of our targets have halo-like kinematics thanks to the Gaia mission \citep{GaiaCollaboration2016}. 
We adopt proper motions provided in Gaia data release 2 \citep[Gaia DR2; ][]{GaiaCollaboration2018a} and radial velocity measured from our spectra (see Section \ref{sec:observation}).
Distances can be estimated from either astrometric parallax or asteroseismic parallax.
For the former, we use parallax measurements in Gaia DR2 \citep{Lindegren2018a} after correcting for the systematic offset of 0.052 mas \citep{Zinn2019a}. 
For the asteroseismic parallax, we first compute the stellar radius using the asteroseismic scaling relations with the $\Delta\nu$ correction as described in the previous section.
Combined with effective temperature from our high-resolution spectra (see Section~\ref{sec:abundance}), we obtain the luminosity of the stars through $L\propto R^2T_{\rm eff}^4$.
To derive parallax, this luminosity is then compared with the bolometric magnitude that is based on The Two Micron All-Sky Survey \citep[2MASS; ][]{Skrutskie2006a} $K_s$ band photometry and the bolometric correction provided by \citet{Casagrande2014}.
Interstellar extinction is corrected for using the 3D extinction map provided by \citet{Green2018a}. 
The two parallaxes are provided in Table \ref{tab:gaia} and are compared in Figure \ref{fig:distance}.
Note that Table~\ref{tab:gaia} lists astrometric parallax without the $0.052\,\mathrm{mas}$ correction, but the correction is applied in Figure~\ref{fig:distance}.
The agreement between the two sets of values is good with a weighted average difference of $\langle \pi_{\rm Gaia}-\pi_{\rm seis} \rangle= 0.028 \pm 0.007\,\mathrm{mas}$ ($\sigma=0.038$).
For the calculation of kinematics, we adopt asteroseismic parallax for all the stars for consistency.

Table \ref{tab:kinematics} shows radial, azimuthal and vertical components of the Galactocentric velocity; $v_R,\,v_\phi,\,v_z$.
We adopt $R_0=8.2\,\mathrm{kpc}$ \citep{McMillan2017a} and $z_0=0.025\,\mathrm{kpc}$ \citep{Juric2008a} for the solar position, and $(v_x,v_y,v_z)_{\odot}=(11.10,\,247.97,\,7.25)\,\mathrm{km\,s^{-1}}$ for the solar velocity relative to the Galactic center, where $v_x$ is the velocity toward the Galactic center, $v_y$ is in the direction of Galactic rotation, and $v_z$ is toward north. 
The $v_x$ and $v_z$ come from \citet{Schoenrich2010a}, and $v_y$ comes from the proper motion measurement of the Sgr A$^\star$ \citep{Reid2004a} and $R_0$. 
Coordinate transformation from observed quantities to the Galactocentric Cartesian system was conducted with the \texttt{astropy.coordinates} package.
We note that $v_{\phi}$ is taken positive toward the Galactic rotation direction.
Uncertainties are estimated by Monte Carlo sampling. 
We note that all the stars are treated as single, although some show radial velocity variations.
While the presence of a binary companion might lead to inaccurate astrometric measurements in Gaia DR2, the Renormalized unit weight error (\texttt{ruwe}), which is an indicator of the goodness of the astrometric solution in Gaia DR2, is smaller than 1.4 except for KIC5858947, for which \texttt{ruwe} is $\sim 1.7$ \footnote{see \url{https://gea.esac.esa.int/archive/documentation/GDR2/Gaia_archive/chap_datamodel/sec_dm_main_tables/ssec_dm_ruwe.html} for more details about the \texttt{ruwe}.}.

\begin{figure*}
\plotone{\figureloc 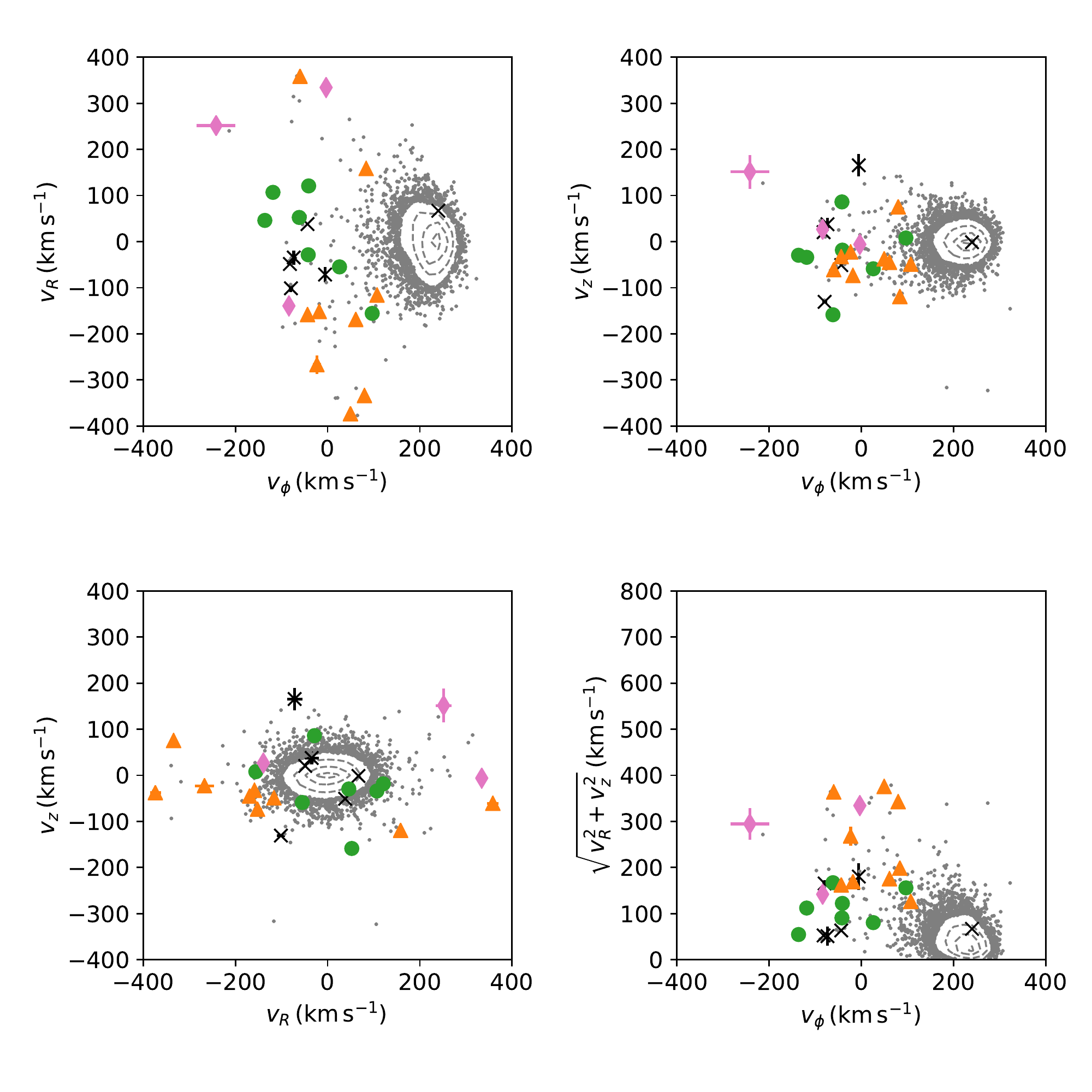}
\caption{Large symbols (black crosses, green circles, orange triangles and pink diamonds) show velocities of the program stars with errorbars, while grey points and contours are the distribution of stars in the crossmatched catalog of LAMOST DR5 \citep{Zong2018a} and \citet{Yu2018a}. The different symbols are used according to the Mg and Fe based classification made in Section \ref{sec:abundance} (black crosses: metal-poor, green circles: high-Mg, orange triangles: low-Mg, pink diamonds: very low-Mg; see Figure~\ref{fig:abtrend0}). For the LAMOST sample, we calculated velocities in exactly the same way as for the program stars. \label{fig:kinematics}}
\end{figure*}

Figure \ref{fig:kinematics} shows the velocity distribution of stars. 
It is clear that most of the program stars do not follow the motion of the majority of the stars in the \textit{Kepler} field, which are shown with grey contours and dots, i.e., they have very different velocities than the Galactic disk stars.
The distribution of $v_\phi$ is particularly different (see upper two panels and lower right).
This is expected since our selection is partly based on radial velocity and since the \textit{Kepler} field is centered at $l=76.32^\circ$, which provides a strong correlation between $v_{\phi}$ and radial velocity.
The only exception would be again KIC7693833, which also stands out during the sample selection (Figure~\ref{fig:selection}).
It follows the motion of disk stars despite its low metallicity ([{Fe}/{H}]$\sim -2.3$), and is separately discussed in Section~\ref{sec:kic769}.

\section{Abundance analysis}\label{sec:abundance}

\subsection{Line list}

\begin{deluxetable}{rlrrrr}
  \tablecaption{Line list and measured equivalent widths. \label{tab:linelist}}
  \tablehead{
\colhead{$\lambda$} & \colhead{species} & \colhead{$\chi$} & \colhead{$\log gf$} & \colhead{Syn\tablenotemark{a}} & \colhead{KIC5184073}\\
\colhead{$(\mathrm{\AA})$} & & \colhead{(eV)} & & & \colhead{$(\mathrm{m\AA})$}
 } 
 \startdata
   4053.821 &   Ti II &  1.893 & -1.070 &      &      80.8  \\
   4056.187 &   Ti II &  0.607 & -3.280 &      &      49.5  \\
   4082.939 &    Mn I &  2.178 & -0.354 &      &   \nodata  \\
   4086.714 &   La II &  0.000 & -0.070 &  syn &      58.9  \\
   4099.783 &     V I &  0.275 & -0.100 &      &      41.6  \\
 \enddata
\tablecomments{The entity of the table is available online. A portion is shown here.}
\tablenotetext{a}{Lines to which spectral synthesis applied are flagged with ``syn'' in this column}
\end{deluxetable}

\begin{figure}
  \plotone{\figureloc 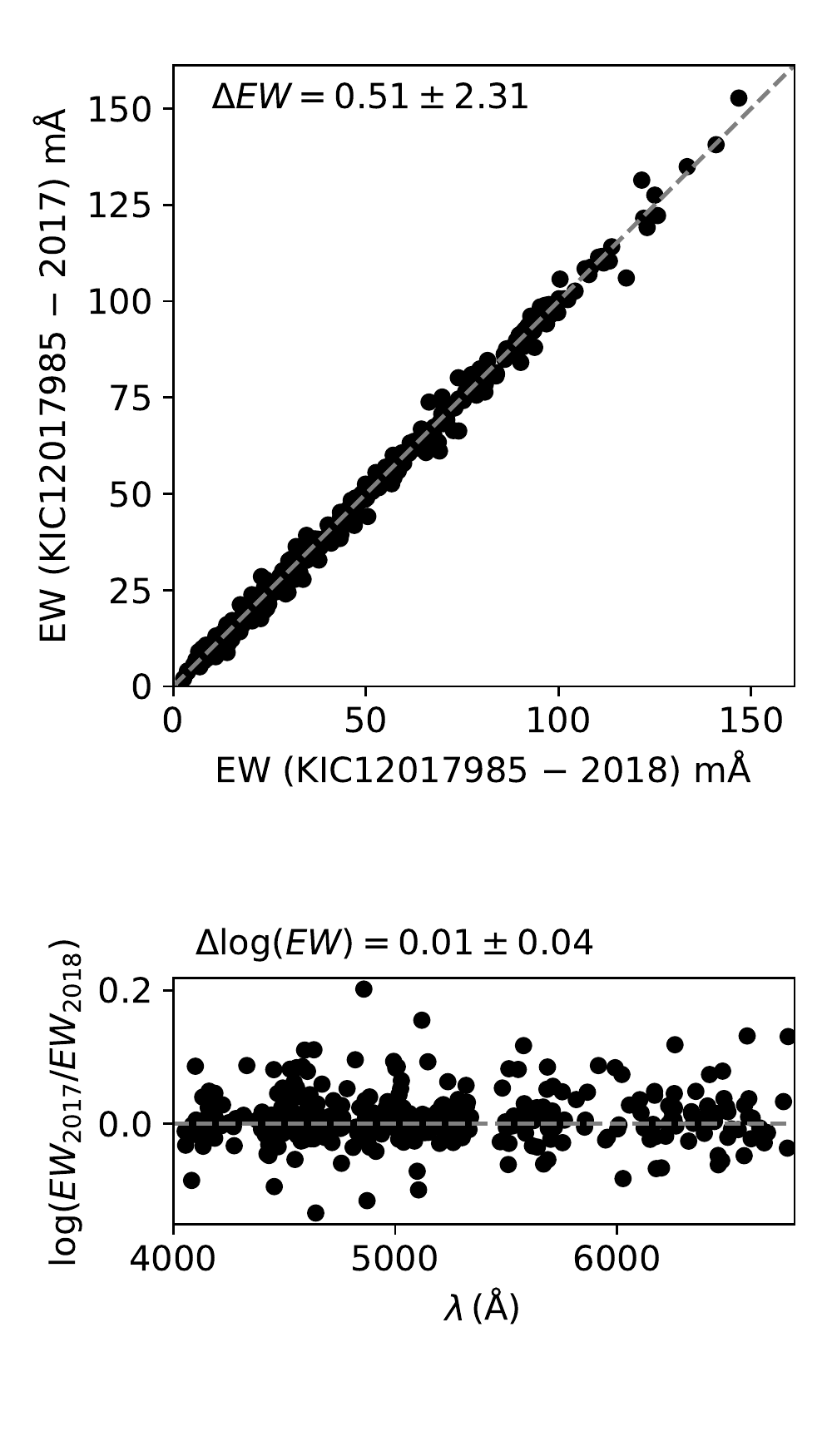}
  \caption{Comparison of equivalent widths for KIC12017985 from 2017 and 2018 observations. Mean difference and line-to-line scatter are shown in the figure. The dashed lines indicate the one-to-one relation. \label{fig:ewcomparison}}
\end{figure}

Table \ref{tab:linelist} shows a list of lines used in this study together with measured equivalent widths. 
Lines were carefully selected by comparing synthetic spectra and a very high-$S/N$ observed spectrum of the archetypal metal-poor red giant HD122563. 
Additional lines were taken from \citet{Matsuno2018a} for analyses of high-metallicity stars.
Hyperfine structure splitting was included for Sc~II, V~I, Mn~I, Co~I, Cu~I, Ba~II, and Eu~II, assuming solar $r$-process abundance ratio for isotopic ratios of neutron capture elements.
Line positions and relative strengths were taken from \citet{McWilliam1998} for Ba, \citet{Ivans2006} for Eu, and Robert L. Kurucz's linelist for the others\footnote{\url{http://kurucz.harvard.edu/linelists.html}}.

Equivalent widths were measured through fitting Gaussian profiles to absorption lines.
Lines were limited to those with reduced equivalent width ($REW=\log(EW/\lambda))$ smaller than $-4.5$ to avoid significant effects of saturation.
Figure~\ref{fig:ewcomparison} compares equivalent widths measured from 2017 and 2018 observations for KIC12017985.
Despite different spectral resolutions, the two measurements show excellent agreement ($EW=0.51\pm2.31\,\mathrm{m\AA}$ and $\Delta\log(EW)=0.01\pm 0.04$; in the direction from 2018 observation minus 2017 observation).

\subsection{Stellar parameter determination}

\begin{figure*}
  \plotone{\figureloc 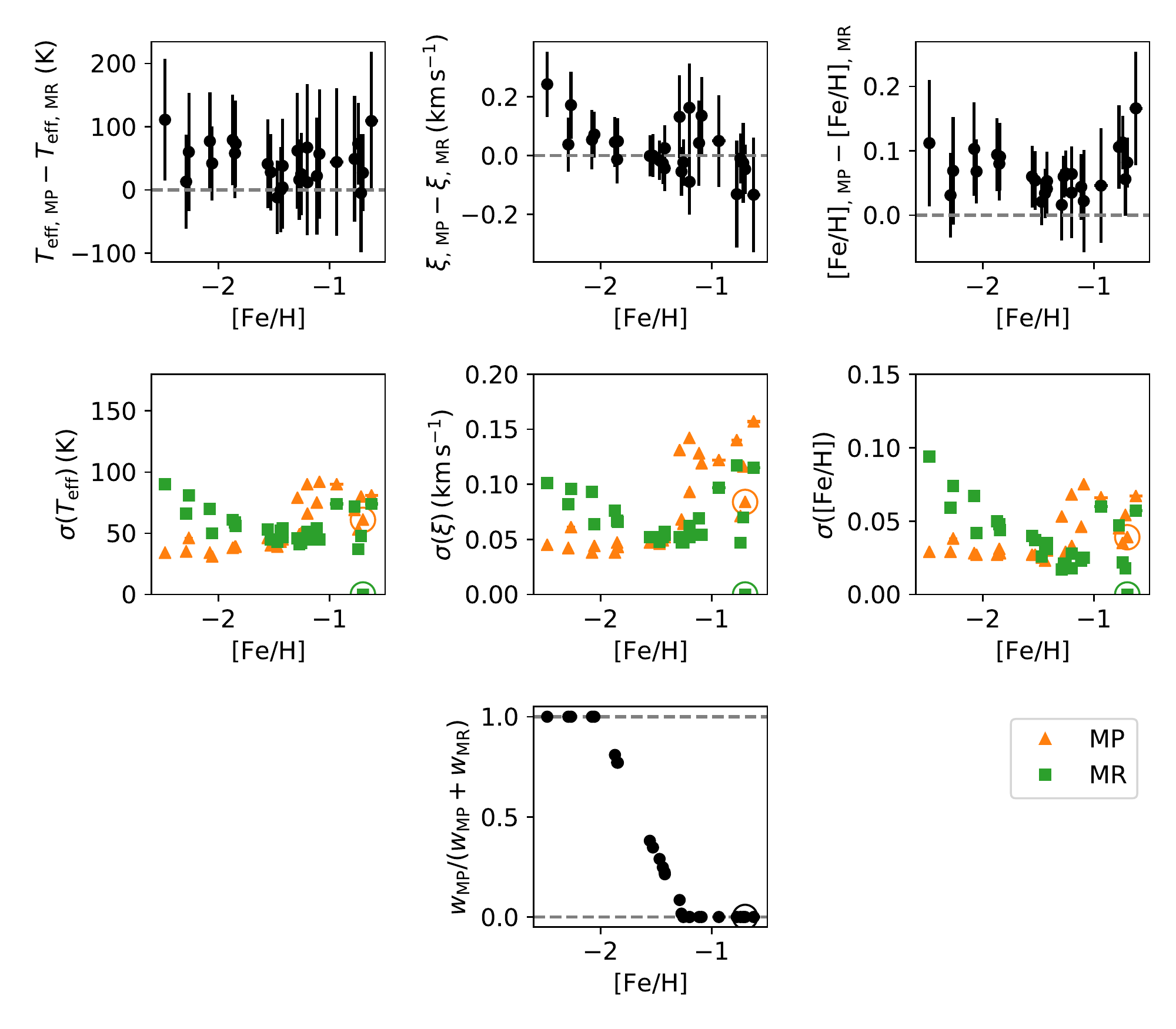}
  \caption{Comparison of stellar parameters from two analyses. One analysis uses the metal-poor standard star (HD122563) and the other uses the metal-rich one (KIC9583607). Since the latter is one of the program stars, we show its location with an open circle in the bottom four panels. Considering the trends in the $\xi$ uncertainties, we adopt stellar parameters with the weights shown in the bottom panel.  \label{fig:paramMPMR}}
\end{figure*}

\begin{deluxetable*}{lrrrrrrrr}
  \tablecaption{Adopted stellar parameters\label{tab:param}}
  \tablehead{
   \colhead{Object} & \colhead{$T_{\rm eff}$}& \colhead{$\sigma_{(T_{\rm eff})}$} & \colhead{$\log g$} & \colhead{$\sigma_{(\log g)}$} &\colhead{$\xi$}& \colhead{$\sigma_{(\xi)}$} &\colhead{$\mathrm{[Fe/H]}$}& \colhead{$\sigma_{(\mathrm{[Fe/H]})}$}    \\  
                    & \colhead{(K)}          & \colhead{(K)}                  & &    &\colhead{$\mathrm{(km\,s^{-1})}$}&\colhead{$\mathrm{(km\,s^{-1})}$} & & 
  }
  \startdata
KIC5184073 &         4864 &               44 &          1.876 &              0.010 &        1.581 &            0.044 &        -1.422 &             0.028 \\
KIC5439372 &         4835 &               34 &          1.716 &              0.016 &        1.915 &            0.045 &        -2.479 &             0.029 \\
KIC5446927 &         5102 &               37 &          2.261 &              0.008 &        1.530 &            0.047 &        -0.742 &             0.022 \\
KIC5698156 &         4644 &               43 &          1.888 &              0.015 &        1.533 &            0.049 &        -1.289 &             0.016 \\
KIC5858947 &         5105 &               72 &          3.149 &              0.004 &        1.221 &            0.117 &        -0.775 &             0.047 \\
KIC5953450 &         5127 &               74 &          3.071 &              0.005 &        1.234 &            0.115 &        -0.623 &             0.057 \\
KIC6279038 &         4761 &               31 &          1.654 &              0.025 &        1.784 &            0.044 &        -2.055 &             0.027 \\
KIC6520576 &         4971 &               35 &          2.169 &              0.014 &        1.602 &            0.042 &        -2.289 &             0.029 \\
KIC6611219 &         4652 &               45 &          1.742 &              0.016 &        1.566 &            0.052 &        -1.201 &             0.018 \\
KIC7191496 &         4903 &               34 &          2.122 &              0.007 &        1.672 &            0.038 &        -2.076 &             0.028 \\
KIC7693833 &         5094 &               46 &          2.422 &              0.006 &        1.664 &            0.061 &        -2.265 &             0.038 \\
KIC7948268 &         5154 &               51 &          3.004 &              0.004 &        1.230 &            0.062 &        -1.199 &             0.028 \\
KIC8350894 &         4797 &               45 &          2.012 &              0.012 &        1.491 &            0.054 &        -1.091 &             0.025 \\
KIC9335536 &         4817 &               32 &          1.951 &              0.026 &        1.527 &            0.038 &        -1.528 &             0.026 \\
KIC9339711 &         4937 &               33 &          2.226 &              0.006 &        1.509 &            0.037 &        -1.469 &             0.020 \\
KIC9583607 &         5059 &          \nodata &          2.322 &          \nodata   &        1.590 &         \nodata  &        -0.700 &          \nodata \\
KIC9696716 &         4962 &               41 &          2.297 &              0.009 &        1.453 &            0.041 &        -1.440 &             0.024 \\
KIC10083815 &         4784 &               74 &          2.161 &              0.013 &        1.476 &            0.097 &        -0.936 &             0.060 \\
KIC10096113 &         4948 &               48 &          2.475 &              0.007 &        1.443 &            0.070 &        -0.717 &             0.018 \\
KIC10328894 &         5018 &               33 &          2.405 &              0.006 &        1.528 &            0.039 &        -1.850 &             0.026 \\
KIC10460723 &         4922 &               40 &          2.275 &              0.010 &        1.466 &            0.046 &        -1.272 &             0.021 \\
KIC10737052 &         4973 &               42 &          2.340 &              0.008 &        1.478 &            0.047 &        -1.255 &             0.021 \\
KIC11563791 &         4974 &               54 &          2.550 &              0.006 &        1.326 &            0.069 &        -1.114 &             0.023 \\
KIC11566038 &         4999 &               38 &          2.413 &              0.005 &        1.451 &            0.046 &        -1.423 &             0.025 \\
KIC12017985 &         4945 &               33 &          2.175 &              0.006 &        1.628 &            0.036 &        -1.844 &             0.024 \\
KIC12017985-17 &         4932 &               33 &          2.175 &              0.007 &        1.637 &            0.034 &        -1.870 &             0.024 \\
KIC12253381 &         4922 &               37 &          2.256 &              0.007 &        1.512 &            0.037 &        -1.556 &             0.027 \\
  \enddata
\end{deluxetable*}

\begin{figure*}
  \plotone{\figureloc 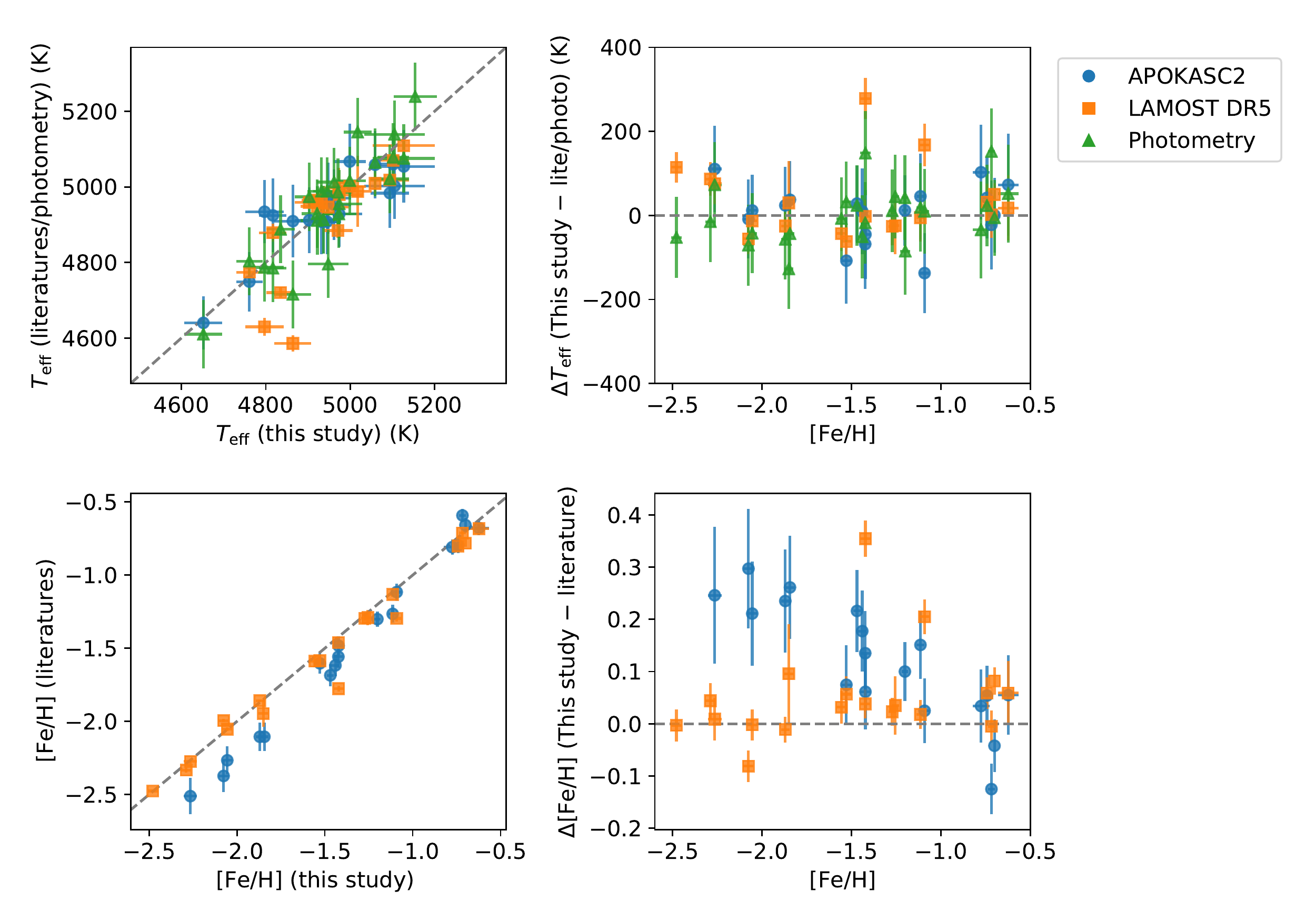}
  \caption{Comparison of stellar parameters with literature values. We note that errorbars only reflect internal uncertainties in our analysis and surveys.  \label{fig:paramcomp}}
\end{figure*}

Stellar parameter determination and subsequent abundance measurements were conducted with a modified version of \texttt{q$^2$} \citep{Ramirez2014,Matsuno2018a}, which utilizes the February 2017 version of \texttt{MOOG} \citep{Sneden1973}.
We obtain models of the structure of atmospheres from interpolation of MARCS stellar model atmospheres with standard chemical composition \citep{Gustafsson2008a}.
In the process of stellar parameter determination, we adopt the line-by-line non-local thermo-dynamical equilibrium (NLTE) corrections for Fe abundances provided by \citet{Amarsi2016}\footnote{The grid is available at \url{http://www.mpia.de/homes/amarsi/index.html}}.
All the subsequent abundance analysis is conducted under one dimensional plane-parallel (1D) and local thermo-dynamical equilibrium (LTE) approximations unless otherwise stated.
The solar chemical abundance is adopted from \citet{Asplund2009}.

Stellar parameters are determined by requiring excitation balance of Fe~I lines, using the asteroseismic scaling relation for $\nu_{\rm max}$, and minimizing the trend between $REW$ and abundance derived from each Fe~I line. 
Basically, each condition is to constrain $T_{\rm eff}$, $\log g$, and microturbulent velocity ($\xi$), respectively.  
Since the asteroseismic scaling relation constrains $\log g$ within a range of $\sim 0.01\,\mathrm{dex}$ at a given temperature, the use of the relation is essentially the same as assuming a tight relation between $T_{\rm eff}$ and $\log g$.
Note that, although the scaling relation might suffer from systematic uncertainties (see Section~\ref{sec:dis_seis}), $\log g$ is robustly constrained.   
We also used a prior on $\xi$ as a function of $\log g$ (see Appendix \ref{appendix1}).

To achieve high-precision and to minimize the effects of departures from 1D approximations and those of uncertain atomic data, we adopt a line-by-line differential abundance analysis.
Since our targets span over $\sim$2 dex in metallicity, we repeated the analysis adopting two different standard stars.
One is the well-studied metal-poor star HD~122563 with [{Fe}/{H}]$\sim-2.6$. 
This star is the only metal-poor giant whose $T_{\rm eff}$ and $\log g$ have been measured with high accuracy through interferometric measurements \citep[$T_{\rm eff}=4636\,\mathrm{K}$;][]{Karovicova2018a} and asteroseismology \citep[$\log g=1.418$;][]{Creevey2019a}, respectively.
The other standard star is one of the program stars, KIC9583607, which is more metal-rich ([{Fe}/{H}]$\sim-0.7$) and was also observed by APOGEE.
Since APOGEE is carefully calibrated at this metallicity and since this star has an asteroseismic $\log g$ constraint, the use of this star as a standard star ensures us that our parameters are not systematically biased.
We adopt the value from the APOGEE DR14 for the temperature of this star ($T_{\rm eff}=5059\,\mathrm{K}$) and an asteroseismic estimate for the surface gravity ($\log g=2.322$).
The microturbulent velocities and metallicities of the standard stars are obtained from a standard analysis of individual iron lines.
Using the aforementioned $T_{\rm eff}$ and $\log g$, we minimize the trend between the $REW$s and the abundances obtained from individual neutral iron lines to determine the microturbulent velocity.
The average abundance from Fe~II lines is adopted as the metallicity of the standard star since it is less sensitive to the choice of stellar parameters than that from Fe~I lines.
Stellar parameters of other stars determined relative to HD122563 and those relative to KIC9583607 are denoted as $X_{\rm MP}$ and $X_{\rm MR}$, respectively. 

Each star has two sets of parameters, $X_{\rm MP}$ and $X_{\rm MR}$ (see Appendix~\ref{appendix:MPMR}), which are combined in such a way that we obtain precise parameters without introducing significant systematic offsets. 
We here combine the two results following the equation,
\begin{equation}
X = X_{\rm MP}*w_{\rm MP} + X_{\rm MR}*w_{\rm MR}, \label{eq:weights}
\end{equation}
where $w_{\rm MP}$ and $w_{\rm MR}$ are the weights.
We determine $w_{\rm MP}$ and $w_{\rm MR}$ from the comparison of the two sets of analyses as described below.

The comparison is provided in Figure~\ref{fig:paramMPMR} for parameters and their uncertainties as a function of metallicity.
When a star and the standard star have large metallicity difference, there are difficulties in accurate parameter determination.
One is that departures from 1D approximations might not act in the same way.
Another difficulty is that the number of common lines becomes smaller as the metallicity difference becomes larger.
This is because absorption lines of the more metal-rich one suffer from saturation or blending while those of the more metal-poor one might be too weak to be detected.
These effects are recognisable in Figure~\ref{fig:paramMPMR}. 
We define the following parameters to compute the weights in Eq.~\ref{eq:weights}:
\begin{eqnarray}
x_{\rm MP} =& 1.0 - ([\mathrm{Fe/H}]_{\rm MP}-[\mathrm{Fe/H}]_{\rm standard})/s_{\rm MP} \\ 
x_{\rm MR} =& 1.0 - ([\mathrm{Fe/H}]_{\rm standard}-[\mathrm{Fe/H}]_{\rm MR})/s_{\rm MR},
\end{eqnarray}
where $s_{\rm MP}$ and $s_{\rm MR}$ are free parameters, and define $w_{\rm MP}=max\{0.0,\,min\{1.0,\,x_{\rm MP}\}\}$ (similarly for $w_{\rm MR}$).
The $w_{\rm MP}$ and $w_{MR}$ are then scaled so that their sum becomes 1.
The two parameters, $s_{\rm MP}$ and $s_{\rm MR}$ are chosen to be 1.4, but results are insensitive to the exact choice of these parameters. 
The $w_{\rm MP}$ and $w_{\rm MR}$ are also shown in Figure~\ref{fig:paramMPMR} and the adopted parameters are listed in Table~\ref{tab:param}.
The parameters in Table~\ref{tab:param} are used in asteroseismic analysis, which includes mass estimates, and in figures throughout this study.

Uncertainties provided in this paper only reflect random uncertainties that are obtained through a MCMC method.
It is important to take systematic uncertainties into consideration when one tries to quantitatively compare our results with other studies.
Sources of systematic uncertainties include the uncertainties in stellar parameters and abundances of the standard stars, possible blending with unknown weak lines, and the NLTE and/or 3D effects. 
Among these, the impact of the first source is studied in Appendix \ref{appendix3}.

The adopted stellar parameters are compared to the results obtained from spectroscopic surveys (APOKASC2 catalog, which is based on APOGEE, and LAMOST) and photometric temperatures in Figure~\ref{fig:paramcomp}.
The latter were derived implementing Gaia $BP, RP$ \citep{Evans2018a} and 2MASS $J, H, K_s$ photometry into the InfraRed Flux Method \citep{Casagrande2010,Casagrande2014a}, with reddening from \citet{Green2019a}.
There are generally good agreements between this study and literature, which indicates that our parameters do not have large systematic offsets.
The exceptions are the outliers found in the comparison with LAMOST and the metallicity-dependent systematic offset found in the metallicity comparison with APOGEE.
The two outliers found in the comparison with LAMOST are KIC5184073 and KIC8350894. 
Although the reasons for the discrepancy are unclear, we note that these stars do not stand out from the overall trend when compared with APOGEE.
The systematic metallicity offset from APOGEE at low metallicity demonstrates the difficulty of determining abundances in the absolute scale.
For example, \citet{Amarsi2016a} reported [{Fe}/{H}]$=-2.49\pm 0.11(\mathrm{stat})\pm 0.14(\mathrm{sys})$ for the metallicity of HD122563, our metal-poor standard star, from 3D and NLTE modelling of stellar atmospheres. 
Our adopted value of [{Fe}/{H}]$=-2.60$ is well within the range of uncertainties. 
We note that the good agreeement at high metallicity would be mainly thanks to our choice of the effective temperature of the standard star, which was taken from APOGEE.

\subsection{Elemental abundances}

\begin{figure}
  \plotone{\figureloc 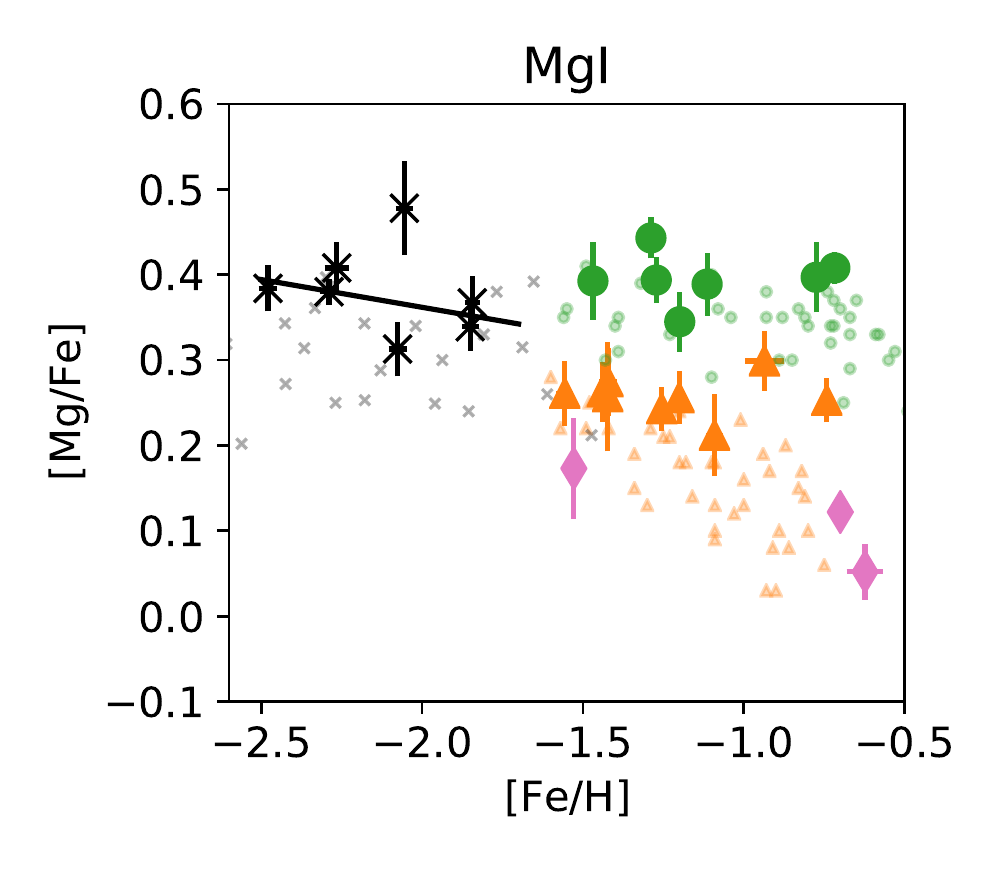}
  \caption{
[{Mg}/{Fe}] versus [{Fe}/{H}] with measurements from Mg~I. 
Our targets are shown with large symbols with errorbars and are divided into four sub groups depending on their Mg and Fe abundance: metal-poor (black crosses), $\alpha$-rich (green circles), $\alpha$-poor (orange triangles) and very $\alpha$-poor (pink diamonds). 
The solid black line is the linear fit for the metal-poor subsample. 
Small green circles and orange triangles are high-/low-$\alpha$ populations from \citet{Nissen2010}. 
Small grey crosses are stars from \citet{Reggiani2017a}.
\label{fig:abtrend0}
}
\end{figure}

\begin{figure}
  \plotone{\figureloc 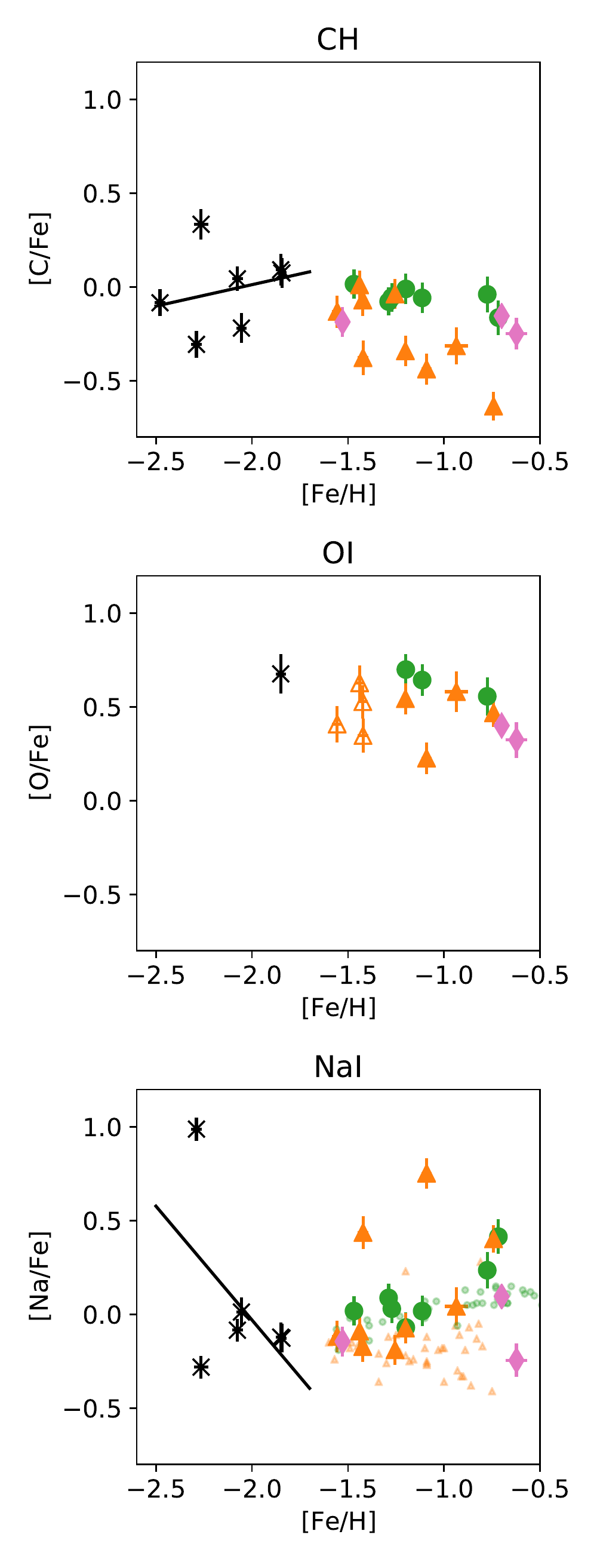}
  \caption{Same as Figure~\ref{fig:abtrend0} but for light elements, namely, C (CH), O (O~I), and Na (Na~I). 
Open symbols indicate that while the abundance should be derived by weighting two analyses with different standard stars, only one of them can be used since the other standard does not have common lines with the target star. 
 \label{fig:abtrend1}}
\end{figure}

\begin{figure*}
  \plotone{\figureloc 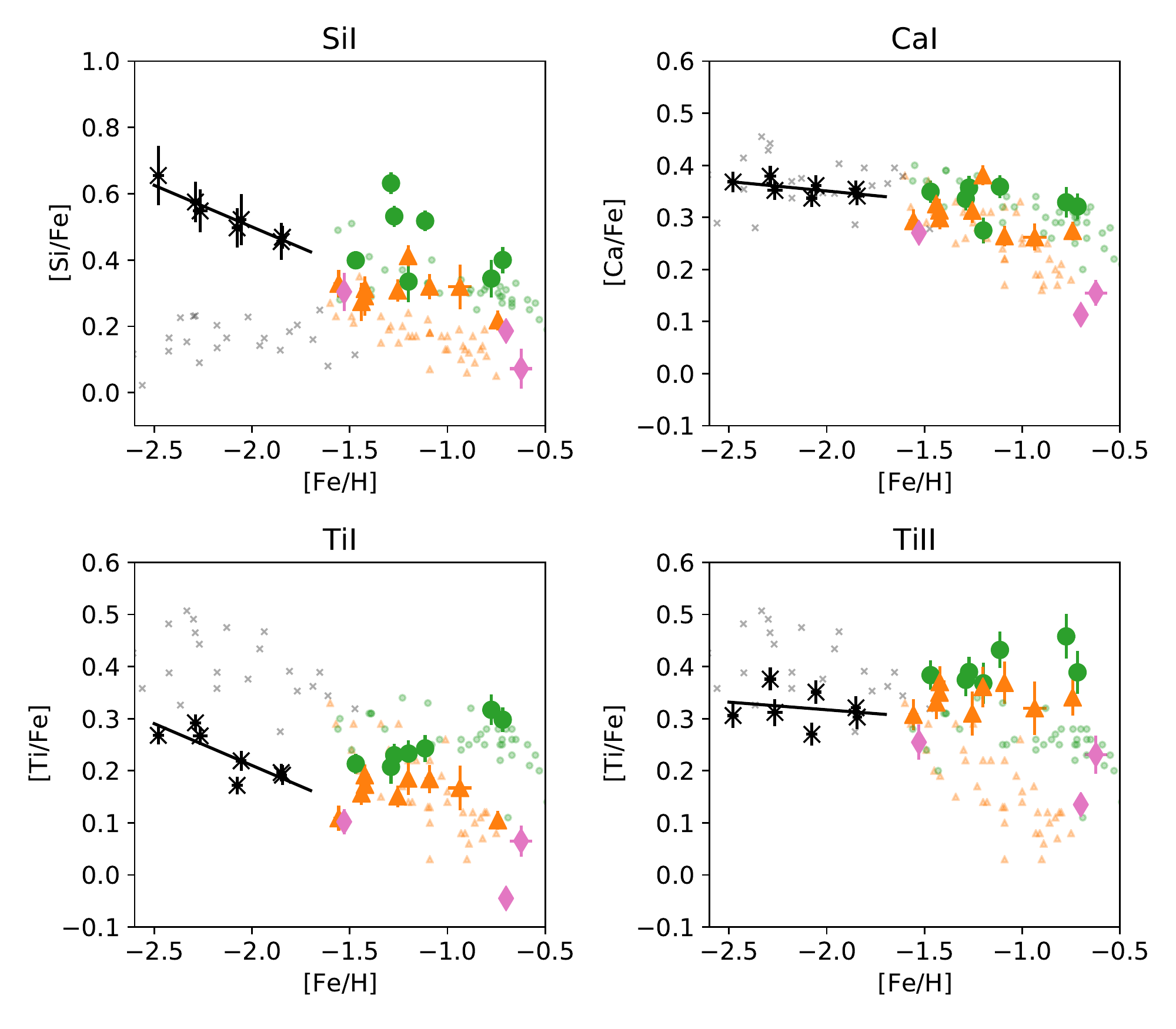}
  \caption{Same as Figure~\ref{fig:abtrend0} but for other $\alpha$-elements, namely, Si (Si~I), Ca (Ca~I), and Ti (Ti~I and Ti~II). Note that the vertical scale is different only for Si. \label{fig:abtrend2}}
\end{figure*}

\begin{figure*}
  \plotone{\figureloc 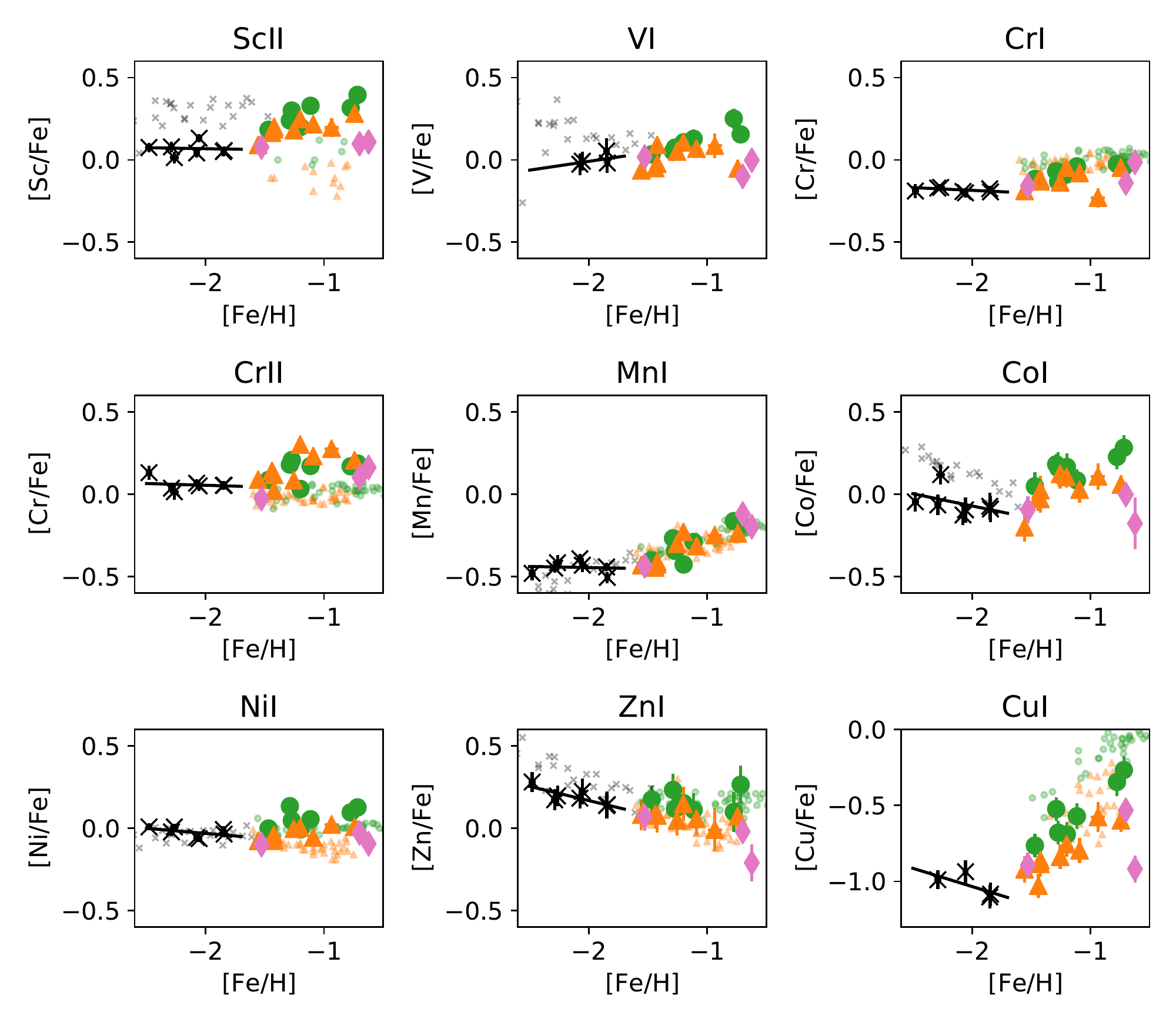}
  \caption{Same as Figure~\ref{fig:abtrend1} but for iron-peak elements, namely, V (V~I), Cr (Cr~I and Cr~II), Mn (Mn~I), Co (Co~I), Ni (Ni~I), Zn (Zn~I) and Cu (Cu~I). Note that the vertical scale is different only for Cu. \label{fig:abtrend3}}
\end{figure*}

\begin{figure*}
  \plotone{\figureloc 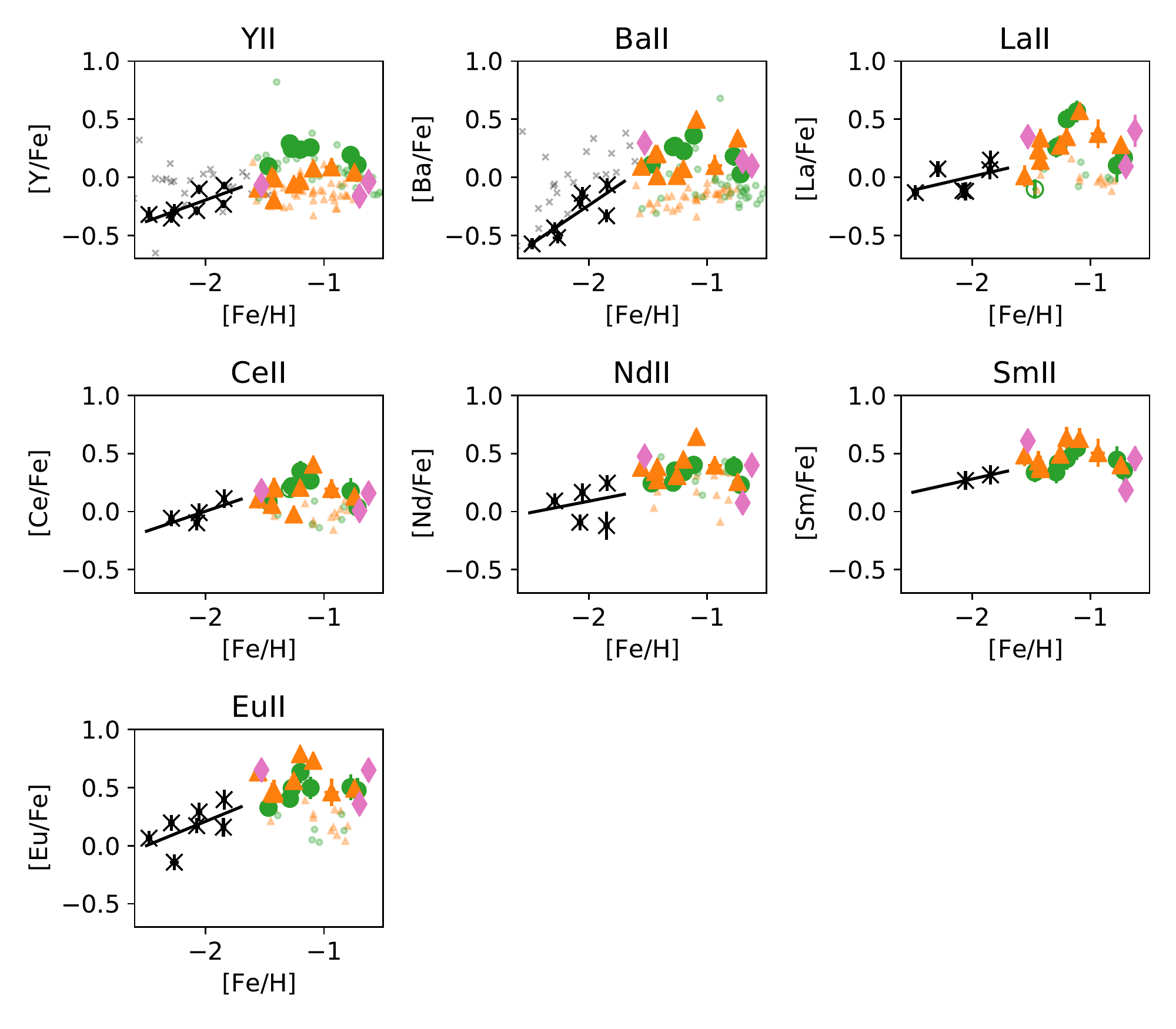}
  \caption{Same as Figure~\ref{fig:abtrend0} but for neutron-capture elements, namely, Y (Y~II), Zr (Zr~I), Ba (Ba~II), La (La~II), Ce (Ce~II), Nd (Nd~II), Sm (Sm~II), and Eu (Eu~II).  \label{fig:abtrend4}}
\end{figure*}

\begin{deluxetable}{lrrrrr}
  \tablecaption{Abundances\label{tab:abundance}}
  \tablehead{
            \colhead{Object} & \colhead{Species} & \colhead{[X/H]} & \colhead{$\sigma_{(\mathrm{[X/H]})}$}& \colhead{[X/Fe]} & \colhead{$\sigma_{(\mathrm{[X/Fe]})}$}
            }
  \startdata
KIC5184073 &  CH    &     -1.92 &      0.12 &     -0.38 &      0.09 \\
 KIC5184073 &  O~I   &     -1.20 &      0.09 &      0.35 &      0.09 \\
 KIC5184073 &  Na~I  &     -1.10 &      0.09 &      0.44 &      0.09 \\
 KIC5184073 &  Mg~I  &     -1.26 &      0.05 &      0.27 &      0.04 \\
 KIC5184073 &  Si~I  &     -1.23 &      0.02 &      0.31 &      0.03 \\
\enddata
\tablecomments{A portion is shown here. The entity is available online.}
\end{deluxetable}
 
\begin{figure*}
  \plotone{\figureloc 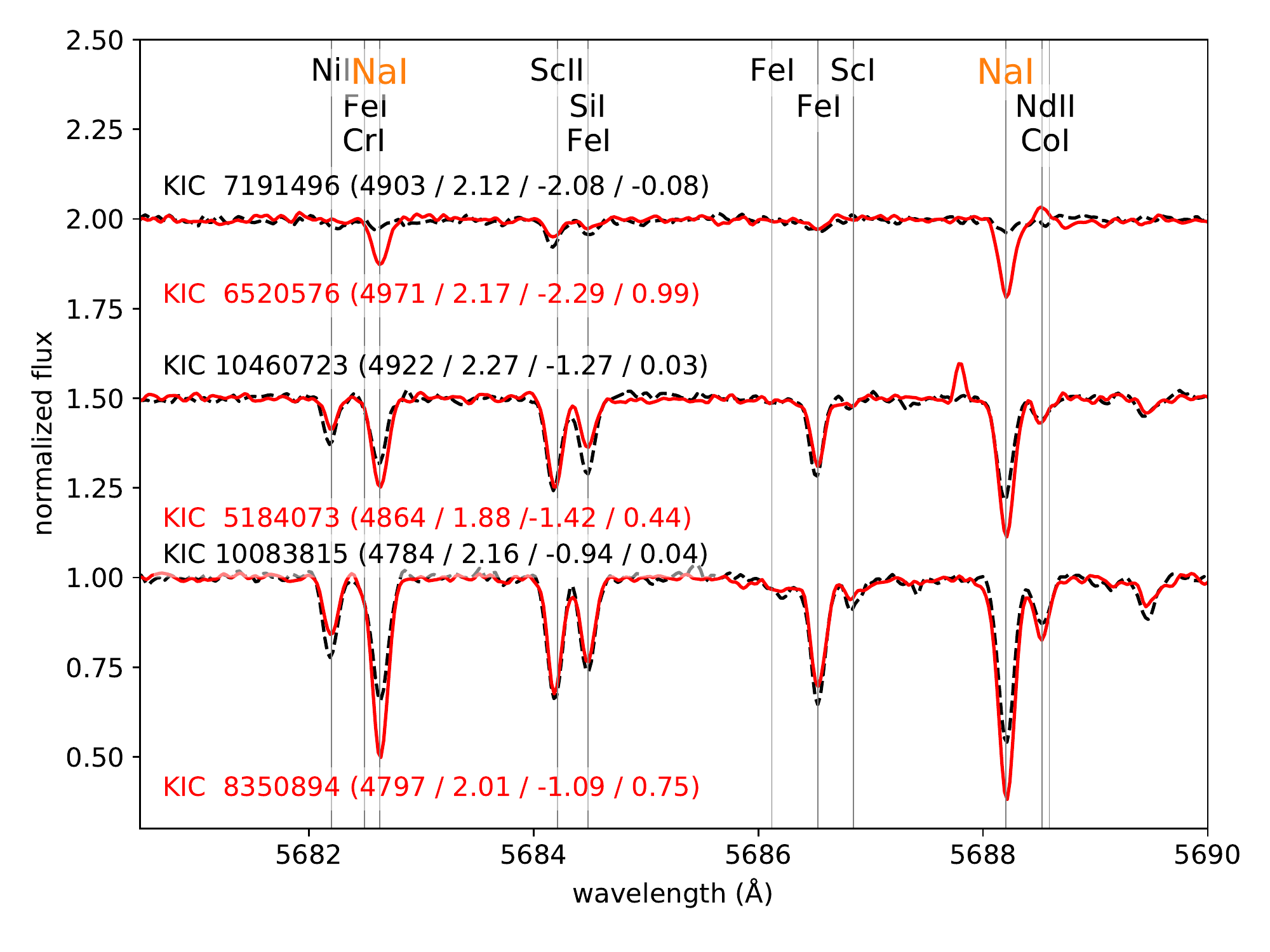}
  \caption{Spectra of Na-enhanced objects (red solid) and those of comparison stars (black dashed) with vertical offsets for visualization purposes. Stellar parameters and 1D LTE abundances are shown as $(T_{\rm eff}\,/\,\log g\,/\,[\mathrm{Fe/H}]\,/\,[\mathrm{Na/Fe}])$.  \label{fig:na}}
\end{figure*}

Elemental abundances are derived based on equivalent widths for most of the elements studied in the present work.
We first derive abundances for the standard stars, then derive abundances of the other stars differentially through a line-by-line analysis.  
Thanks to this approach, we achieve high precision in relative abundances, although the absolute scale could be affected by systematic uncertainties such as those due to the NLTE effect.

As in the case for the stellar parameters, we obtain two sets of chemical abundance for each star through two analyses with different standard stars.
We combine these abundances for each species with the Eq. \ref{eq:weights}.
However, there are cases where a star has no common line with one of the standard stars for some elements.
In this case, we have to rely on the abundance derived from the analysis with the other standard star.
Unless the other analysis has a weight of 1, the abundance of the star has to be taken with a caution.
We show these cases with open symbols in the figures. 

Here we take into account two sources of uncertainty in abundance measurements.
One is due to the noise present in the spectra, which affects measured equivalent widths. 
We denote this component as $\sigma_1$ and estimate it from the line-to-line scatter ($\sigma_{\rm sct}$) in derived abundance for each species as $\sigma_1 = \sigma_{\rm sct}/\sqrt{N}$, where $N$ is the number of lines used for the abundance measurements.
When $N$ is smaller than 4 and when $\sigma_{\rm sct}$ is smaller than $\sigma_{\rm sct}$ for Fe~I lines, we substitute $\sigma_{\rm sct}$ with the latter.
The other source is the uncertainties of stellar parameters ($\sigma_2$).
This component is estimated by repeating analyses changing stellar parameters by the same amount as their uncertainties. 
We quadratically sum $\sigma_1$ and $\sigma_2$ to obtain final uncertainty estimates.

Abundances from CH (molecule), and O~I, Na~I, Mg~I, Si~I, Co~I, Cu~I, Zn~I, Ba~II, La~II, Ce~II, Nd~II, Sm~II, and Eu~II are analysed through spectral synthesis.
The spectral synthesis is necessary for species that produce significantly asymmetric lines, such as CH, Co and Eu.
In addition, these elements have small available numbers of lines and the abundance measurement has to rely on partially blended lines for which we need spectral synthesis. 
Using the line list from VALD3\footnote{\url{http://vald.astro.uu.se/}}, series of synthetic spectra were generated with varying the abundance of the single element of interest to determine the best-fit abundance by minimizing $\chi^2$.
After the abundance is determined, we measure the equivalent width of the line produced by the element of interest by creating a synthetic spectrum without considering other lines into account.
To estimate the measurement uncertainty due to the noise and the stellar parameters, the measured equivalent width is then fed into the same process as that applied for species analysed through equivalent widths analysis.

\subsection{Abundances: results \label{sec:abres}} 

Abundances are provided in Table~\ref{tab:abundance} and shown as a function of metallicity in Figures~\ref{fig:abtrend0}, \ref{fig:abtrend1}, \ref{fig:abtrend2}, \ref{fig:abtrend3}, and \ref{fig:abtrend4}.
Our stars are compared with turn-off halo stars studied by \citet{Nissen2010,Nissen2011}, \citet{Fishlock2017a}, and \citet{Reggiani2017a} to qualitatively compare our results with already studied halo populations.
It is necessary to account for systematic uncertainties in addition to those reported in this study when our abundances are quantitatively compared with other studies (see Appendix \ref{appendix3}). 
We also note that there could be systematic offsets in the absolute abundance scale because of differences in the linelist or in spectral types of the targets.

Following \citet{Nissen2010}, we divide our sample using Mg abundance (Figure~\ref{fig:abtrend0}).
Firstly, our sample is divided by the metallicity at [{Fe}/{H}]$\sim-1.7$ (metal-poor/others).
For the metal-rich sample, we further divide the sample by [{Mg}/{Fe}]: stars having [{Mg}/{Fe}] comparable to the metal-poor subsample (high-$\alpha$), those having lower [{Mg}/{Fe}] (low-$\alpha$), and those having even lower [{Mg}/{Fe}] (very low-$\alpha$). 
While [{Mg}/{Fe}] spreads among each of the three metal-rich subsamples are small (0.03 and 0.02 dex for high- and low- $\alpha$ subsamples) and comparable to the measurement uncertainties (typically 0.03 dex), differences in average [{Mg}/{Fe}] between different subsamples are significantly larger than measurement uncertainties (Table~\ref{tab:absignificance})\footnote{We note that the low-$\alpha$-stars in Table~\ref{tab:absignificance} and in subsequent similar tables do not include the very low-$\alpha$ stars.}.
However, this division is arbitrary, and we quantify the difference and investigate if the three subsamples also show differences in other element abundances or in kinematics in Section \ref{sec:dis_pop}.

There are a handful of Na-enhanced objects at [{Fe}/{H}]$<-1$ (bottom panel of Figure~\ref{fig:abtrend1}; KIC8350894, KIC5184073, and KIC6520576), for which spectra around Na I 5688 $\mathrm{\AA}$ are shown in Figure~\ref{fig:na}.
The Na enhancements are clear and cannot be attributable to cosmic ray or bad pixels.
While tabulated Na abundances in Table~\ref{tab:abundance} are in 1D/LTE, it is known that Na suffers from large NLTE effect \citep{Lind2011a}.
NLTE corrections (NLTE $-$ LTE) from \citet{Lind2011a} \footnote{\url{http://www.inspect-stars.com}} are $-0.114,\, -0.141\,\mathrm{and}\,-0.232$ dex for KIC6520576, KIC5184073, and KIC8350894, while those for their comparison stars in Figure~\ref{fig:na} are $-0.06,\,-0.115,\,\mathrm{and}\,-0.145$ dex, respectively.
Therefore, the NLTE effect cannot be the cause of the large Na abundance.
The origin of these stars are discussed in the Section~\ref{sec:dis_na}.

Other two objects (KIC5446927 and KIC10096113) at [{Fe}/{H}]$\sim -0.7$ also have [{Na}/{Fe}]$\sim 0.4$, which is similar to the value found for one of the previous mentioned Na-enhanced objects. 
However, their [{Na}/{Fe}] values are not distinctly high compared to other stars at similar metallicity.
It is still interesting that these two stars are significantly more massive than the rest of the stars in our sample (see Section~\ref{sec:massive}.

The Si abundance ratio exhibits large scatter and an increasing trend toward low metallicity (top left panel of Figure~\ref{fig:abtrend2}), which is not seen in previous studies \citep[e.g., ][]{Zhang2011a}.  
The NLTE corrections for Si are not expected to be large for giants according to \citet{Shi2009a}, who derived exactly the same abundance in LTE and NLTE calculations for a metal-poor giant.
The high Si abundance at low metallicity would be related to the Si abundance of HD122563 adopted in this work.
Our adopted value of [{Si}/{Fe}]$=0.557$ falls within the range reported in literature, which spans from 0.30 to 0.72 dex after correcting for different solar abundance assumed \citep{Cayrel2004a,Honda2004a,Johnson2002a,Westin2000a,Aoki2005a,Mishenina2001a,Barbuy2003a,Aoki2007a,Fulbright2000a,Roederer2014,Sakari2018a}\footnote{Data are collected via the SAGA database \citep{Suda2008}}.
On the other hand, stars with high Si abundance at [{Fe}/{H}]$\sim-1.2$ are not affected by the choice of the Si abundance of HD122563.
Typically their Si abundance relies on four Si~I lines and all the lines consistently provide high Si abundance.
Therefore, their Si abundance could be astrophysical rather than artificial.
\citet{FernandezTrincado2020a} recently suggested that there is a population of stars in the inner halo with enhanced [{Si}/{Fe}] ratios and that the population is remnants of globular clusters being disrupted.

Small offsets between our results and the literature are also found for other elements, e.g., in Ti, Sc and Ni (Figures~\ref{fig:abtrend2} and \ref{fig:abtrend3}).
The offsets might reflect systematic uncertainties in both our study and the literature. 
Our discussion is not significantly affected by the systematic uncertainties, since our main interest is to compare abundance ratios of different stellar populations at a given metallicity within our sample.

\begin{figure*}
\plotone{\figureloc 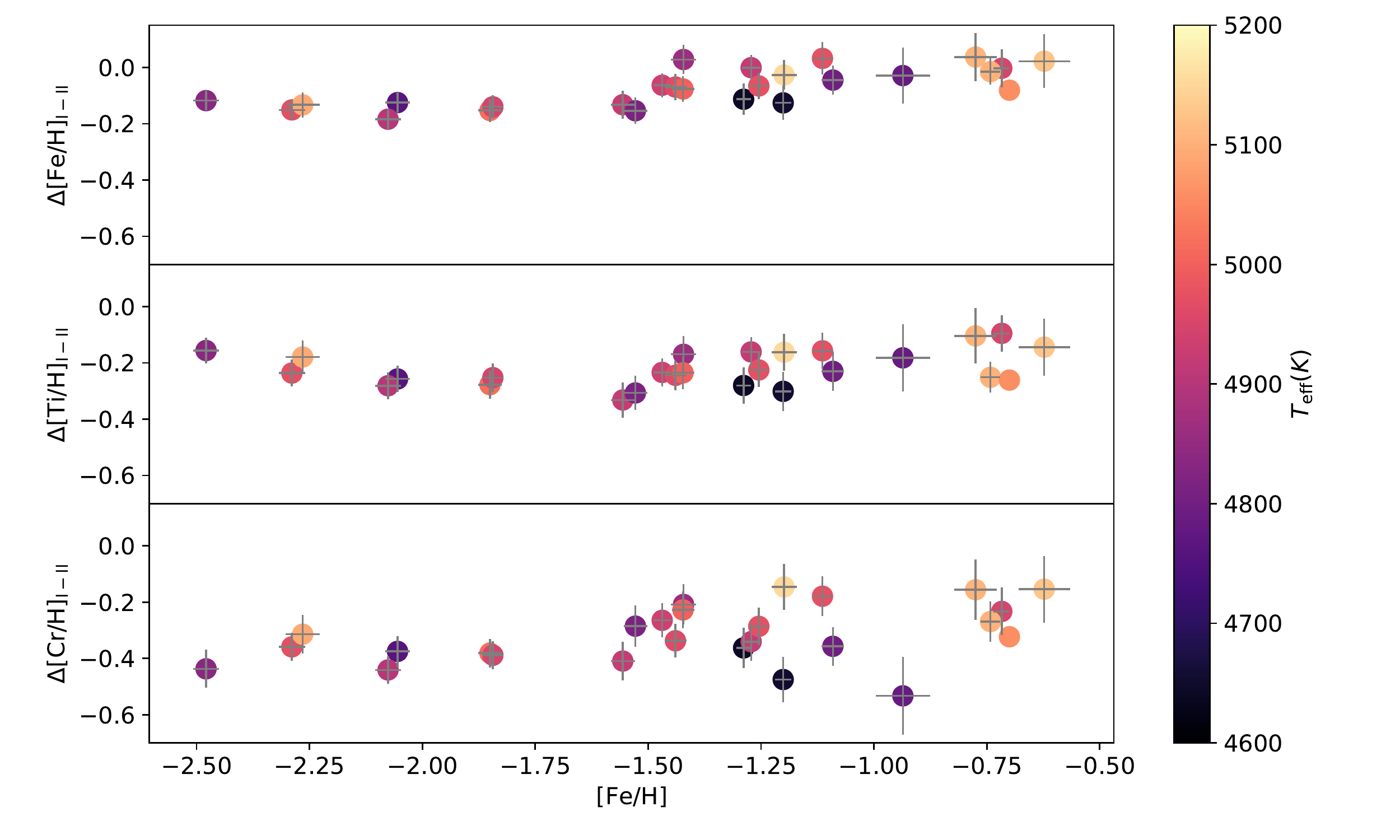}
\caption{Abundance difference between those derived from neutral species and ionized species. The color shows the temperature of the stars. \label{fig:ion}}
\end{figure*}

We investigate ionization balance for elements for which absorption lines of both neutral and singly-ionized species are detected, namely Ti, Cr, Fe (Figure~\ref{fig:ion}).
The derived abundance difference, $[\mathrm{X/H}]_{\rm I}-[\mathrm{X/H}]_{\rm II}$, shows non-zero offsets for all the three elements.
The offsets are in the sense that abundances derived from neutral species are smaller than those from ionized species, which might indicate they could be due to the NLTE effect.
The differences correlate with metallicity, indicating that the amplitude of the NLTE effect mainly depends on the metallicity.
These features are consistent with results of NLTE calculations \citep[e.g.,][]{Bergemann2010a,Bergemann2011a,Amarsi2016a}, which predict that neutral species are over-abundant in LTE calculations and that lower abundances are obtained from neutral species than from ionized species.
While grids of NLTE corrections are available for a subset of stars, most of them do not cover the entire range of stellar parameters spanned by our targets.
We therefore stick to 1D/LTE analysis in order to keep consistency within our sample. 

\section{Discussion}\label{sec:discussion}

\subsection{Halo subpopulations\label{sec:dis_pop}}

\begin{deluxetable*}{lrrrrrrrrrr}
  \tablecaption{Abundance differences \label{tab:absignificance}}
  \tablehead{
 & \multicolumn{3}{c}{high-$\alpha$} & & \multicolumn{3}{c}{low-$\alpha$} & & &\\\cline{2-4}\cline{6-8}
Element & $\langle[\mathrm{X/Fe}]\rangle$  & $\sqrt{\rm Var}$ & $N$ & & $\langle[\mathrm{X/Fe}]\rangle$ & $\sqrt{\rm Var}$ & $N$ & $\Delta \langle[\mathrm{X/Fe}]\rangle$ & $\sigma([\mathrm{X/Fe}])_{\rm median}$ & $p$-value}
\startdata
MgI & 0.404 & 0.030 & 7 & &0.258 & 0.021 & 9 &0.146 & 0.035 & 0.000\\
CH & -0.052 & 0.054 & 7 & &-0.260 & 0.228 & 9 &0.207 & 0.082 & 0.027\\
OI & 0.644 & 0.068 & 3 & &0.443 & 0.157 & 4 &0.201 & 0.083 & 0.079\\
NaI & 0.086 & 0.153 & 7 & &0.112 & 0.341 & 9 &-0.025 & 0.081 & 0.845\\
SiI & 0.479 & 0.105 & 7 & &0.310 & 0.063 & 9 &0.169 & 0.036 & 0.004\\
CaI & 0.335 & 0.028 & 7 & &0.305 & 0.039 & 9 &0.030 & 0.022 & 0.097\\
TiI & 0.245 & 0.040 & 7 & &0.152 & 0.035 & 9 &0.093 & 0.024 & 0.000\\
TiII & 0.395 & 0.029 & 7 & &0.340 & 0.025 & 9 &0.055 & 0.035 & 0.002\\
ScII & 0.275 & 0.078 & 7 & &0.193 & 0.054 & 9 &0.082 & 0.036 & 0.039\\
VI & 0.101 & 0.069 & 7 & &0.017 & 0.069 & 9 &0.085 & 0.051 & 0.029\\
CrI & -0.073 & 0.039 & 7 & &-0.123 & 0.050 & 9 &0.050 & 0.032 & 0.042\\
CrII & 0.149 & 0.060 & 7 & &0.162 & 0.090 & 9 &-0.013 & 0.041 & 0.730\\
MnI & -0.302 & 0.100 & 7 & &-0.330 & 0.082 & 9 &0.028 & 0.028 & 0.561\\
CoI & 0.166 & 0.079 & 7 & &0.030 & 0.093 & 9 &0.136 & 0.080 & 0.007\\
NiI & 0.075 & 0.055 & 7 & &-0.027 & 0.039 & 9 &0.102 & 0.019 & 0.002\\
ZnI & 0.163 & 0.059 & 7 & &0.074 & 0.035 & 9 &0.089 & 0.096 & 0.006\\
CuI & -0.567 & 0.177 & 7 & &-0.810 & 0.147 & 9 &0.242 & 0.083 & 0.013\\
YII & 0.198 & 0.083 & 7 & &-0.030 & 0.091 & 9 &0.229 & 0.042 & 0.000\\
BaII & 0.213 & 0.107 & 7 & &0.171 & 0.165 & 9 &0.042 & 0.064 & 0.548\\
LaII & 0.329 & 0.174 & 6 & &0.294 & 0.156 & 9 &0.034 & 0.089 & 0.704\\
CeII & 0.196 & 0.109 & 6 & &0.177 & 0.121 & 9 &0.019 & 0.081 & 0.757\\
NdII & 0.310 & 0.073 & 7 & &0.398 & 0.138 & 9 &-0.088 & 0.061 & 0.125\\
SmII & 0.409 & 0.076 & 7 & &0.476 & 0.099 & 9 &-0.067 & 0.093 & 0.147\\
EuII & 0.459 & 0.099 & 7 & &0.577 & 0.129 & 9 &-0.119 & 0.083 & 0.056\\
\enddata
\end{deluxetable*}

We divided the sample into four subsamples in Section \ref{sec:abundance}.
We investigate kinematics and abundances of the four subsamples in this subsection.

We first focus on our high-$\alpha$ and low-$\alpha$ subsamples.
There are hints of abundance differences between the high-$\alpha$ and low-$\alpha$ subsamples in [{X}/{Fe}] of many elements (C, O, Na, Si, Ca, Sc, Ti, V, Co, Ni, Cu, Zn, and Y; Figures~\ref{fig:abtrend1}--\ref{fig:abtrend4}).
The abundance differences and their significance are quantified in Table~\ref{tab:absignificance}.
We note that, although the abundance differences listed in Table~\ref{tab:absignificance} may be compared with the Table~5 of \citet{Nissen2011}, a direct comparison is not possible due to the difference in metallicity coverage.
While the majority of our our sample is at [{Fe}/{H}]$\lesssim-1.1$, \citet{Nissen2011} defines the difference at $-1.1<[\mathrm{Fe/H}]<-0.7$.
This relatively lower metallicity of our sample leads to smaller abundance difference in general (e.g., $\Delta[\mathrm{Mg/Fe}]=[\mathrm{Mg/Fe}]_{\rm high-\alpha}-[\mathrm{Mg/Fe}]_{\rm low-\alpha}$ is 0.145 and 0.219 in our study and in their study, respectively). 
 
For most of these elements except for O, Na and Ca, the probability that the two subsamples have the same mean abundance is small ($P<0.05$; Table~\ref{tab:absignificance}), where the $P$-value is calculated by the $t$-test for means of two independent samples.
The limited sample size of O abundances (middle panel of Figure~\ref{fig:abtrend1}), which is due to telluric absorption lines, prevents us from establishing the statistical significance. 
The effect of the outlier is evident for Na (bottom panel of Figure~\ref{fig:abtrend1}) in Table~\ref{tab:absignificance}, where the square root of variance among each subsample is much larger than the typical measurement uncertainties.
The Ca abundance difference is not statistically significant nor as clear as that seen in \citet{Nissen2010} (top right panel of Figure~\ref{fig:abtrend2} and Table~\ref{tab:absignificance}).
Although Ca abundance difference is clear in \citet{Nissen2010}, \citet{Nissen2011} reported the difference to be small ($\Delta[\mathrm{Ca/Fe}]=0.074$ at $-1.1\lesssim [\mathrm{Fe/H}]\lesssim -0.7$).
Recall that our sample provides smaller $\Delta[\mathrm{X/Fe}]$ for other elements because of lower metallicity.

Other elements do not exhibit clear differences. 
Cr and Mn abundance differences (Figure~\ref{fig:abtrend3}) are reported to be small (0.03--0.04 dex) by \citet{Nissen2011}; therefore, the lack of difference would be due to the lack of intrinsic difference.
The abundance differences in neutron-capture elements (Ba, La, Ce, Nd, Sm) are not detectable (Figure~\ref{fig:abtrend4}) partly because of the large intrinsic dispersion in each of the two subsamples as can be seen in the square root of variances and small intrinsic abundance difference ($\Delta\langle[\mathrm{X/Fe}]\rangle$) in Table~\ref{tab:absignificance}.

These results are consistent with abundance differences between turn-off high-$\alpha$ and low-$\alpha$ populations reported in the literature \citep{Nissen2010,Nissen2011,Nissen2014a}.
This suggests that our high-/low-$\alpha$ subsamples correspond to the two distinct halo populations reported by \citet{Nissen2010}.

These two subsamples also differ in stellar kinematics (Figure~\ref{fig:kinematics}).
The radial component of the velocity ($v_R$) of stars in our low-$\alpha$ subsample shows completely different distributions having large absolute values when compared to the high-$\alpha$ subsample.
This suggests that the low-$\alpha$ subsample has a more radial orbit, which is a signature of Gaia Sausage \citep{Belokurov2018a}.
Since Gaia Sausage is considered to correspond to the low-$\alpha$ population of \citet{Nissen2010}, kinematics also support the correspondence of our low-$\alpha$ subsample and the low-$\alpha$ population of \citet{Nissen2010}. 
Although \citet{Nissen2010} reported that the high-$\alpha$ and low-$\alpha$ populations tend to be prograde and retrograde, respectively, we do not detect such difference in $v_\phi$.
However, we caution that the $v_\phi$ distribution is strongly affected by our sample selection.
The selection is partly based on the observed heliocentric radial velocity, which highly correlates with $v_\phi$ because of the galactic longitude of the \textit{Kepler} field.
Our selection has a bias against stars on prograde orbit.

We here discuss the very low-$\alpha$ subsample.
This subsample consists of three stars, KIC5953450, KIC9335536, and KIC9583607, among which KIC9335536 is located at [{Fe}/{H}]$\sim -1.5$ and the others are at [{Fe}/{H}]$\sim -0.6$ (Figure~\ref{fig:abtrend0}).
While these three stars are selected as very low-$\alpha$ subsample based on Mg abundance, other element abundances also seem to behave differently from other subsamples with a tendency of being extreme cases of the low-$\alpha$ subsample discussed above.
The most metal-poor star among this subsample, KIC9335536, shows large retrograde motion as well as a relatively large $\sqrt{v_R^2+v_z^2}$.
Its metallicity, low Mg and Ca abundances (Figure~\ref{fig:abtrend0} and top right panel of Figure~\ref{fig:abtrend2}), and kinematics suggest its association with Sequoia \citep{Myeong2019a,Matsuno2019a}, a kinematic substructure suggested to be a signature of a dwarf galaxy accretion.
Kinematics of the other two stars characterized by high $v_{R}$ basically follow the trend of the low-$\alpha$ subsample (Figure~\ref{fig:kinematics}).
While these results indicate that the very low-$\alpha$ subsample is clearly different from the high-$\alpha$ population, it is still unclear if it can be regarded as a separate component from the low-$\alpha$ subsample.

The metal-poor subsample is chemically homogeneous to some extent; there are tight trends of [{X}/{Fe}] with [{Fe}/{H}] for many of the elements studied (Figures~\ref{fig:abtrend0}--\ref{fig:abtrend4}).
The dispersions around the linear fit in [{X}/{Fe}]--[{Fe}/{H}] are significant at more than 3$\sigma$ only for CH (probably because of evolutionary effect), Na (because of the outlier) in Figure~\ref{fig:abtrend1}, and neutron capture elements, Y, Ba, Nd, and Eu (Figure~\ref{fig:abtrend4}).
This is consistent with the result reported by \citet{Reggiani2017a}, who conducted high-precision abundance analysis for metal-poor turn-off stars concluding that there is no significant scatter in abundances in most of the elements for the main halo population.

It is not clear from chemical abundances if the metal-poor subsample corresponds to the low metallicity extension of other subsamples. 
On the other hand, kinematics are more similar to the high-$\alpha$ subsample, which might be related to the fraction of metal-poor stars in the low-$\alpha$ or high-$\alpha$ populations. 
We note, however, that radial velocity was not taken into consideration in the sample selection for the lowest metallicity stars (Figure~\ref{fig:selection}), which would introduce a bias in their distribution in the velocity space.

\subsection{How reliable is the asteroseismic mass? \label{sec:dis_seis}}

\begin{figure*}
  \plotone{\figureloc 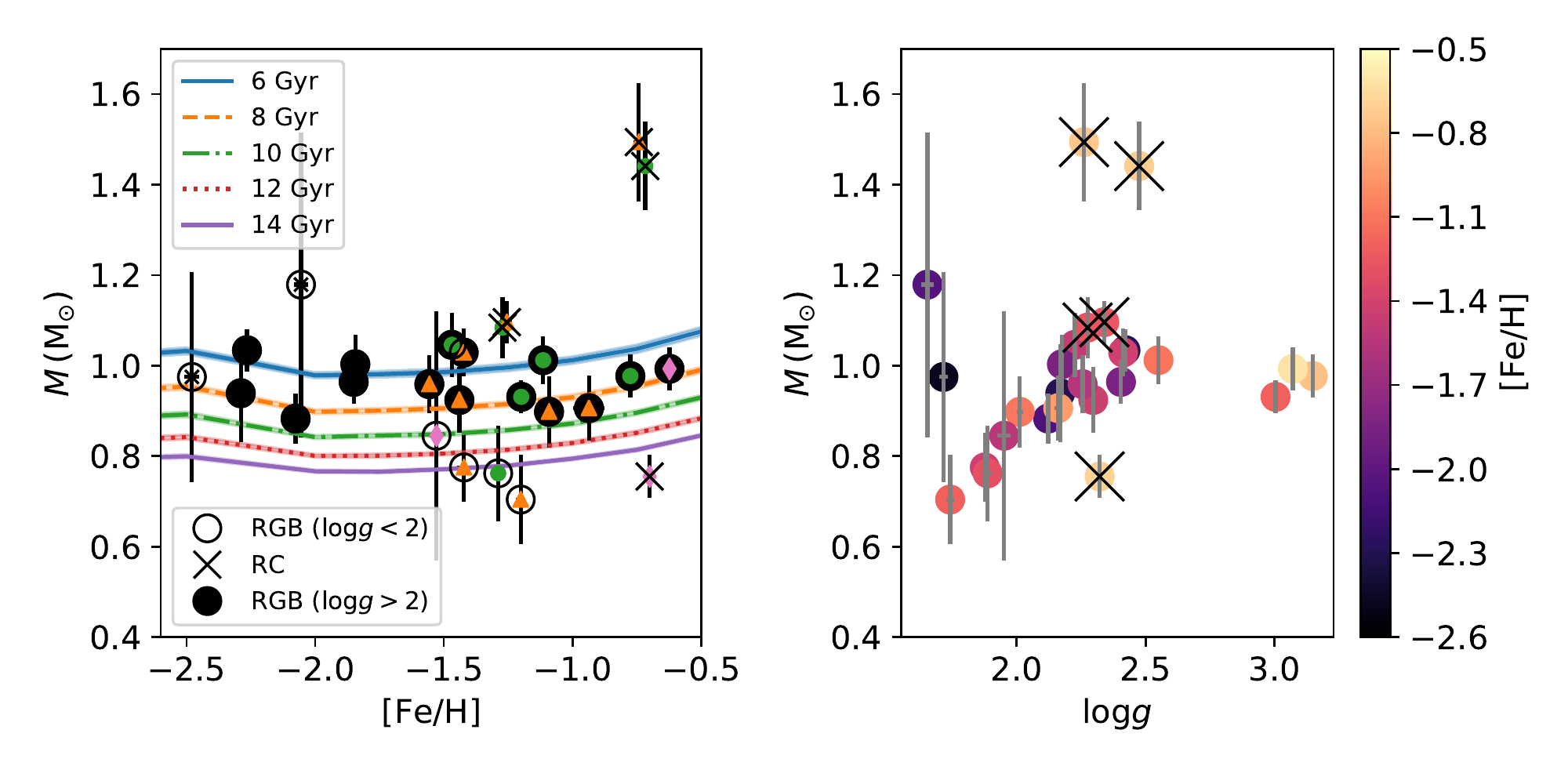}
  \caption{(\textit{Left:}) Mass as a function of metallicity. Red clump stars, luminous stars, and the other red giant branch stars are separated. 
Small symbols are over plotted according to the Mg and Fe classification in Section~\ref{sec:abres}.
We present initial mass range for stars to be on the red giant branch at given age and metallicity using MIST isochrones \citep[][ see text for more details.]{Dotter2016a,Choi2016a}. Note that while each theoretical line has a width reflecting the mass range of stars along the red giant branch, the width is too narrow to be visible in most cases. (\textit{Right:}) Mass as a function of surface gravity (the points are color-coded by the metallicity). Cross symbols are over-plotted for red clump stars.\label{fig:mass_feh}}
\end{figure*}

In this subsection, we discuss reliability of mass estimates from the scaling relations of asteroseismology. 
Previous studies on this for low-metallicity stars include \cite{Epstein2014a}, \citet{Casey2018a}, \citet{Miglio2016a}, and \citet{Valentini2019a}.
Here we have 26 halo stars, of which five are below [{Fe/H}]$<-2$ and additional 16 are below [{Fe/H}]$<-1$.
As far as we know, this is the largest sample of metal-poor stars for which asteroseismology and high-resolution spectroscopy are combined.
We take advantage of our sample to re-visit the asteroseismology at low-metallicity.
We firstly study possible systematic bias in the estimated mass that is dependent on evolutionary status by separating stars to luminous giants ($\log g<2$), less luminous giants ($\log g>2$), and red clump stars.
We also discuss how accurately the scaling relations can estimate stellar masses by comparing the derived masses and the expected masses for our sample using theoretical stellar evolution models.

In Figure~\ref{fig:mass_feh}, we investigate if the estimated mass correlates with metallicity or surface gravity.
We also present the range of initial stellar masses within which stars are on the red giant branch phase at given age and metallicity using MESA Isochrones and Stellar Tracks \citep[MIST; ][]{Dotter2016a,Choi2016a}, which is based on Modules for Experiments in Stellar Astrophysics \citep[MESA; ][]{Paxton2011a}. 
Since halo stars are generally considered to be old from independent studies \citep[$\gtrsim 10\,\mathrm{Gyr}$; e.g.,][]{Jofre2011a,Schuster2012,VandenBerg2013a,Carollo2016a,Kilic2019a}, we present the mass as a function of metallicity for $6,\,8,\,10,\,12,\mathrm{and}\,14\,\mathrm{Gyr}$ isochrones.  
Red giant stars are selected based on surface gravity ($\log g<3.3$) and the \texttt{phase} parameter as they are on either of red giant branch phase, core He burning phase, early AGB phase, or thermal-pulsing AGB phase (\texttt{phase}$=2,\,3,\,4,\mathrm{and}\,5$).
The range of initial mass of red giant stars is measured for each isochrone and shown in Figure~\ref{fig:mass_feh}.
Although the mass loss is not considered, its effect remains small ($\lesssim 0.01M_{\odot}$) until a hydrogen shell-burning low-mass star reaches $\log g=2.0$ while ascending the red giant branch in the case of MIST stellar evolutionary tracks, which assume $\eta=0.1$ as the Reimer's mass-loss efficiency parameter.
The mass lost remains small ($<0.015\,\mathrm{M_{\odot}}$) 
even when considering another stellar evolution model, BaSTI \citep{Pietrinferni2004a,Pietrinferni2006a}, which assumes higher mass-loss efficiency, $\eta=0.4$.
This is because mass-loss is assumed to become important only when stars evolved close to the tip of the RGB.

In the left panel of Figure~\ref{fig:mass_feh}, the large scatter and/or the presence of outliers are evident in particular at high metallicity ([{Fe}/{H}]$>-1.7$), while, at low metallicity, the scatter seems to be mostly due to the measurement uncertainties\footnote{This can be confirmed by calculating reduced $\chi^2$ ($\chi_\nu$). We obtain $\chi_\nu=0.73$ (7 stars) for the metal-poor sample and $\chi_\nu=1.65$ (14 stars) for the metal-rich stars. Red clump stars are excluded in this $\chi_\nu$ calculations.}.
Although the scatter might indicate a presence of a significant age dispersion, it could also be caused by other effects, such as systematic bias as a function of evolutionary status.

In order to inspect if this spread is due to the evolutionary status, the right panel of Figure~\ref{fig:mass_feh} visualizes the correlation between mass and surface gravity. 
This panel suggests that masses of luminous giants ($\log g\lesssim 2.0$, equivalently $\Delta \nu\lesssim 2.0\,\mathrm{\mu\,Hz}$) are systematically lower or underestimated ($\langle M\rangle=0.78\,\mathrm{M_{\odot}}$) than less luminous hydrogen shell-burning red giant branch stars ($\langle M\rangle=0.97\,\mathrm{M_{\odot}}$), although uncertainties are large for some luminous stars due to low oscillation frequencies.
Red clump stars also show a large dispersion in mass including the two obviously over-massive stars, whose origin remains unclear (see Section~\ref{sec:massive}).
Possible systematic mass offsets for these luminous giants and red clump stars are also suggested by, e.g., \citet{Pinsonneault2018a}.
This offset could be due to mass loss or systematic uncertainties that only affect these stars. 

\begin{figure}
  \plotone{\figureloc 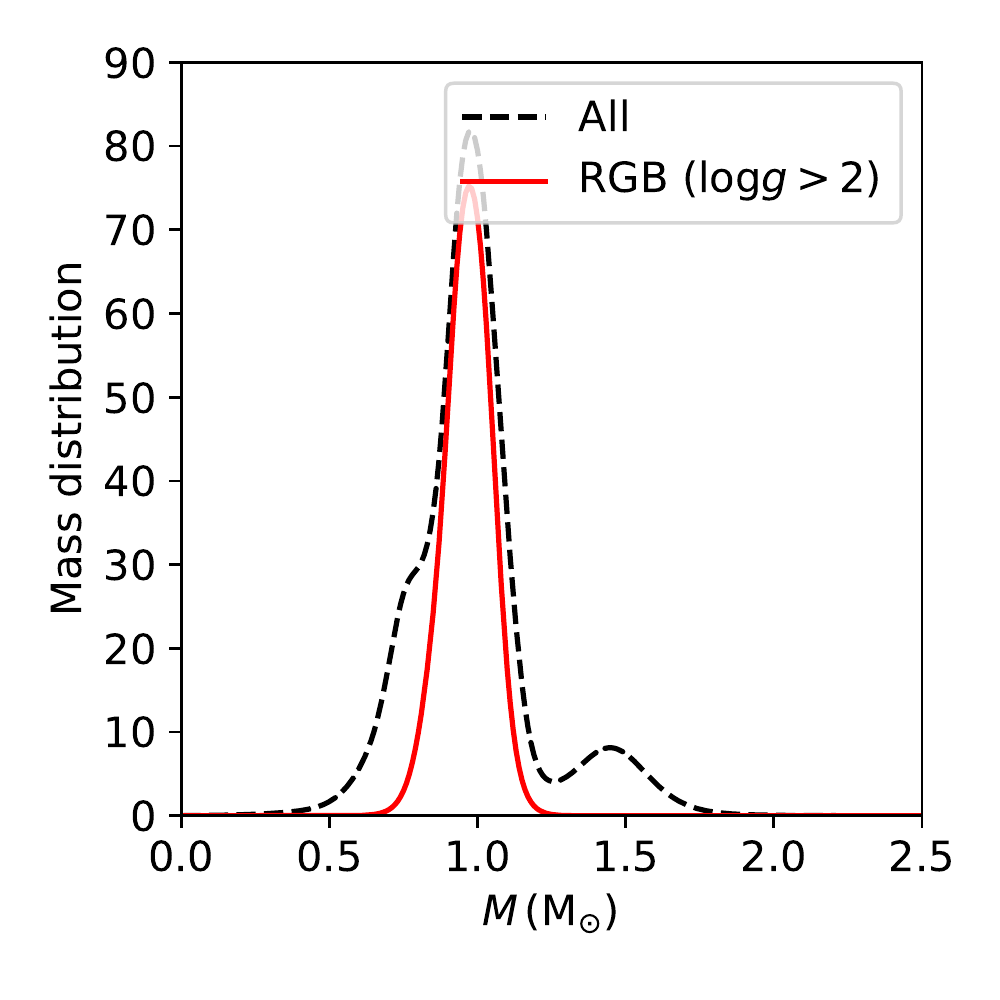}
  \caption{The distribution of mass of our program stars. The distribution functions are the sum of gaussian functions, each of which corresponds to one star. Each gaussian is normalized to 1 and centered to the median estimate of the mass, and has a width of the corresponding uncertainty.\label{massdf}}
\end{figure}

Considering these possible effects of evolutionary status on the mass estimates and uncertainties in modelling of mass loss, we separate red clump stars and luminous giants ($\log g<2.0$) from the other 15 red giant branch stars.
The estimated masses of less luminous red giant branch stars show smaller dispersion.
The $\chi^2$-test for the 15 stars indicates that the mass dispersion is insignificant ($P=0.65$), whereas there is a significant dispersion in mass ($P<0.0001$) when we consider all the stars (26 stars). 
The distribution of mass is also shown in Figure~\ref{massdf}, which illustrates a large scatter when all the stars are considered and a small scatter when we focus on less luminous red giant branch stars.
These results indicate that the scaling relations provide consistent mass estimates within less luminous giants ($\log g>2$); otherwise, the dispersion should be significantly larger than measurement uncertainties.
We hereafter focus on less luminous giants when discussing stellar masses otherwise stated and consider that a comparison within this subset is not significantly affected by the systematic bias influenced by evolutionary status.

We now discuss the absolute accuracy of our estimated masses. 
Even though mass dispersion disappears by restricting the sample to less luminous giants, the average mass ($0.97\,\mathrm{M_{\odot}}$) is still higher by $\sim10\%$ than the value expected for the age of halo stars ($0.8-0.9\,\mathrm{M_{\odot}}$; see the left panel of Figure~\ref{fig:mass_feh}).
This offset cannot be attributed to stellar parameters.
In order to reduce the derived mass by 10\%, we would need to decrease $T_{\rm eff}$ by 15\%, which corresponds to $\sim750\,\mathrm{K}$.
Recall that our stellar parameters were obtained in a differential manner with respect to well-calibrated stars, and the median uncertainties in $T_{\rm eff}$ are $40\,\mathrm{K}$ for this subset of stars.
Although the mass loss does not affect the discussion, it makes the situation worse if taken into account.

We note that the offset found in the present study might not be unique to low-metallicity stars.
Several studies reported that the asteroseismic scaling relations provide systematically larger masses for red giants in eclipsing binary systems than dynamical mass estimates \citep{Brogaard2016a,Gaulme2016a,Themesl2018a,Hekker2020a}.
\citet{Gaulme2016a} found that the asteroseismic mass estimates are systematically larger by 13-17\% even when corrections, including the one used in the present study, are applied to the scaling relations.
Although the correction suggested by \citet{Rodrigues2017a} is not discussed in \citet{Gaulme2016a}, we note that we find consistent mass between \citet{Valentini2019a}, who make use of the correction of \citet{Rodrigues2017a}, and our study when the same frequencies are used.
\citet{Themesl2018a} suggested to use a lower reference large frequency separation of $\Delta\nu=130.8\,\mathrm{\mu Hz}$ instead of the measured solar value of $\Delta\nu_\odot = 135.1\,\mathrm{\mu Hz}$ from an empirical approach, which would also lower the derived masses by $\sim14\%$.
If our derived masses are lowered by $\sim 14\%$, the average mass would be consistent with the expected mass for red giants in the Milky Way halo.
These results do not contradict with \citet{Zinn2019b}, who also showed that the scaling relations overestimate stellar masses for low-metallicity stars compared to the mass estimates from the Eq. \ref{eq:delnu}, where $R$ is from $L\propto R^2T_{\rm eff}$ and $L$ is partly based on the Gaia DR2 parallax.

Although there seems to be a systematic offset, the small scatter in masses obtained for stars limited to less luminous red giants indicates that the mass of low-metallicity stars can be estimated with high internal precision.
Therefore, we should be able to explore the history of the Milky Way halo with accurate stellar ages of red giants once the systematic offset is resolved by future studies. 

\subsection{Formation timescales of the halo}
In this subsection, we discuss formation timescales of the Galactic halo focusing on our high-$\alpha$ and low-$\alpha$ subsamples. 
We first discuss constraints from chemical abundances and then discuss those from stellar mass.
Note that the chemical abundance is discussed with the whole sample, while the stellar mass is discussed based on a subsample of stars because of possible evolutionary-status dependent offsets in the measurements.
Appendix \ref{sec:RC} verifies that chemical abundance is not correlated with evolutionary status, which ensures our use of the entire sample for its discussion.
Since we have shown that these two subsamples correspond to the low-$\alpha$ and high-$\alpha$ populations of \citet{Nissen2010}, we refer each subsample as a population.

\subsubsection{Constraints from abundances}

\begin{figure*}
  \plotone{\figureloc 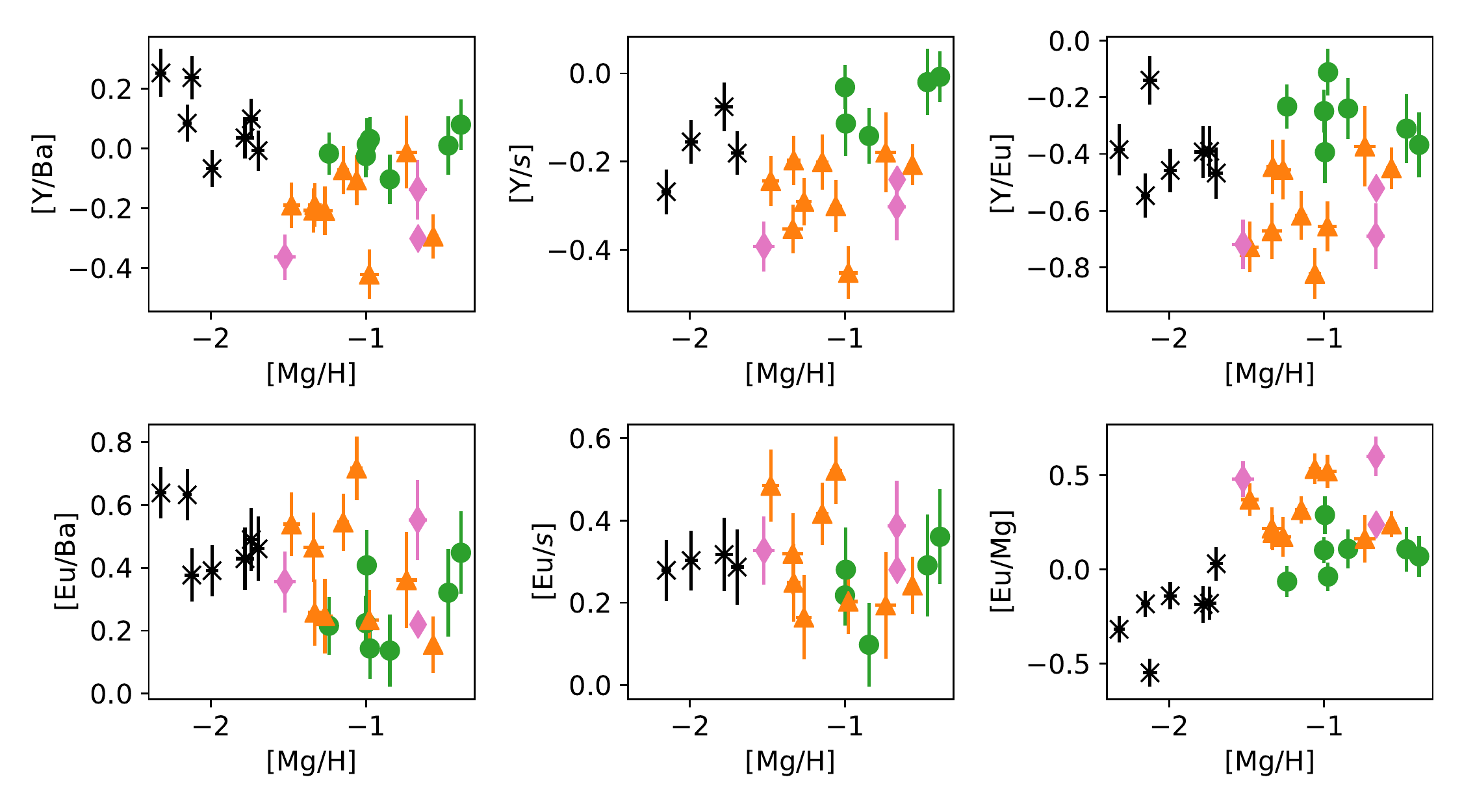}
  \caption{Abundance trends of neutron-capture elements as a function of [{Mg}/{H}]. Here $s$ is the average of Ba, La, Ce, and Nd. The predicted solar abundance is [{Eu}/{Ba}]$=0.94$ and [{Eu}/$s$]$=0.70$ for pure $r$-process and [{Eu}/{Ba}]$=-1.26$ and [{Eu}/$s$]$=-1.20$ for pure $s$-process \citep{Prantzos2019a} \label{fig:ncap}}
\end{figure*}

\begin{figure*}
  \plotone{\figureloc 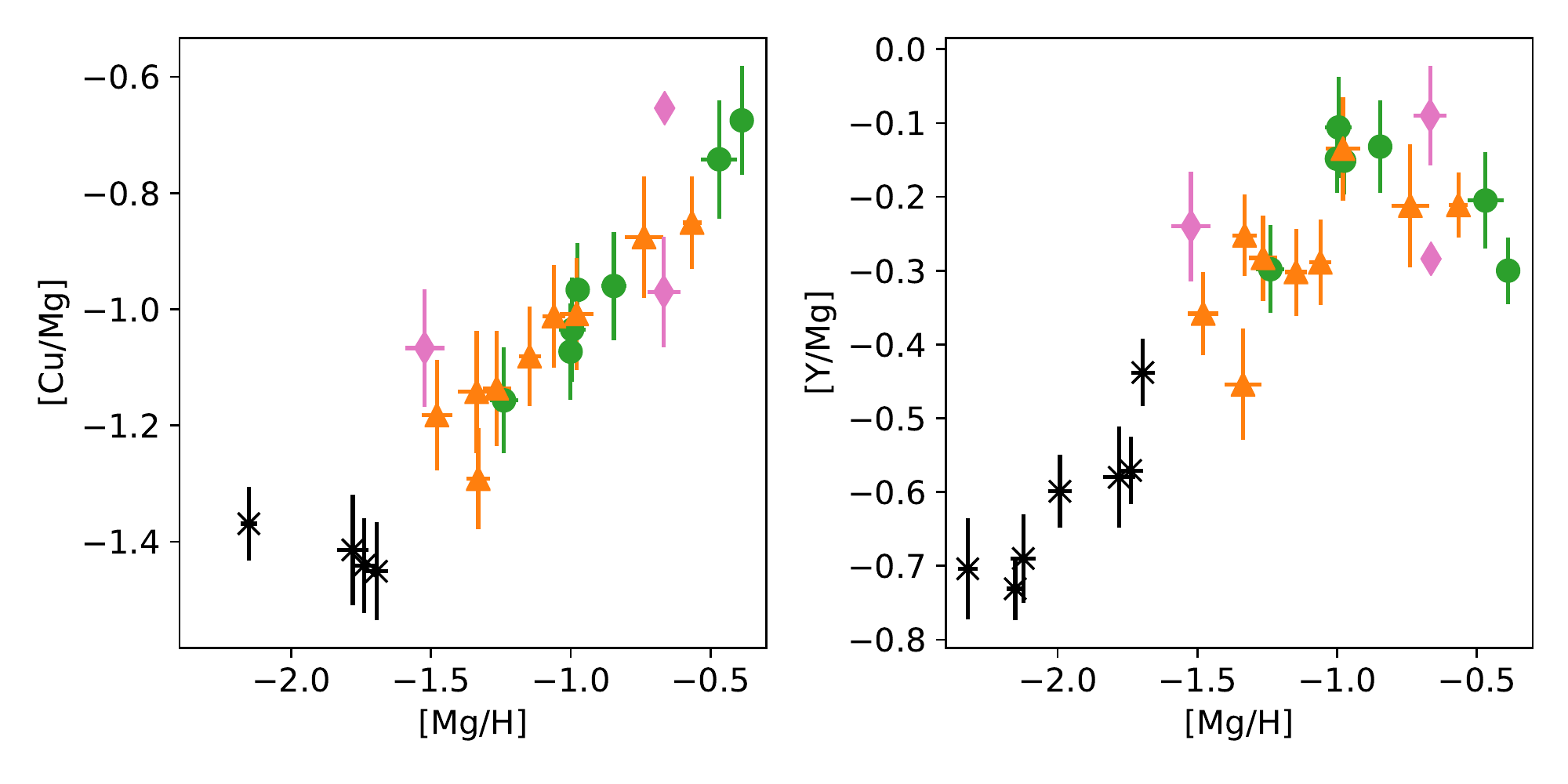}
  \caption{Abundance trends of Cu and Y as a function of [{Mg}/{H}]. \label{fig:CuY}}
\end{figure*}

\begin{deluxetable*}{lrrrrrrrrrr}
  \tablecaption{Differences of abundance ratios in Figure~\ref{fig:ncap} \label{tab:absignificance2}}
  \tablehead{
 & \multicolumn{3}{c}{high-$\alpha$} & & \multicolumn{3}{c}{low-$\alpha$} & & &\\\cline{2-4}\cline{6-8}
Ratio & $\langle[\mathrm{X/Y}]\rangle$  & $\sqrt{\rm Var}$ & $N$ & & $\langle[\mathrm{X/Y}]\rangle$ & $\sqrt{\rm Var}$ & $N$ & $\Delta \langle[\mathrm{X/Y}]\rangle$ & $\sigma([\mathrm{X/Y}])_{\rm median}$ & $p$-value}
\startdata
$[\mathrm{Y/Ba}]  $& -0.003 & 0.055 & 7 & &-0.202 & 0.111 & 9 &0.198 & 0.081 & 0.000\\
$[\mathrm{Y}/s]   $& -0.057 & 0.060 & 5 & &-0.272 & 0.087 & 9 &0.215 & 0.058 & 0.000\\
$[\mathrm{Y/Eu}]  $& -0.251 & 0.094 & 7 & &-0.592 & 0.148 & 9 &0.341 & 0.093 & 0.000\\
$[\mathrm{Eu/Ba}] $& 0.250 & 0.115 & 7 & &0.393 & 0.197 & 9 &-0.143 & 0.103 & 0.092\\
$[\mathrm{Eu}/s]  $& 0.237 & 0.093 & 5 & &0.322 & 0.133 & 9 &-0.085 & 0.097 & 0.187\\
$[\mathrm{Eu/Mg}] $& 0.065 & 0.117 & 7 & &0.321 & 0.140 & 9 &-0.256 & 0.091 & 0.001\\
\enddata
\end{deluxetable*}

The [$\alpha$/{Fe}] difference, such as seen in Figure~\ref{fig:abtrend0}, is usually interpreted as a result of different contributions from Type~Ia supernovae (SNe~Ia), which is a result of different star formation timescale \citep{Nissen2010}.
Differences in other elemental abundances could also be by SNe~Ia contributions.
For example, Na and Sc in the bottom panel of Figure~\ref{fig:abtrend1} and in the top left panel of Figure~\ref{fig:abtrend3}, which are not $\alpha$-elements, are mostly synthesized by massive stars and ejected by Type~II supernovae (SNe~II) having similar nucleosynthesis origins as $\alpha$-elements.
On the other hand, some other elements, especially neutron-capture elements, could have different nucleosynthesis origins, which would deliver independent information for estimates of formation time scale of stellar systems.

Figure~\ref{fig:ncap} shows abundance ratios between neutron capture elements \citep[see also ][]{Fishlock2017a} as a function of [{Mg}/{H}].
Associated abundance difference and significance are quantified in Table~\ref{tab:absignificance2} in the similar manner as in Table~\ref{tab:absignificance}.
Y is a light neutron capture element, which is considered to be formed by the weak $s$-process in massive stars \citep[e.g., ][]{Pignatari2010a}.
The [{Y}/{Fe}] ratio is generally low in the low-$\alpha$ population (top left of Figure~\ref{fig:abtrend4}).
Abundance ratios relative to other elements, [{Y}/{Ba}], [{Y}/{$s$}]\footnote{The $s$ abundance is the average of Ba, La, Ce, and Nd.}, and [{Y}/{Eu}] (top row of Figure~\ref{fig:ncap}), are also low, indicating underproduction of Y within the progenitor of the low-$\alpha$ population.
Similar behavior is also observed at low metallicity in dwarf galaxies \citep{Skuladottir2019a}.

The low abundance of Y as well as that of Cu (bottom right of Figure~\ref{fig:abtrend3}) in the low-$\alpha$ population is interpreted as a result of low-efficiency of the weak-$s$ process as discussed in \citet{Nissen2011}. 
The $^{22}$Ne($\alpha$,$n$)$^{25}$Mg reaction is the source of neutrons in the weak $s$-process.
Since the $^{22}$Ne is produced from CNO elements through H burning and $\alpha$ captures to $^{14}$N, this process is dependent on CNO abundance and only efficient at [{C,N,O}/{H}]$\gtrsim -2$ \citep[e.g.,][]{Prantzos1990a}.
Hence, the products of the weak $s$-process like Cu and Y have a secondary nature.
Figure~\ref{fig:CuY} shows chemical evolution of these two elements in relation to Mg.
We here choose Mg instead of Fe because both CNO and Mg are mostly produced by SNe~II, whereas Fe is also produced by SNe~Ia.
Figure~\ref{fig:CuY} demonstrates that Cu and Y abundances show tighter correlations with Mg abundance than with Fe abundance.
In addition, abundance ratios of both low-$\alpha$ and high-$\alpha$ stars distribute on the same line.
These results suggest that the low Cu and Y abundances of the low-$\alpha$ population could be due to their low yield in the weak $s$-process, which may be the result of the low CNO and Mg abundances at a given [{Fe}/{H}]\footnote{We note that we have not measured N abundances for the targets. We assumed that relative [{N}/{Fe}] values between the low-/high-$\alpha$ populations follow the trend of [{C}/{Fe}] and [{O}/{Fe}].}.
The two elements show different behavior at high metallicity ([{Mg}/{H}]$\gtrsim -1$), where [{Y}/{Mg}] seems to be constant or have turn-over while [{Cu}/{Mg}] keeps increasing.
This behavior might be explained by the primary-like behavior of Y at high metallicity when nucleosynthesis by rotating massive stars are considered. \citep{Prantzos2018a}.
Therefore, we tentatively conclude that the star formation timescales of the low-/high-$\alpha$ populations and abundance differences of Cu and Y in Figures~\ref{fig:abtrend3}, \ref{fig:abtrend4} and \ref{fig:ncap} could be indirectly related through different [{C},{N},{O}/{Fe}] abundances.

We note that Cu abundance can significantly be affected by the NLTE effect \citep{Shi2018a,Andrievsky2018a}.
For example, \citet{Shi2018a} reported that the [{Cu}/{Fe}] abundance of HD122563 is underestimated in LTE analysis by $0.3-0.4$ dex compared to NLTE analysis when the absorption line at $5105\,\mathrm{\AA}$ is used. 
Although the trend with the metallicity could change when studied with a NLTE analysis, the discussion presented here is mostly based on the relative abundance difference between the low-$\alpha$ and high-$\alpha$ populations at a given metallicity, which should be less affected by the NLTE effect \citep[e.g.,][]{Yan2016a}.

We now discuss heavy neutron-capture elements presented in Figure~\ref{fig:abtrend4}, starting from an $r$-process element, Eu, and then moving to $s$-process elements.
Eu is an almost pure $r$-process element.
Eu is significantly enhanced relative to Y or Mg in the low-$\alpha$ population by $\sim 0.3\,\mathrm{dex}$ (bottom row of Figure~\ref{fig:ncap}, and Table~\ref{tab:absignificance2}), as suggested in \citet{Fishlock2017a} and \citet{Ishigaki2013}. 
The difference of Eu abundance ratios between the low-$\alpha$ and high-$\alpha$ populations could be understood if there is a delay time in $r$-process enrichments; the long star formation timescale of the low-$\alpha$ population might have allowed the delayed $r$-process enrichments to contribute to the chemical evolution.
The delay time of $r$-process events is, however, still under debate \citep{Hotokezaka2018a,Cote2019a,Skuladottir2020a,Lin2020a}. 
While neutron star mergers, which are associated with short gamma ray bursts (GRBs), are promising $r$-process sites, these studies suggested that there should be more $r$-process production events in the early Universe than the rate predicted based on the observations of GRBs.
An additional $r$-process source with shorter delay time, such as rare type of supernova, is suggested as a solution.
Since we discuss low metallicity stars, this ``quick'' source would matter.
Our results might indicate the presence of non-negligible delay time for the quick source.
Although the uncertain delay-time of the $r$-process enrichments prevents us from obtaining a quantitative constraint on star formation timescales, the results demonstrate that the star formation timescale is constrained from elements whose origin is totally different from $\alpha$-elements or Fe.

At solar metallicity, the dominant nucleosynthesis site of heavy $s$-process elements, including Ba, La, Ce, and Nd, are considered to be low-to-intermediate mass AGB stars \citep[so-called main-$s$;][]{Prantzos2019a}.
Since these AGB progenitors have long main-sequence lifetimes, their contributions are expected to occur at late times.
Abundances of these elements could consequently differ between systems with different star formation timescales.
However, the heavy $s$-process element abundances, especially relative to Eu, do not show differences between our high-$\alpha$/low-$\alpha$ populations (Figures~\ref{fig:abtrend4} and \ref{fig:ncap}, see also Table~\ref{tab:absignificance2}). 
This suggests that contribution of low-to-intermediate mass AGB stars would be small for both populations. 
Almost flat trend of [{Eu}/$s$] among the whole target stars supports the lack of low-to-intermediate mass star contributions.
Previous studies \citep[e.g., ][]{Ishigaki2013,Fishlock2017a} also pointed out similar lack of [{Eu}/$s$] abundance difference between the outer/inner halo or low-/high-$\alpha$ halo. 

It is worth noting that the [{Eu}/$s$] ratio is slightly lower than the pure $r$-process ratio (bottom middle panel of Figure~\ref{fig:ncap}).
This might indicate that massive stars contribute to the heavy $s$-process element enrichments.
It has been suggested that massive stars with high rotation speed can produce heavy $s$-process elements \citep[e.g., ][]{Frischknecht2012a,Choplin2018a}.

It is surprising that the $s$-process contribution is small in both populations while SNe~Ia appear to have already started significant contribution at least for the low-$\alpha$ population.
The maximum initial mass of white dwarf progenitors is expected to be higher than that of AGB progenitors that produce significant amount of heavy $s$-process elements.
However, SNe~Ia would need additional time after the formation of first white dwarfs.
Our results suggest that the minimum delay time for white dwarfs to explode might be shorter than the timescale of the evolution of low-to-intermediate mass stars.
There are observational and theoretical results supporting that the minimum delay time of SNe~Ia is shorter than a few $\times 100\,\mathrm{Myr}$ \citep[e.g., ][]{Totani2008a,Maoz2014a}.

Although the low-$\alpha$ population is considered to be an accreted massive dwarf galaxy, the behavior of $s$-process elements, i.e., the lack of significant $s$-process enrichments, is different from dwarf galaxies currently found around the Milky Way.
They tend to show signatures of heavy $s$-process enhancements \citep[e.g., ][]{Hill2019a,Letarte2010a}.
This suggests that the low-$\alpha$ population has experienced faster chemical evolution than those of stars in the dwarf galaxies.

We note that our low-$\alpha$ population do not include many stars at [{Fe}/{H}]$\gtrsim -1.0$ (see Figure~\ref{fig:abtrend0}), where the low-$\alpha$ population has been known to show main $s$-process contribution \citep{Ishigaki2013,Fishlock2017a}. 
We expect more dedicated observations of halo stars at high metallicity in the \textit{Kepler} field would reveal the $s$-process contribution in the low-$\alpha$ population.
Discussions presented here are applicable only to the metallicity range well-covered in the present study ($-1.5\lesssim [\mathrm{Fe/H}]\lesssim -1.0$).

We have investigated contributions of SNe~Ia, SNe~II, weak $s$-process in massive stars, $r$-process, and heavy $s$-process in the low-$\alpha$ and high-$\alpha$ populations. 
They give constraints on the star formation timescale of the halo populations.
Firstly, SNe~Ia need at least $\sim 100\,\mathrm{Myr}$ after star formation, which corresponds to the lifetime of the most massive white dwarf progenitor \citep{Maoz2014a}.
This sets a lower limit on the star formation timescale of the low-$\alpha$ population as it shows signatures of SNe~Ia contribution (e.g., in low [{Mg}/{Fe}] in Figure~\ref{fig:abtrend0}). 
The lack of main $s$ contribution provides an additional constraint (bottom middle panel of Figure~\ref{fig:ncap}).
Heavy $s$-process elements are mainly produced by stars less massive than $\sim3\,\mathrm{M_{\odot}}$ \citep{Straniero2014a}. 
The lifetime of a $3\,\mathrm{M_{\odot}}$ star is $\sim 200-300\,\mathrm{Myr}$. 
Therefore, both the low-$\alpha$ and the high-$\alpha$ populations have to be evolved up to [{Mg}/{H}]$\sim -1$ within a few $\times 100\,\mathrm{Myr}$ before $\sim 3\,\mathrm{M_{\odot}}$ stars start to contribute to the chemical evolution.
This constraint imposes an upper limit on the star formation timescale of a few $\times 100\,\mathrm{Myr}$.
In principle, the $r$-process enhancement of the low-$\alpha$ population can also give an independent constraint.
However, since there is a large uncertainty in the delay time of $r$-process enrichments, we do not aim to put quantitative constraints from $r$-process abundances \citep{Hotokezaka2018a,Cote2019a}.

Based on these estimates, we conclude that the star formation timescale $\tau$ can be constrained to $100\lesssim \tau/\mathrm{Myr} \lesssim 300$ for the low-$\alpha$ population. 
Although the timescale is not well constrained for the high-$\alpha$ population in the present study, it is clear from the smaller amount of SNe Ia contribution that it has a shorter timescale than the low-$\alpha$ population.

\subsubsection{Constraints from mass}

\begin{figure*}
  \plotone{\figureloc 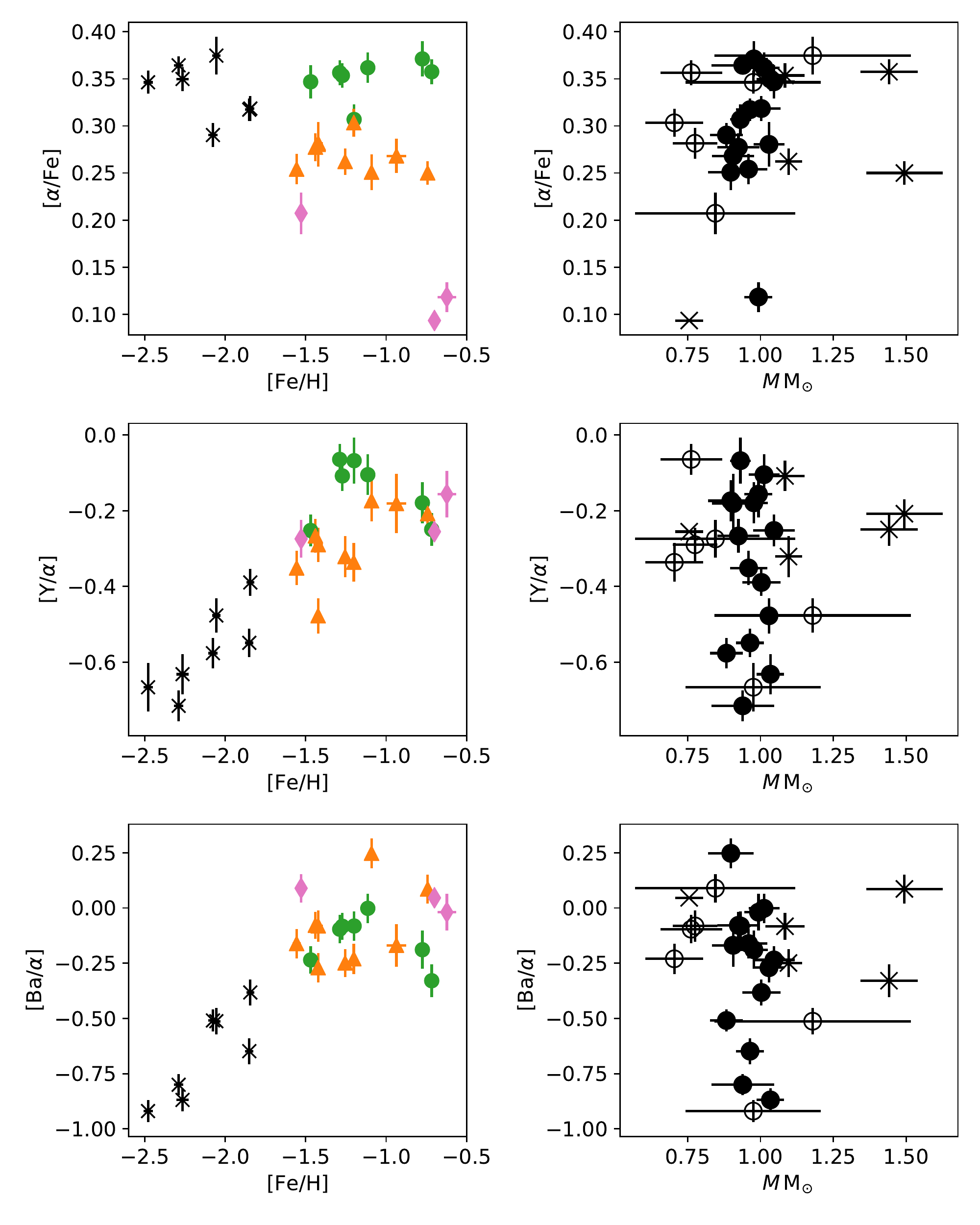}
  \caption{Investigation of correlations between mass and abundances. Symbols follow Figure~\ref{fig:mass_feh}. \label{fig:mass_abundance}}
\end{figure*}

Stellar age is a fundamental parameter when we retrieve the information about the Galaxy formation history from stars. 
Since we have systematic offsets in mass determination, we refrain from using age directly. 
However, as shown in Figure~\ref{fig:mass_feh}, the red giant mass does not depend on the metallicity significantly.
Therefore, we here use stellar mass as an indicator of relative stellar age.
Discussions presented in this subsection are based on the sample of red giant branch stars with $\log g>2.0$ following Section~\ref{sec:dis_seis}.
The numbers of stars are 15 for the entire halo population, and 5 and 4 for high-$\alpha$ and low-$\alpha$ populations, respectively.

We obtained the average age of $6-7\,\mathrm{Gyr}$ for the entire halo population from 15 less luminous giants with $\log g>2$.
This contradicts with the known old age ($\gtrsim 10\,\mathrm{Gyr}$) of halo stars \citep[e.g., ][]{MarinFranch2009a,Forbes2010a,Dotter2011a,Jofre2011a,Schuster2012,Carollo2016a,Kilic2019a}.
As discussed in Section~\ref{sec:dis_seis}, this is likely due to the systematic offset in the scaling relations of asteroseismology.
In this subsection we focus on the age spread of halo stars and age differences between the two halo populations.
These are equivalent to studying the relative age among halo stars, which is feasible even in the presence of systematic offsets in estimated ages if the internal precision is sufficiently high (see Section~\ref{sec:dis_seis}). 
In addition, the information obtained from age estimates is complementary to chemical abundances since it does not rely on theoretical nucleosynthesis yields and since it enables us to discuss not only the formation timescale but also difference of the formation epoch between systems.

Although chemical abundances and kinematics span a significantly wide range with an amplitude much larger than the measurement uncertainties, we do not expect significant correlations between masses and other properties of stars since we did not find significant dispersion in mass.
We have, in fact, explored possible correlations between masses and abundances, and those between masses and kinematics, and found no correlation. 
For example, abundance ratios such as [$\alpha$/{Fe}], or [{Y}/{Mg}] are suggested to show correlations with stellar age for disk stars at around solar metallicity \citep[][; but see also SilvaAguirre et al. 2018]{daSilva2012a,Nissen2016a,TucciMaia2016a,Feltzing2017a,Spina2018a}.
Figure~\ref{fig:mass_abundance} shows the results of our investigation on [$\alpha$/{Fe}], [{Y}/{$\alpha$}], [{Ba}/{$\alpha$}]\footnote{Here we define $\alpha$ abundance as the average of Mg, Ca, and Ti. For Ti, we adopt the average of Ti~I and Ti~II.}.
The absence of correlation indicates that chemical evolution has to proceed rapidly compared to the uncertainty of the age estimate, which is not inconsistent with the discussion from chemical abundances.

The lack of mass dispersion among our subsample of halo stars with $\log g>2$ (Figure~\ref{massdf}) provides an upper limit on intrinsic mass dispersion, which can be translated into dispersion of relative age.
Currently the weighted sample standard deviation of the mass is $0.05\,\mathrm{M_{\odot}}$ (or $5$\% of the average mass) for the 15 stars on the lower red giant branch phase.
Since this dispersion can be explained by measurement uncertainties (Section~\ref{sec:dis_seis}), the intrinsic mass dispersion should be smaller than this value.
The age of red giant branch stars is roughly proportional to  $\sim M^{-3.5}$ in this metallicity and mass range according to the MIST isochrones.
Therefore, we obtain an upper limit on relative age dispersion of $<18$\% for the entire halo population.
Considering typical ages of halo stars provided in literature of 10 Gyr\footnote{We note that the average mass we obtained are converted to $6-7\,\mathrm{Gyr}$ if the systematic offset are not considered. However, since this apparent younger age is likely due to the systematic offset present in the mass estimates from the scaling relations, we stick to the conventional old age of the stellar halo.}, $18$\% corresponds to $\sim 2\,\mathrm{Gyr}$.

The mass difference between our high-$\alpha$ and low-$\alpha$ stars is also insignificant ($\langle M\rangle = 0.958$ and $0.972\,\mathrm{M_{\odot}}$ respectively; the probability that these two values are different is $0.87$).
The $1\sigma$ upper limits on the mass difference is $0.04\,\mathrm{M_{\odot}}$ (or 4\% of the average).   
Following the same argument as for the dispersion, the relative age difference between the two populations is $<1.5\,\mathrm{Gyr}$ (15\%; $1\sigma$ upper limit).

This result is not inconsistent with the study of the age difference of halo stars but for nearby turn-off stars by \citet{Schuster2012}, who reported the difference is $2-3\,\mathrm{Gyr}$.
We note that they obtained smaller age difference at lower metallicity bin ([{Fe}/{H}]$<-1.2$).
A similar conclusion is also obtained by \citet{Hawkins2014a}, who statistically study age-metallicity relations for the two populations using a low-resolution spectroscopic data following \citet{Jofre2011a}.
\citet{Das2020a} reached a conclusion that the age of metal-poor stars are similar irrespective of their abundance ratios, which is also consistent with our conclusion.

The non-detection of any age dispersion or age difference is due to the combined effect of limited age (mass) precision, the metallicity range of the targets, and limited sample size.
Improvements in age precision are definitely desired.
Although there is room for improvements for modelling of stellar oscillations as we saw in Section~\ref{sec:dis_seis}, it might be difficult to achieve revolutionary improvements in terms of precision.
We note that most of the uncertainties in derived mass stem from the uncertainties in oscillation frequencies.
We used oscillation frequencies that are measured from the best available light curves obtained from the about four years of continuous observation by the \textit{Kepler} mission.
No near-future mission is planned to obtain higher-quality or longer-term light curves.
One of the ways to improve the mass measurement precision is to take luminosity into consideration.
If the luminosity is measured with sufficiently high precision, it enables us to achieve about twice smaller uncertainty \citep{Rodrigues2017a}.
This is becoming possible thanks to the Gaia mission.
On the other hand, it is necessary to resolve the systematic mass offsets found in Section~\ref{sec:dis_seis}.
Unless we confirm there is no systematic offsets, combining asteroseismic information with luminosity does not necessarily lead to a narrower posterior distribution.
Individual frequencies modelling recently conducted for a halo star by \citet{Chaplin2020a} would help to resolve the issue.

The metallicity range is another issue.
In \citet{Schuster2012} and in \citet{Hawkins2014a}, the age difference between the high-$\alpha$ and low-$\alpha$ stars is larger in the bins of higher metallicity. 
Since our targets are on more metal-poor side compared to their study, age differences might be detected using asteroseismology by focusing on the high metallicity end of the halo.
Increasing the sample size in general would allow us to more tightly constrain the average mass and age for each population, which would enable us to detect a small age difference, if present.
For example, three times larger sample would enable us to detect age difference of $\sim 0.8\,\mathrm{Gyr}$.

\subsection{Peculiar objects}

\subsubsection{Na-enhanced stars\label{sec:dis_na}}

\begin{figure}
\plotone{\figureloc 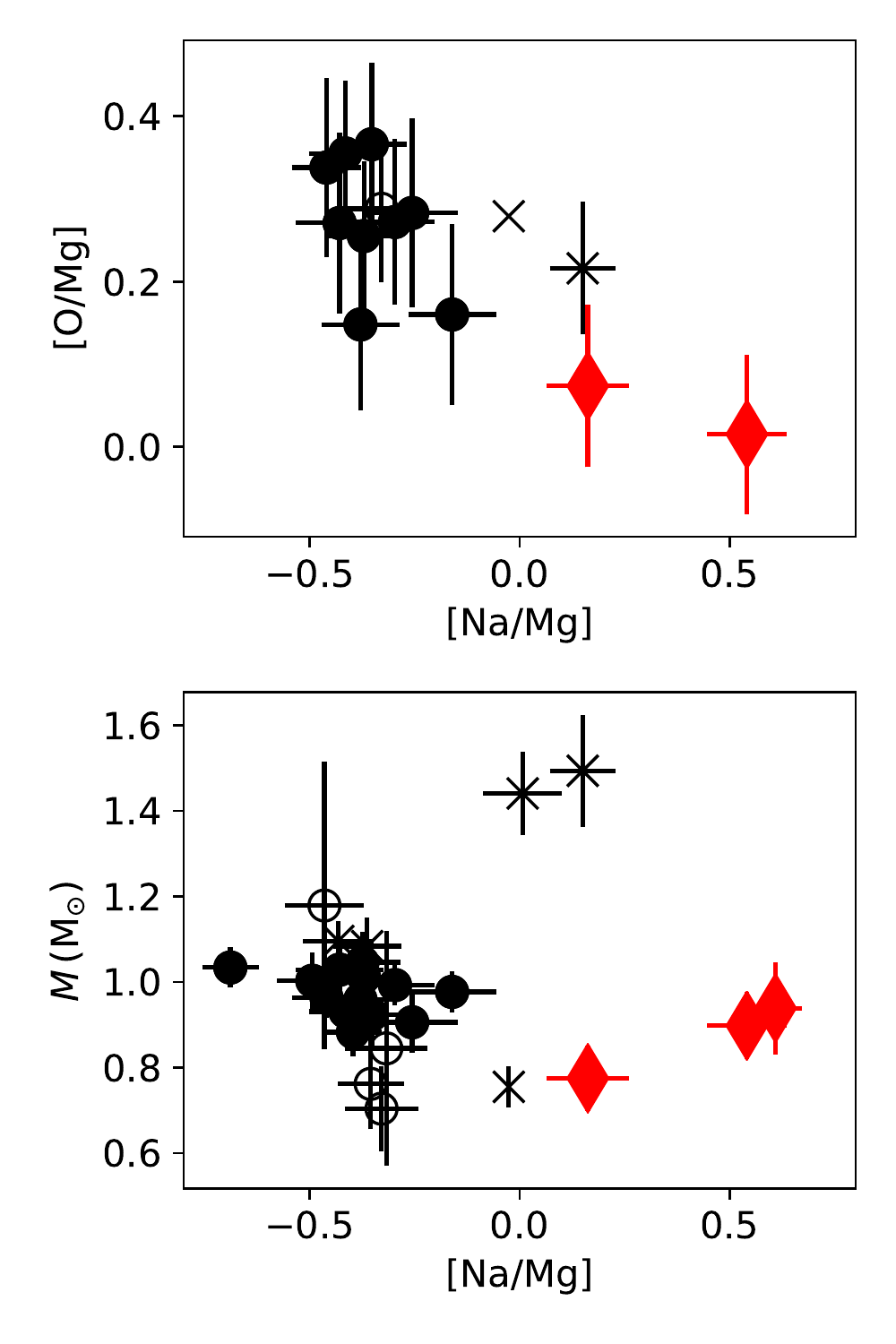}
\caption{The relation between [{O}/{Mg}] and [{Na}/{Mg}] (upper) and that between mass and [{Na}/{Mg}]. Na enhanced stars are shown with red diamonds while others are shown with black symbols. The shape of the symbols follow Figure~\ref{fig:mass_feh}. \label{fig:nacorr}}
\end{figure}

As noted in Section~\ref{sec:abundance}, there are three Na-enhanced objects (KIC8350894, KIC5184073, and KIC6520576). 
There is no anomaly in their elemental abundances except for possible low O abundances found in two stars (Figure~\ref{fig:nacorr}; KIC8350894 and KIC5184073) and high $s$-process element abundance in one star (KIC8350894; Ba, La, Ce, and Nd abundances are 0.50, 0.57, 0.40, and 0.64 in [{X}/{Fe}]).
The O absorption [{O~I}] $6300\,\mathrm{\AA}$ is blended with telluric lines for KIC6520576.
The mass of the three stars are low (Figure~\ref{fig:nacorr}), and KIC5184073 and KIC8350894 are in the red giant branch phase.
Unfortunately, KIC6520576 is the only star among our sample whose evolutionary status is not available in \citet{Yu2018a}.
KIC5184073 has $\log g<2.0$, while the other two have $\log g>2.0$. 
We also note that no radial velocity variations have been detected for the Na-enhanced objects (Figure~\ref{fig:rv} and Table~\ref{tab:obs}).

Large Na enhancements are sometimes seen in globular clusters \citep[e.g., ][]{Carretta2009a} and were used to search for metal-poor stars ejected from globular clusters \citep{Pereira2019a}.
Those Na enhanced objects in clusters usually accompany O deficiency \citep[known as Na-O anti-correlation; ][]{Carretta2009a}.
The possibly low O abundances of the two stars might indicate their globular cluster origin.
Further investigation on Mg-Al anti-correlation, Li, C, and N abundances are highly desired to confirm their globular cluster origin.

Internal mixing might also cause Na-enhancements especially in AGB stars or in massive stars.
However, the low mass and the evolutionary phase might not be compatible with this hypothesis.

KIC6520576 resembles CD$-$23$^{\circ}$16310, which was reported to have [{Na}/{Fe}]$=+1.09$ at [{Fe}/{H}]$=-1.93$ \citep{Pereira2019a}.
Both giants have large ($\sim 1$ dex) Na enhancements at low metallicity, but do not show peculiar abundance pattern including C.
The most notable difference is the luminosity or surface gravity; while KIC6520576 has $\log g=2.17$, CD-23$^{\circ}$16310 has $\log g=0.9$.
\citet{Pereira2019a} suspect CD-23$^{\circ}$16310 is in the early AGB phase.
It is not yet clear at this stage if KIC6520576 can also be interpreted as in the early AGB phase.

The $s$-process enhancement of KIC8350894 might be related to its Na-enhancement. 
Although $s$-process enhancements are often ascribed to a presence of an evolved companion, this star has not shown any sign of binarity. 
In addition, it is worth noting that the C is not enhanced ([{C}/{Fe}]$=-0.44$)

\subsubsection{Over-massive stars\label{sec:massive}}

Among our sample, two stars are obviously over-massive (KIC5446927 and KIC10096113; $1.49\pm0.13$ and $1.44\pm0.10\,\mathrm{M_{\odot}}$, respectively).
The APOKASC-2 catalog also provides similar masses for these objects, $1.451$ and $1.396\,\mathrm{M_{\odot}}$, respectively.
Even though these stars are on the metal-rich side in our halo stars ($\mathrm{[Fe/H]}=-0.74$ and $-0.72$), their metallicity is still sufficiently low as we can assume they are very old and low-mass.
They also have super-solar [{$\alpha$}/{Fe}] and large relative velocity to the Sun, supporting their old age. 

Therefore, these two stars are too massive for their chemical and kinematic properties.
Similar stars have been found among the Galactic disk \citep[so-called young $\alpha$-rich stars; ][]{Martig2015a,Chiappini2015}.
The young $\alpha$-rich stars are massive but their chemical abundance are just the same as normal old disk stars \citep{Yong2016,Matsuno2018a}.
An important property of the young $\alpha$-rich stars is that they show a high fraction of radial velocity variation \citep{Jofre2016,Matsuno2018a}, suggesting binary interaction is the key for the formation of these stars.
Theoretical work also supports this scenario \citep{Izzard2018}.

Radial velocity measurements of KIC5446927 and KIC10096113 do not reveal their binary nature at this stage (Figure~\ref{fig:rv} and Table~\ref{tab:obs}).
KIC5446927 actually has precise radial velocity measurements by our observation, APOGEE, and Gaia, all of which perfectly agree.
Although not every young $\alpha$-rich star shows radial velocity variation, the absence of radial velocity variation might indicate the origin of these stars be different from the majority of young $\alpha$-rich stars. 
These stars could in fact be evolved blue stragglers, which are found to be quite common among kinematically selected halo stars \citep{Casagrande2020a}, and a few are thus expected in our sample.
Further monitoring of radial velocities of these stars is obviously welcomed to draw robust conclusion about the origin of these stars.

It is interesting to note that the Na abundance of the two objects appears enhanced (Figure~\ref{fig:nacorr}), although the level of the enhancements is not as significant as the three stars studied in the previous subsection considering the Galactic chemical evolution (bottom panel of Figure~\ref{fig:abtrend1}).
\citet{Smiljanic2016a} showed that massive giants have high Na abundance due to the internal mixing. 
However, they concluded that the effects appear in stars more massive than $2.0\,\mathrm{M_{\odot}}$, which is larger than the mass of our two over-massive stars.
Another interesting aspect of these stars is their evolutionary status; both of the two stars are red clump stars, which would support the binary merger scenario \citep{Miglio2020a}.
Additional mixing near the tip of red giant branch might affect Na abundance of these stars \citep{Weiss2000a,Fujimoto1999a}.

\subsubsection{KIC7693833\label{sec:kic769}}

As noted in Section \ref{sec:kinematics}, KIC7693833 is one of the most metal-poor star whilst its velocity is similar to disk stars.
A population of stars with disk-like kinematics are found among very metal-poor stars \citep{Sestito2019a,Carollo2019a,Sestito2020a,Venn2020a}, whose formation mechanism is still under debate.
The property of such stars including their ages would shed light on their origin.
KIC7693833 would belong to this population; therefore, we here briefly summarize its property.

Compared to stars with similar metallicity in our sample, the chemical abundance of KIC7693833 is characterized by high [{C}/{Fe}], low [{Na}/{Fe}], high [{Co}/{Fe}], and low [{Eu}/{Fe}]. 
We also note that other heavy neutron-capture elements except for Ba are not detected.
The mass of this star is $1.04\pm 0.05\,\mathrm{M_{\odot}}$, which is higher than the average. 
Although these chemical abundance and mass might be peculiar, a study of large sample of stars is clearly needed to make a further interpretation.

\section{Conclusion}\label{sec:summary}

We obtain precise stellar parameters and precise chemical abundances for 26 halo stars in the \textit{Kepler} field, for which asteroseismic information is available.
The sample is selected based on radial velocity and metallicity estimated from spectroscopic surveys.
The kinematics of the selected stars are later investigated with the Gaia DR2 data, confirming that they have halo-like kinematics.
Stellar parameters are determined precisely and accurately by adopting a differential abundance analysis and using standard stars with accurate $T_{\rm eff}$ and $\log g$.
Subsequently we obtain precise chemical abundances.

Our study is so far the largest sample of low-metallicity stars with asteroseismic information and precise stellar parameters.
We investigate the reliability of asteroseismic mass estimates at low metallicity with this sample.
The average mass obtained from asteroseismology is $1.03\,\mathrm{M_{\odot}}$ (without correction) and $0.96\,\mathrm{M_{\odot}}$ (with $\Delta\nu$ correction).
This result indicates that, although correction to the $\Delta\nu$ scaling relation helps to reduce inconsistency between asteroseismic and expected masses ($\sim 0.8\,\mathrm{M_{\odot}}$), it is not capable of completely resolving the issue.

We also show that luminous red giant stars and red clump stars could suffer from systematic uncertainties in asteroseismic masses.
After excluding these stars, there is no significant mass dispersion among our sample. 
This fact indicates that, despite the systematic offset, the mass of halo stars can be estimated with high-precision if we focus on less-evolved red giants.

The precise chemical abundances allow us to separate our targets into high-$\alpha$ and low-$\alpha$ halo populations based on the Mg abundance.
The two populations also differ in other element abundances consistent with \citet{Nissen2010,Nissen2011} and \citet{Fishlock2017a}.
The low-$\alpha$ population shows low values of [{X}/{Fe}], where X is C, O, Na, Ca, Sc, Ti, V, Ni, Cu, Zn, and Y.
Most of these elements are ejected from massive stars, and thus the low values are understood as a result of SNe~Ia contribution.
The Y abundance in the low-$\alpha$ population is lower compared to heavy $s$-elements or Eu.
This, together with the Cu abundance, is understood as a result of their secondary nature in the production by the weak $s$-process. 
There is no significant difference in Eu-to-heavy $s$-process abundance ratio, indicating that the main $s$-process from low-to-intermediate mass AGB stars does not contribute significantly to neither of the low-$\alpha$ or high-$\alpha$ populations.
Eu abundance relative to Mg seems enhanced in the low-$\alpha$ population, suggesting that the delay time of $r$-process enrichments plays a role.

These chemical abundances provide constraints on the timescale of star formation ($\tau$).
Since the low-$\alpha$ population is enriched from SNe~Ia, $\tau$ should be longer than $\gtrsim 100\,\mathrm{Myr}$, which comes from the lifetime of the most massive white dwarf progenitor.
The lack of the main $s$-process contribution, on the other hand, provides an upper limit on the timescale as $\tau\lesssim 300\,\mathrm{Myr}$.
The low-$\alpha$/high-$\alpha$ populations are now understood as accretion and in-situ origin, respectively \citep{Nissen2010,Helmi2018a}. 
Therefore, our study indicates the accreted population formed with a timescale of $100\lesssim \tau/\mathrm{Myr}\lesssim 300$, and the in-situ population formed with a shorter timescale.

The asteroseismic information independently constrains the formation timescale and additionally provides a constraint on the relative formation epochs of the two populations.
The lack of significant mass dispersion among less luminous red giant branch stars gives an upper limit on the intrinsic mass dispersion of $<0.05\,\mathrm{M_{\odot}}$.
This can be translated into an age dispersion of $\lesssim 2\,\mathrm{Gyr}$.
The average mass of less luminous red giant branch stars is $0.958$ and $0.972\,\mathrm{M_{\odot}}$ for the high-$\alpha$ and low-$\alpha$ populations. 
This difference is not significant and constrains a relative age difference of $<1.5\,\mathrm{Gyr}$.
These results do not contradict with previous works \citep{Schuster2012,Hawkins2014a,Das2020a} or with chemical abundances.
The star formation timescale in the early Galaxy is too short for asteroseismology to provide a meaningful constraint at this stage.

Our study is the first that combines chemical abundances and stellar mass (age) estimates for more than 10 field halo stars beyond the solar neighbourhood.
However, we did not detect age spreads among our sample or age difference between populations unlike \citet{Schuster2012}.
The reason of the non-detection would be limited precision, metallicity range, and/or sample size.
A factor of 2--3 improvements in the precision would make it comparable to the precision that was achieved for nearby turn-off stars \citep{Schuster2012}.
This level of improvement might be achieved by incorporating luminosities into mass estimates \citep{Rodrigues2017a}, after resolving the systematic offset in masses we obtained in this study. 
Instead, a study that focuses on the high metallicity end of the halo with a larger sample size might be able to reveal the age difference.
Such a sample will be provided by space-based photometric monitoring observations with a wider field coverage, such as K2, TESS, or PLATO.

\newpage
\acknowledgments
We are grateful to the anonymous referee for carefully reviewing the draft and providing comments that greatly improved the clarity.
We thank Evan Kirby, who provided helpful comments on the early version of the manuscript.
We also acknowledge useful discussions with Jie Yu.
Paula Jofr\'e and Joel Zinn provided comments that improved discussions presented in this paper.
This work was partially supported by JSPS - CAS Joint Research Program.  
TM is supported by Grant-in-Aid for JSPS Fellows (Grant number 18J11326) and by the course-by-course education program of SOKENDAI.
WA was supported by JSPS KAKENHI Grant Numbers 16H02168.
LC acknowledges support from the Australian Research Council Future Fellowship FT160100402.
HNL is supported by NSFC grant No. 11973049 and the Strategic Priority Research Program of Chinese Academy of Sciences, Grant No. XDB34020205. 
TS was supported by JSPS KAKENHI Grant Numbers 20HP8012 and 16H02166.

\software{Asfgrid \citep{asfgrid},
          Astropy \citep{Astropy},
          Corner \citep{corner},
          emcee \citep{emcee},
          Matplotlib \citep{Matplotlib},
          NumPy \citep{numpy}, 
          Pandas \citep{pandas},
          q2 \citep{Ramirez2014},
          Scipy \citep{Scipy}}
          
\newpage

\appendix

\section{Microturbulent velocity prior as a function of $\log g$\label{appendix1}}

The microturbulent velocity is determined spectroscopically such that abundances derived from individual neutral iron lines do not show correlation with line strengths.
However, due to the wide parameter range of our sample, microturbulent velocities do not converge well for some of the stars.
Poorly constrained microturbulent velocity affects the precision of the temperature and the abundances.
In order to mitigate this problem, we use a non-flat prior on microturbulent velocity during stellar parameter determination.

\begin{figure}
\plottwo{\figureloc 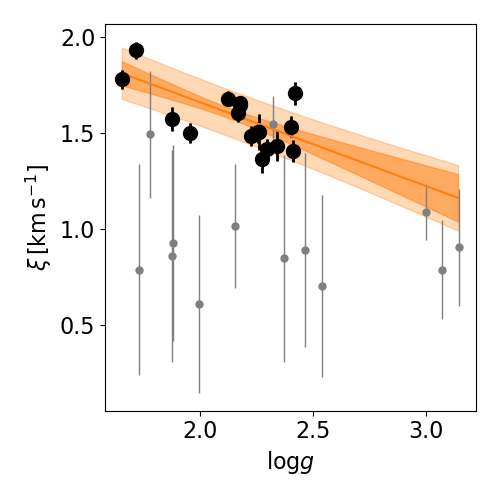}{\figureloc 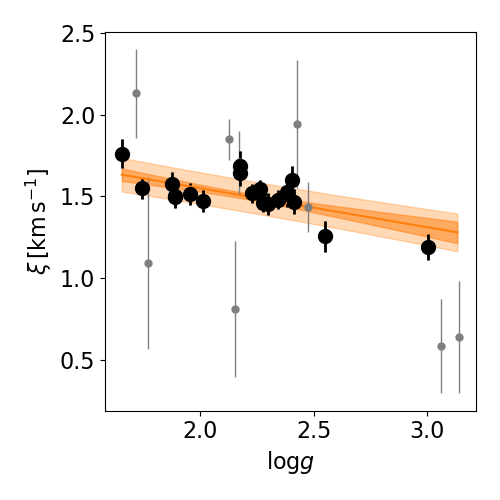}
\caption{The relation between microturbulent velocity and surface gravity. Black large symbols are used to determine the prior, and grey small circles are other objects. The orange solid line, thick filled region, and thin filled region are the best fit linear relation, uncertainties in the fit, and $1\sigma$ interval of the prior at given $\log g$. The left and right panels show results of analysis relative to HD122563 and KIC9583607, respectively. \label{fig:vtprior}}
\end{figure}

Here we describe how we chose the $\xi$ prior (Figure \ref{fig:vtprior}).
We first run a stellar parameter determination processes with flat prior on microturbulent velocity. 
We then carry out a linear fit to the relation between the microturbulent velocity and the surface gravity using stars whose microturbulent velocity is determined with high precision ($\sigma(\xi)<0.1\,\mathrm{km\,s^{-1}}$).
Note that the microturbulent velocity of the other stars with large uncertainties agree with the best fit linear relation at $<2\sigma$ level.
Based on this fit we put prior on $\xi$ as 
\begin{equation}
p(\xi|\log g) = \frac{1}{\sqrt{2\pi\sigma^2}}\exp(-(\xi-\mu(\log g))^2/2\sigma^2),
\end{equation} 
where $\mu(\log g)$ is the expected value of $\xi$ at given $\log g$ from the linear fit, $\sigma$ is obtained from $\sigma=\sigma_{\rm fit}^2(\log g) +\sigma_{\rm resid}^2$, and the $\sigma_{\rm fit}(\log g)$ is the expected standard deviation at given $\log g$ due to the uncertainties in the fitting parameters.
The $\sigma_{\rm resid}$ is a constant expressed as $\sigma_{\rm resid}^2=\sum_i (\xi_i - \mu(\log g_i))^2/(N-\nu)$, where $N$ and $\nu$ are the number of stars used for the fit and the degree of freedom ($\nu=2$ in this case), respectively.

The inclusion of $\xi$ prior does not affect stars whose $\xi$ is determined precisely in the first step.
For the other stars, it provides a better convergence in the stellar parameter determination.

\section{Stellar parameters from analyses with different standard stars\label{appendix:MPMR}}

In this study, we carried out two sets of the analyses with two reference stars, HD122563 and KIC9583607.
Table \ref{tab:paramMPMR} shows stellar parameters derived from each of the analyses.
Final stellar parameters adopted for, e.g., mass estimates, are the combination of two results following the weights ($w_{\rm MP}$ and $w_{\rm MR}$) provided in the Table.

\begin{deluxetable}{lrrrrrrrrrrrrrrr}
  \tablecaption{Stellar parameters from the two analyses\label{tab:paramMPMR}}
  \tablehead{
                    & \multicolumn{7}{c}{Analysis with HD122563 (MP)} & & \multicolumn{7}{c}{Analysis with KIC958367 (MR)} \\\cline{2-8}\cline{10-16}
   \colhead{Object} & \colhead{$T_{\rm eff}$}& \colhead{$\sigma_{(T_{\rm eff})}$} &\colhead{$\xi$}& \colhead{$\sigma_{(\xi)}$} &\colhead{$\mathrm{[Fe/H]}$}& \colhead{$\sigma_{(\mathrm{[Fe/H]})}$} & \colhead{$w_{\rm MP}$} && \colhead{$T_{\rm eff}$}& \colhead{$\sigma_{(T_{\rm eff})}$} &\colhead{$\xi$}& \colhead{$\sigma_{(\xi)}$} &\colhead{$\mathrm{[Fe/H]}$}& \colhead{$\sigma_{(\mathrm{[Fe/H]})}$} & \colhead{$w_{\rm MR}$}   \\  
                    & \colhead{(K)}          & \colhead{(K)}                      &\colhead{$\mathrm{(km\,s^{-1})}$}&\colhead{$\mathrm{(km\,s^{-1})}$} & & && &\colhead{(K)}          & \colhead{(K)}                      &\colhead{$\mathrm{(km\,s^{-1})}$}&\colhead{$\mathrm{(km\,s^{-1})}$}
  }
  \startdata
KIC5184073 &    4894 &        51 &  1.600 &     0.056 & -1.380 &      0.030 &  0.213 &   & 4856 &        54 &  1.576 &     0.054 & -1.433 &      0.035   &  0.787  \\ 
KIC5439372 &    4835 &        34 &  1.915 &     0.045 & -2.479 &      0.029 &  1.000 &   & 4724 &        90 &  1.673 &     0.101 & -2.591 &      0.094   &  0.000  \\ 
KIC5446927 &    5175 &        53 &  1.519 &     0.071 & -0.629 &      0.035 &  0.000 &   & 5102 &        37 &  1.530 &     0.047 & -0.742 &      0.022   &  1.000  \\ 
KIC5698156 &    4701 &        79 &  1.653 &     0.131 & -1.274 &      0.053 &  0.085 &   & 4639 &        46 &  1.522 &     0.052 & -1.290 &      0.017   &  0.915  \\ 
KIC5858947 &    5154 &        69 &  1.089 &     0.140 & -0.669 &      0.045 &  0.000 &   & 5105 &        72 &  1.221 &     0.117 & -0.775 &      0.047   &  1.000  \\ 
KIC5953450 &    5236 &        81 &  1.100 &     0.157 & -0.457 &      0.067 &  0.000 &   & 5127 &        74 &  1.234 &     0.115 & -0.623 &      0.057   &  1.000  \\ 
KIC6279038 &    4761 &        31 &  1.784 &     0.044 & -2.055 &      0.027 &  1.000 &   & 4719 &        50 &  1.713 &     0.064 & -2.123 &      0.042   &  0.000  \\ 
KIC6520576 &    4971 &        35 &  1.602 &     0.042 & -2.289 &      0.029 &  1.000 &   & 4958 &        66 &  1.565 &     0.082 & -2.320 &      0.059   &  0.000  \\ 
KIC6611219 &    4719 &        90 &  1.728 &     0.142 & -1.166 &      0.068 &  0.000 &   & 4652 &        45 &  1.566 &     0.052 & -1.201 &      0.018   &  1.000  \\ 
KIC7191496 &    4903 &        34 &  1.672 &     0.038 & -2.076 &      0.028 &  1.000 &   & 4826 &        70 &  1.619 &     0.093 & -2.179 &      0.067   &  0.000  \\ 
KIC7693833 &    5094 &        46 &  1.664 &     0.061 & -2.265 &      0.038 &  1.000 &   & 5034 &        81 &  1.493 &     0.096 & -2.334 &      0.074   &  0.000  \\ 
KIC7948268 &    5166 &        66 &  1.140 &     0.093 & -1.135 &      0.033 &  0.000 &   & 5154 &        51 &  1.230 &     0.062 & -1.199 &      0.028   &  1.000  \\ 
KIC8350894 &    4854 &        92 &  1.626 &     0.119 & -1.069 &      0.075 &  0.000 &   & 4797 &        45 &  1.491 &     0.054 & -1.091 &      0.025   &  1.000  \\ 
KIC9335536 &    4835 &        40 &  1.526 &     0.051 & -1.493 &      0.027 &  0.347 &   & 4807 &        45 &  1.527 &     0.052 & -1.547 &      0.037   &  0.653  \\ 
KIC9339711 &    4928 &        39 &  1.497 &     0.046 & -1.454 &      0.025 &  0.290 &   & 4940 &        43 &  1.514 &     0.048 & -1.475 &      0.026   &  0.710  \\ 
KIC9583607 &    5086 &        61 &  1.543 &     0.084 & -0.618 &      0.039 &  0.000 &   & 5059 &     \nodata &  1.590 &    \nodata& -0.700 &     \nodata  &  1.000  \\ 
KIC9696716 &    4962 &        43 &  1.433 &     0.049 & -1.414 &      0.023 &  0.248 &   & 4962 &        52 &  1.460 &     0.052 & -1.448 &      0.031   &  0.752  \\ 
KIC10083815 &    4828 &        90 &  1.525 &     0.122 & -0.890 &      0.066 &  0.000 &  &  4784 &        74 &  1.476 &     0.097 & -0.936 &      0.060  &  1.000  \\ 
KIC10096113 &    4943 &        80 &  1.417 &     0.116 & -0.661 &      0.054 &  0.000 &  &  4948 &        48 &  1.443 &     0.070 & -0.717 &      0.018  &  1.000  \\ 
KIC10328894 &    5031 &        39 &  1.525 &     0.047 & -1.832 &      0.031 &  0.771 &  &  4973 &        59 &  1.539 &     0.067 & -1.912 &      0.048  &  0.229  \\ 
KIC10460723 &    4938 &        49 &  1.412 &     0.068 & -1.213 &      0.025 &  0.017 &  &  4922 &        41 &  1.467 &     0.047 & -1.273 &      0.021  &  0.983  \\ 
KIC10737052 &    4998 &        50 &  1.454 &     0.064 & -1.190 &      0.029 &  0.000 &  &  4973 &        42 &  1.478 &     0.047 & -1.255 &      0.021  &  1.000  \\ 
KIC11563791 &    4996 &        75 &  1.368 &     0.128 & -1.070 &      0.046 &  0.000 &  &  4974 &        54 &  1.326 &     0.069 & -1.114 &      0.023  &  1.000  \\ 
KIC11566038 &    5002 &        45 &  1.417 &     0.052 & -1.391 &      0.030 &  0.223 &  &  4998 &        47 &  1.461 &     0.057 & -1.432 &      0.031  &  0.777  \\ 
KIC12017985 &    4962 &        39 &  1.639 &     0.043 & -1.823 &      0.028 &  0.770 &  &  4889 &        56 &  1.591 &     0.066 & -1.914 &      0.044  &  0.230  \\ 
KIC12017985-17 &    4947 &     38 &  1.646 &     0.038 & -1.852 &      0.027 &  0.809 &  &  4868 &        61 &  1.601 &     0.076 & -1.946 &      0.050  &  0.191  \\ 
KIC12253381 &    4947 &        46 &  1.511 &     0.047 & -1.519 &      0.027 &  0.380 &  &  4906 &        53 &  1.513 &     0.052 & -1.579 &      0.040  &  0.620  \\ 
HD122563\tablenotemark{a}       & 4636 &\nodata & 2.05 & \nodata & -2.599 & \nodata & \nodata & & 4434 & 62 & 3.613 & 0.060 & -2.851 & 0.029 & \nodata \\
  \enddata
\tablenotetext{a}{HD122563 is included for completeness. It is not used for the other figures.}
\end{deluxetable}

\section{Uncertainties of the stellar parameters of the standard stars \label{appendix3}}

Abundances and stellar parameters are precisely determined in this study adopting differential abundance analyses.
While the relative abundance/parameters among our sample should not be significantly affected by the uncertainties of stellar parameters of the standard stars, the absolute scale totally depends on the abundance/stellar parameters of the standard stars.
In this section, we explore how the change in stellar parameters of the standard stars affects the absolute scale of our program stars.
One has to take into account these effects when trying to quantitatively compare our results  with different studies. 

\begin{deluxetable}{l*{18}{r}}
  \tablecaption{Change of the stellar parameters/abundances due to the stellar parameters of HD122563\label{tab:B1}}
  \tablehead{ & & & \multicolumn{4}{c}{$\Delta T_{\rm eff}=36\,\mathrm{K}$\tablenotemark{a}}   &  \multicolumn{4}{c}{$\Delta\log g =0.007$\tablenotemark{a}} &  \multicolumn{4}{c}{$\Delta\xi =0.08$\tablenotemark{b}} &  \multicolumn{4}{c}{$\Delta[\mathrm{Fe/H}] =0.016$\tablenotemark{b}} \\
              & \colhead{$n_{\rm line}$\tablenotemark{c}} & \colhead{$\sigma_{\rm sct}$\tablenotemark{c}} & \colhead{ref} & \colhead{(1)} & \colhead{(2)} & \colhead{(3)} & \colhead{ref} & \colhead{(1)} & \colhead{(2)} & \colhead{(3)} & \colhead{ref}& \colhead{(1)} & \colhead{(2)} & \colhead{(3)} & \colhead{ref}& \colhead{(1)} & \colhead{(2)} & \colhead{(3)}
            }
  \startdata
$T_{\rm teff}$    & & & 36 & 48 & 51 & 46 & 0 & 3 & 4 & 0 & 0 & -5 & -1 & 2 & 0 & 5 & 3 & 1 \\
$\log g$          & & & 0.00 & 0.00 & 0.00 & 0.00 & 0.01 & -0.00 & 0.00 & 0.00 & 0.00 & 0.00 & 0.00 & 0.00 & 0.00 & 0.00 & 0.00 & 0.00 \\
$\xi$             & & & 0.00 & 0.00 & 0.01 & 0.02 & 0.00 & 0.00 & 0.01 & 0.00 & 0.08 & 0.09 & 0.06 & 0.06 & 0.00 & 0.00 & 0.00 & 0.00 \\
$[\mathrm{Fe/H}]$ & & & 0.00 & 0.01 & 0.00 & 0.00 & 0.00 & 0.00 & 0.00 & -0.00 & 0.00 & -0.00 & -0.01 & -0.01 & 0.02 & 0.02 & 0.02 & 0.02 \\
CH   &   1 & 0.00 & 0.06 & 0.09 & 0.07 & 0.05 & -0.00 & 0.01 & 0.01 & -0.00 & -0.00 & -0.01 & -0.00 & -0.01 & 0.01 & 0.02 & 0.01 & 0.01 \\
NaI  &   1 & 0.00 & 0.02 & \nodata & 0.03 & 0.03 & 0.00 & \nodata & 0.00 & 0.00 & -0.00 & \nodata & -0.00 & 0.00 & 0.00 & \nodata & 0.00 & -0.00 \\
MgI  &   4 & 0.10 & 0.02 & 0.03 & 0.04 & 0.04 & -0.00 & 0.00 & 0.00 & -0.00 & -0.01 & -0.01 & -0.01 & -0.01 & -0.00 & 0.00 & 0.00 & 0.00 \\
SiI  &   4 & 0.12 & 0.02 & 0.02 & 0.02 & 0.02 & 0.00 & 0.00 & 0.00 & 0.00 & 0.00 & -0.00 & -0.00 & -0.00 & 0.00 & 0.00 & 0.00 & 0.00 \\
CaI  &  28 & 0.07 & 0.03 & 0.03 & 0.03 & 0.04 & 0.00 & 0.00 & 0.00 & 0.00 & -0.00 & -0.00 & -0.01 & -0.01 & 0.00 & 0.00 & 0.00 & -0.00 \\
ScII &  14 & 0.10 & 0.01 & 0.03 & 0.01 & 0.01 & 0.00 & 0.00 & -0.00 & 0.00 & -0.00 & -0.00 & -0.01 & -0.01 & 0.00 & 0.00 & 0.00 & 0.00 \\
TiI  &  46 & 0.08 & 0.06 & 0.07 & 0.07 & 0.07 & 0.00 & 0.00 & 0.01 & -0.00 & -0.00 & -0.01 & -0.01 & -0.01 & -0.00 & 0.00 & 0.00 & -0.00 \\
TiII &  26 & 0.08 & 0.00 & 0.01 & 0.00 & 0.01 & 0.00 & 0.00 & 0.00 & -0.00 & -0.00 & -0.01 & -0.01 & -0.01 & 0.00 & 0.00 & 0.00 & 0.00 \\
VI   &   6 & 0.12 & 0.06 & \nodata & 0.08 & 0.07 & 0.00 & \nodata & 0.01 & 0.00 & 0.00 & \nodata & -0.00 & 0.00 & -0.00 & \nodata & 0.00 & -0.00 \\
CrI  &  13 & 0.10 & 0.05 & 0.06 & 0.06 & 0.06 & -0.00 & 0.00 & 0.00 & 0.00 & -0.00 & -0.01 & -0.01 & -0.01 & -0.00 & 0.00 & 0.00 & -0.00 \\
CrII &  10 & 0.10 & -0.01 & -0.01 & -0.01 & -0.01 & 0.00 & 0.00 & -0.00 & 0.00 & -0.00 & -0.00 & -0.00 & -0.01 & 0.00 & 0.00 & 0.00 & 0.00 \\
MnI  &   7 & 0.05 & 0.04 & 0.04 & 0.05 & 0.05 & -0.00 & 0.00 & 0.00 & -0.00 & -0.00 & -0.01 & -0.00 & -0.01 & -0.00 & 0.00 & 0.00 & 0.00 \\
FeI  & 158 & 0.12 & 0.04 & 0.05 & 0.05 & 0.05 & 0.00 & 0.00 & 0.00 & 0.00 & -0.01 & -0.02 & -0.02 & -0.02 & 0.00 & 0.00 & 0.00 & 0.00 \\
FeII &  35 & 0.05 & -0.01 & 0.00 & -0.01 & -0.01 & 0.00 & -0.00 & -0.00 & 0.00 & -0.00 & -0.00 & -0.01 & -0.02 & 0.00 & 0.00 & 0.00 & 0.00 \\
CoI  &   2 & 0.03 & 0.06 & 0.06 & 0.07 & 0.06 & 0.00 & 0.00 & 0.00 & 0.00 & -0.00 & -0.01 & -0.00 & 0.00 & -0.00 & 0.01 & 0.00 & 0.00 \\
NiI  &  27 & 0.06 & 0.03 & 0.04 & 0.04 & 0.04 & -0.00 & 0.00 & 0.00 & 0.00 & -0.00 & -0.01 & -0.00 & -0.00 & -0.00 & 0.00 & -0.00 & 0.00 \\
CuI  &   1 & 0.00 & 0.05 & \nodata & 0.07 & 0.06 & 0.00 & \nodata & 0.01 & 0.00 & 0.00 & \nodata & -0.00 & 0.00 & 0.00 & \nodata & 0.00 & 0.00 \\
ZnI  &   2 & 0.01 & 0.00 & 0.02 & 0.01 & 0.00 & 0.00 & 0.00 & 0.00 & 0.00 & -0.00 & -0.00 & -0.01 & -0.01 & 0.00 & 0.00 & 0.00 & 0.00 \\
YII  &   9 & 0.09 & 0.01 & 0.03 & 0.02 & 0.01 & 0.00 & 0.00 & 0.00 & -0.00 & 0.00 & 0.00 & -0.00 & -0.01 & 0.00 & 0.00 & 0.00 & 0.00 \\
BaII &   3 & 0.08 & 0.02 & 0.03 & 0.02 & 0.02 & 0.00 & 0.00 & 0.00 & -0.00 & -0.00 & -0.02 & -0.04 & -0.03 & 0.00 & 0.01 & 0.00 & 0.01 \\
LaII &   1 & 0.00 & 0.02 & 0.03 & 0.02 & 0.02 & 0.00 & 0.00 & 0.00 & 0.00 & -0.00 & -0.01 & -0.01 & -0.01 & 0.00 & 0.01 & 0.01 & 0.00 \\
CeII &   1 & 0.00 & 0.02 & \nodata & 0.02 & 0.02 & 0.00 & \nodata & 0.00 & 0.00 & 0.00 & \nodata & -0.00 & -0.01 & 0.00 & \nodata & 0.00 & 0.01 \\
NdII &   2 & 0.01 & 0.02 & \nodata & 0.03 & 0.02 & 0.00 & \nodata & 0.00 & 0.00 & -0.00 & \nodata & -0.00 & -0.01 & 0.00 & \nodata & 0.00 & 0.01 \\
SmII &   1 & 0.00 & 0.02 & \nodata & 0.02 & 0.02 & 0.00 & \nodata & 0.00 & 0.00 & 0.00 & \nodata & -0.00 & -0.00 & 0.00 & \nodata & 0.01 & 0.01 \\
EuII &   1 & 0.00 & 0.02 & 0.03 & 0.03 & 0.02 & 0.00 & 0.00 & 0.00 & 0.00 & 0.00 & -0.00 & -0.00 & -0.00 & 0.00 & 0.01 & 0.01 & 0.00 \\
  \enddata
\tablenotetext{}{(1): KIC5439372 ($[\mathrm{Fe/H}]=-2.53$), (2): KIC12017985 ($[\mathrm{Fe/H}]=-1.87$), (3): KIC11566038 ($[\mathrm{Fe/H}]=-1.36$). For the elemental abundances we show the results for [X/H].}
\tablenotetext{a}{\citet{Karovicova2018a} and \citet{Creevey2019a}.}
\tablenotetext{b}{Derived in this study. We consider the effects of the uncertainties of the other parameters.}
\tablenotetext{c}{The number of lines used and the scatter among abundances derived from individual lines.} 
\end{deluxetable}

\begin{deluxetable}{l*{18}{r}}
  \tablecaption{Change of the stellar parameters/abundances due to the stellar parameters of KIC9583607\label{tab:B2}}
  \tablehead{ & & & \multicolumn{4}{c}{$\Delta T_{\rm eff}=89\,\mathrm{K}$\tablenotemark{a}}   &  \multicolumn{4}{c}{$\Delta\log g =0.010$\tablenotemark{b}} &  \multicolumn{4}{c}{$\Delta\xi =0.09$\tablenotemark{b}} &  \multicolumn{4}{c}{$\Delta[\mathrm{Fe/H}] =0.065$\tablenotemark{b}} \\
              & \colhead{$n_{\rm line}$\tablenotemark{c}} & \colhead{$\sigma_{\rm sct}$\tablenotemark{c}} & \colhead{ref} & \colhead{(1)} & \colhead{(2)} & \colhead{(3)} & \colhead{ref} & \colhead{(1)} & \colhead{(2)} & \colhead{(3)} & \colhead{ref}& \colhead{(1)} & \colhead{(2)} & \colhead{(3)} & \colhead{ref}& \colhead{(1)} & \colhead{(2)} & \colhead{(3)}
            }
  \startdata
$T_{\rm teff}$   & & & 89 & 85 & 84 & 79 & 0 & 5 & 1 & 9 & 0 & 6 & -5 & -2 & 0 & -1 & 0 & -5 \\
$\log g$         & & & 0.00 & 0.00 & 0.00 & 0.00 & 0.01 & 0.00 & 0.00 & 0.00 & 0.00 & 0.00 & 0.00 & 0.00 & 0.00 & 0.00 & 0.00 & 0.00 \\
$\xi$            & & & 0.00 & 0.00 & 0.02 & -0.01 & 0.00 & 0.01 & 0.00 & 0.01 & 0.09 & 0.08 & 0.07 & 0.07 & 0.00 & 0.00 & -0.00 & -0.02 \\
$[\mathrm{Fe/H}]$& & & 0.00 & 0.00 & 0.00 & 0.00 & 0.00 & 0.00 & 0.00 & 0.00 & 0.00 & 0.01 & 0.02 & 0.03 & 0.06 & 0.07 & 0.06 & 0.06 \\
CH   &   1 & 0.00 & 0.07 & 0.06 & 0.10 & 0.10 & 0.00 & 0.00 & 0.00 & 0.01 & -0.00 & 0.01 & 0.01 & 0.01 & 0.04 & 0.06 & 0.04 & 0.03 \\
OI   &   1 & 0.00 & 0.02 & 0.02 & 0.04 & \nodata & 0.00 & 0.00 & 0.00 & \nodata & -0.00 & 0.00 & 0.00 & \nodata & 0.02 & 0.02 & 0.02 & \nodata \\
NaI  &   1 & 0.00 & 0.06 & 0.08 & 0.05 & 0.04 & -0.00 & 0.00 & 0.00 & 0.01 & -0.02 & -0.02 & -0.01 & -0.00 & -0.00 & 0.00 & -0.00 & -0.00 \\
MgI  &   5 & 0.16 & 0.04 & 0.05 & 0.04 & 0.03 & 0.00 & 0.00 & 0.00 & 0.01 & -0.00 & 0.01 & 0.01 & -0.00 & 0.00 & 0.00 & -0.00 & -0.00 \\
SiI  &   4 & 0.01 & 0.03 & 0.02 & 0.04 & 0.03 & 0.00 & 0.00 & 0.00 & 0.00 & -0.01 & -0.00 & -0.01 & 0.00 & 0.00 & 0.01 & 0.00 & -0.00 \\
CaI  &  25 & 0.14 & 0.08 & 0.09 & 0.08 & 0.06 & 0.00 & 0.00 & 0.00 & 0.01 & -0.03 & -0.02 & -0.02 & -0.00 & -0.01 & -0.01 & -0.01 & -0.01 \\
ScII &   9 & 0.07 & -0.01 & -0.00 & 0.01 & 0.00 & 0.00 & -0.00 & 0.00 & 0.00 & -0.03 & -0.02 & -0.01 & 0.01 & 0.01 & 0.01 & 0.01 & 0.01 \\
TiI  &  41 & 0.17 & 0.13 & 0.15 & 0.12 & 0.11 & 0.00 & 0.00 & 0.00 & 0.01 & -0.02 & -0.01 & -0.01 & 0.01 & -0.00 & -0.00 & -0.01 & -0.01 \\
TiII &  12 & 0.17 & 0.00 & 0.01 & 0.02 & 0.01 & 0.00 & -0.00 & 0.00 & 0.00 & -0.02 & -0.02 & 0.00 & 0.01 & 0.01 & 0.02 & 0.02 & 0.01 \\
VI   &  10 & 0.16 & 0.13 & 0.14 & 0.12 & 0.11 & 0.00 & 0.01 & 0.00 & 0.01 & -0.00 & 0.01 & -0.00 & 0.01 & -0.00 & 0.00 & -0.01 & -0.01 \\
CrI  &   9 & 0.07 & 0.13 & 0.14 & 0.12 & 0.11 & 0.00 & -0.00 & 0.00 & 0.01 & -0.06 & -0.05 & -0.03 & -0.01 & -0.01 & -0.01 & -0.01 & -0.01 \\
CrII &  11 & 0.24 & -0.04 & -0.04 & -0.01 & -0.03 & 0.00 & 0.00 & 0.00 & -0.00 & -0.03 & -0.03 & 0.00 & 0.00 & 0.01 & 0.01 & 0.01 & 0.01 \\
MnI  &   6 & 0.07 & 0.10 & 0.11 & 0.09 & 0.10 & 0.00 & 0.00 & 0.00 & 0.01 & -0.04 & -0.02 & 0.00 & 0.01 & -0.01 & -0.00 & -0.00 & -0.01 \\
FeI  & 121 & 0.14 & 0.10 & 0.10 & 0.10 & 0.09 & 0.00 & 0.00 & 0.00 & 0.01 & -0.03 & -0.02 & -0.01 & -0.00 & -0.00 & 0.00 & -0.00 & -0.01 \\
FeII &  23 & 0.10 & -0.03 & -0.03 & -0.01 & -0.01 & 0.00 & -0.00 & 0.00 & -0.00 & -0.03 & -0.02 & 0.00 & -0.01 & 0.01 & 0.02 & 0.02 & 0.01 \\
CoI  &   3 & 0.14 & 0.10 & 0.11 & 0.10 & 0.10 & 0.00 & 0.00 & 0.00 & 0.01 & 0.00 & 0.01 & 0.00 & 0.00 & 0.00 & 0.01 & -0.01 & -0.01 \\
NiI  &  32 & 0.15 & 0.10 & 0.10 & 0.10 & 0.10 & 0.00 & 0.00 & 0.00 & 0.01 & -0.03 & -0.01 & -0.01 & 0.00 & -0.00 & 0.01 & -0.01 & -0.01 \\
CuI  &   1 & 0.00 & 0.12 & 0.11 & 0.11 & 0.10 & 0.00 & 0.00 & 0.00 & 0.01 & -0.03 & -0.01 & -0.01 & -0.01 & -0.00 & 0.01 & -0.00 & -0.01 \\
ZnI  &   2 & 0.01 & -0.00 & -0.01 & 0.02 & 0.01 & 0.00 & -0.01 & 0.00 & 0.00 & -0.04 & -0.03 & -0.01 & -0.00 & 0.01 & 0.01 & 0.01 & 0.01 \\
YII  &   6 & 0.09 & 0.01 & 0.02 & 0.02 & 0.01 & 0.00 & -0.00 & 0.00 & 0.00 & -0.06 & -0.05 & -0.00 & -0.00 & 0.01 & 0.02 & 0.02 & 0.02 \\
BaII &   3 & 0.09 & 0.02 & 0.03 & 0.04 & 0.04 & 0.00 & -0.00 & 0.00 & 0.01 & -0.07 & -0.05 & -0.04 & -0.06 & 0.01 & 0.02 & 0.03 & 0.00 \\
LaII &   3 & 0.04 & 0.02 & 0.02 & 0.04 & 0.03 & 0.00 & 0.00 & 0.00 & 0.01 & -0.00 & -0.00 & 0.00 & 0.01 & 0.02 & 0.02 & 0.02 & 0.01 \\
CeII &   4 & 0.05 & 0.02 & 0.02 & 0.03 & 0.03 & 0.00 & -0.00 & 0.00 & 0.00 & -0.01 & -0.02 & 0.01 & 0.00 & 0.02 & 0.02 & 0.02 & 0.01 \\
NdII &   4 & 0.02 & 0.03 & 0.04 & 0.05 & 0.04 & 0.00 & -0.00 & 0.00 & 0.01 & -0.01 & -0.03 & 0.00 & 0.00 & 0.02 & 0.02 & 0.02 & 0.01 \\
SmII &   1 & 0.00 & 0.02 & 0.03 & 0.04 & 0.03 & 0.00 & 0.00 & 0.00 & 0.01 & -0.01 & -0.01 & 0.00 & 0.00 & 0.02 & 0.02 & 0.02 & 0.01 \\
EuII &   3 & 0.06 & -0.00 & 0.00 & 0.02 & 0.00 & 0.00 & 0.00 & 0.00 & 0.00 & -0.00 & 0.00 & 0.01 & 0.01 & 0.01 & 0.02 & 0.02 & 0.01 \\
  \enddata
\tablecomments{(1): KIC8350894 ($[\mathrm{Fe/H}]=-0.85$), (2): KIC11566038 ($[\mathrm{Fe/H}]=-1.36$), (3): KIC12017985 ($[\mathrm{Fe/H}]=-1.87$)}
\tablenotetext{a}{From APOGEE DR14 catalog.}
\tablenotetext{b}{Derived in this study. We consider the effects of the uncertainties of the other parameters. For the elemental abundances we show the results for [X/H].}
\tablenotetext{c}{The number of lines used and the scatter among abundances derived from individual lines.} 
\end{deluxetable}

We re-carry out analyses for a subset of stars adopting stellar parameters of the standard stars with offsets that correspond to their uncertainties (Tables \ref{tab:B1} \& \ref{tab:B2}).
We select three stars as the subset for each reference star to cover the wide metallicity range.
It is clear that the absolute scale does depend on the adopted stellar parameters of the standard star.
However, we emphasize that since the change is systematic, we are almost free from this effect as long as we discuss abundance trends among our sample. 
These effects are only important when one tries to compare our results with other studies.

Tables \ref{tab:B1} and \ref{tab:B2} also include the number of lines used to derive the abundance of the reference star and the scatter of abundances derived from individual lines.
Readers may derive uncertainties of the abundance of the reference star by quadratically summing $\sigma_{\rm sct}/\sqrt{N_{\rm line}}$ and shifts in abundances caused by those in stellar parameters (``ref'' columns).
Since the abundance of the reference star determines our abundance scale, readers may add the obtained values to the reported uncertainties in Table \ref{tab:abundance} for a comparison purpose.

\section{Chemical abundances of red giant branch and red clump stars\label{sec:RC}}

\begin{figure*}
\plotone{\figureloc 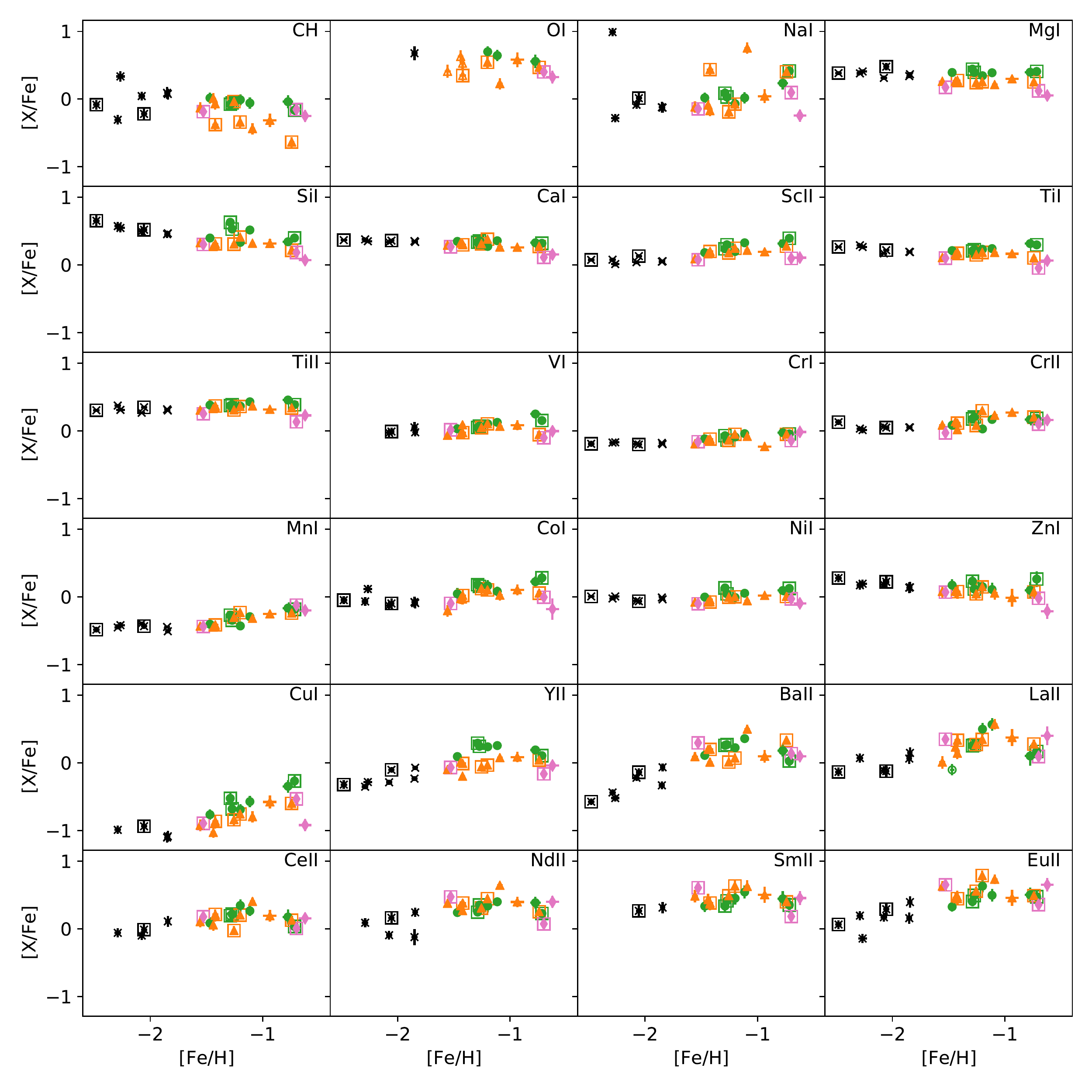}
\caption{Same as Figures~\ref{fig:abtrend0}--\ref{fig:abtrend4} but with red clump stars and luminous ($\log g<2.0$) stars being marked with open squares.\label{fig:abRC}}
\end{figure*}
\begin{figure*}
\plotone{\figureloc 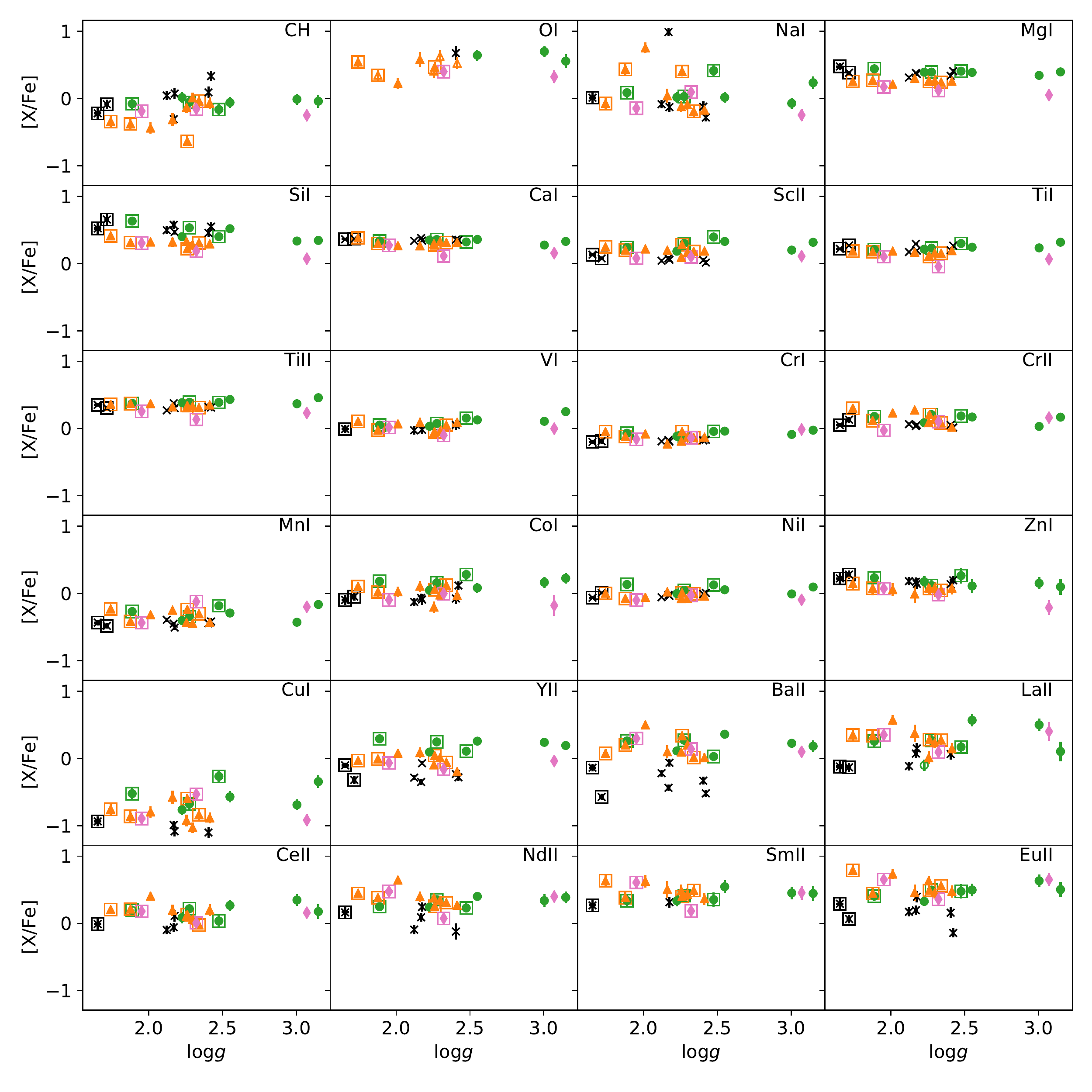}
\caption{Same as Figure~\ref{fig:abRC} but plotted against $\log g$. \label{fig:abRClogg}}
\end{figure*}

\begin{deluxetable*}{lrrrrrrrrrrr}
\tablecaption{Average [X/Fe] for groups of stars\label{tab:abRC}}
\tablehead{
 & \multicolumn{2}{c}{All} & & \multicolumn{2}{c}{Metal-poor} & & \multicolumn{2}{c}{high-$\alpha$} & & \multicolumn{2}{c}{low-$\alpha$} \\
Elem & Group 1\tablenotemark{a} & Group 2\tablenotemark{a} & & Group 1 & Group 2 & & Group 1 & Group 2 & & Group 1 & Group 2 }
\startdata
CH & -0.212 & -0.069 & &-0.151 & 0.048 & &-0.098 & -0.023 & &-0.348 & -0.188 \\
OI & 0.441 & 0.528 & &\nodata & 0.677 & &\nodata & 0.634 & &0.454 & 0.475 \\
NaI & 0.108 & 0.050 & &0.013 & 0.075 & &0.178 & 0.051 & &0.144 & 0.083 \\
MgI & 0.312 & 0.312 & &0.431 & 0.361 & &0.415 & 0.381 & &0.257 & 0.258 \\
SiI & 0.407 & 0.383 & &0.589 & 0.509 & &0.521 & 0.399 & &0.313 & 0.306 \\
CaI & 0.308 & 0.313 & &0.364 & 0.353 & &0.338 & 0.328 & &0.317 & 0.292 \\
ScII & 0.202 & 0.149 & &0.104 & 0.048 & &0.312 & 0.257 & &0.227 & 0.170 \\
TiI & 0.172 & 0.200 & &0.243 & 0.224 & &0.245 & 0.252 & &0.153 & 0.162 \\
TiII & 0.326 & 0.342 & &0.329 & 0.316 & &0.384 & 0.410 & &0.346 & 0.336 \\
VI & 0.027 & 0.050 & &-0.008 & 0.003 & &0.093 & 0.130 & &0.020 & 0.026 \\
CrI & -0.117 & -0.129 & &-0.195 & -0.180 & &-0.082 & -0.067 & &-0.088 & -0.152 \\
CrII & 0.139 & 0.106 & &0.090 & 0.044 & &0.190 & 0.114 & &0.176 & 0.150 \\
MnI & -0.314 & -0.371 & &-0.457 & -0.442 & &-0.266 & -0.321 & &-0.296 & -0.375 \\
CoI & 0.062 & -0.001 & &-0.072 & -0.048 & &0.207 & 0.130 & &0.076 & -0.025 \\
NiI & 0.005 & -0.020 & &-0.028 & -0.023 & &0.103 & 0.036 & &-0.017 & -0.046 \\
ZnI & 0.138 & 0.097 & &0.253 & 0.167 & &0.205 & 0.134 & &0.086 & 0.058 \\
CuI & -0.688 & -0.821 & &-0.936 & -1.058 & &-0.490 & -0.593 & &-0.763 & -0.842 \\
YII & -0.006 & -0.040 & &-0.211 & -0.244 & &0.215 & 0.196 & &-0.015 & -0.024 \\
BaII & 0.082 & 0.020 & &-0.355 & -0.314 & &0.186 & 0.220 & &0.156 & 0.180 \\
LaII & 0.193 & 0.212 & &-0.127 & 0.043 & &0.231 & 0.267 & &0.308 & 0.266 \\
CeII & 0.115 & 0.150 & &-0.010 & -0.015 & &0.149 & 0.219 & &0.130 & 0.192 \\
NdII & 0.293 & 0.280 & &0.164 & 0.030 & &0.277 & 0.342 & &0.348 & 0.406 \\
SmII & 0.407 & 0.451 & &0.267 & 0.315 & &0.368 & 0.445 & &0.477 & 0.483 \\
EuII & 0.457 & 0.410 & &0.178 & 0.156 & &0.458 & 0.490 & &0.571 & 0.551 \\
\hline \# of stars & 11 & 15 && 2& 5 &&3&4&&4&5 \\
\enddata
\tablenotetext{a}{Group 1 is a group of red clump stars and luminous ($\log g<2.0$) red giant branch stars and Group 2 is a group of less luminous ($\log g>2.0$) red giant stars.} 
\end{deluxetable*}

Since while the whole sample is used to discuss chemical abundance, only a subsample is used for the interpretation of asteroseismic data, we here verify that the evolutionary status does not affect the discussion.
There is no astrophysical reason that the evolutionary status affects abundances of most of the elements.
The exceptional elements are light elements, for which the surface abundances can be altered because of internal nucleosynthesis and the mixing.
This could be the case for C and O among the elements investigated in the present study.

Another concern is that the systematic uncertainties in the abundance analysis act differently between red giant branch stars and red clump stars.
For example, \citet{Masseron2017a} reported that the offset between spectroscopic and asteroseismic $\log g$ varies between the two classes of stars.
We note, however, that the measurements by \citet{Takeda2015a} and \citet{Takeda2016a} do not show such difference.
This different result might be due to the different methods for stellar parameter determination; while the spectroscopic stellar parameters used in \citet{Masseron2017a} are derived by globally fitting synthetic spectra to the observed ones, \citet{Takeda2015a} and \citet{Takeda2016a} derived based on equivalent widths of iron lines.
Our approach is more similar to that adopted by \citet{Takeda2015a} and \citet{Takeda2016a}.
We also note that red giant branch stars and red clump stars have almost identical spectra except for the regions that are sensitive to light elements abundances \citep[e.g.,][]{Hawkins2018b}.

Nonetheless, it is worth an investigation to verify that the sample of red clump stars and luminous giants yields a consistent result as less luminous giants before combining the results for the discussion.
Figures~\ref{fig:abRC} and \ref{fig:abRClogg} show abundance trends with metallicity and $\log g$, respectively.
Red clump stars and stars with $\log g<2.0$ are marked with open squares. 
There is no obvious trend between abundances and evolutionary status. 
Table~\ref{tab:abRC} quantifies the consistency of the results.
It shows average [X/Fe] for stars in different evolutionary status and in different sub-samples, and should be compared to Table \ref{tab:absignificance}. 
The abundance differences found in Section \ref{sec:dis_pop}, such as low [$\alpha$/Fe], low [Cu/Fe], and low [Y/Fe] of the low-$\alpha$ populations are confirmed both in a sample of red clump and luminous giants and that of less luminous giants.
These results ensure us to combine stars in both evolutionary status for the discussion of chemical abundances without unnecessarily reducing the sample size.

\bibliographystyle{aasjournal}

\end{document}